\documentclass[aip,amsmath,amssymb,floatfix,citeautoscript,reprint]{revtex4-2}
\usepackage{cancel}
\usepackage{amsmath}
\usepackage{graphicx}
\usepackage{bm}
\usepackage{physics}
\usepackage[version=3]{mhchem}
\usepackage{rotating}

\usepackage{txfonts}

\usepackage{color}
\usepackage[usenames,dvipsnames]{xcolor}
\definecolor{myblue}{rgb}{0,0,1}
\newcommand{\commentthisout}[1]{}
\usepackage[breaklinks=true,colorlinks=true,linkcolor=myblue,urlcolor=myblue,citecolor=myblue]{hyperref}

\begin{document}

\title{Full Configuration Interaction Excited-State Energies in Large Active Spaces
from Subspace Iteration with Repeated Random Sparsification}

%\SectionNumbersOn

\author{Samuel M. Greene}
\affiliation{Department of Chemistry, Columbia University, New York, New York 10027, United States}
\author{Robert J. Webber}
\altaffiliation{Present address: Computing and Mathematical Sciences, California Institute of Technology, Pasadena, California 91125, United States}
\affiliation{Courant Institute of Mathematical Sciences, New York University, New York, New York 10012, United States}
\author{James E. T. Smith}
\affiliation{Center for Computational Quantum Physics, Flatiron Institute, New York, New York 10010, United States}
\author{Jonathan Weare}
\email{weare@nyu.edu}
\affiliation{Courant Institute of Mathematical Sciences, New York University, New York, New York 10012, United States}
\author{Timothy C. Berkelbach}
\email{tim.berkelbach@gmail.com}
\affiliation{Department of Chemistry, Columbia University, New York, New York 10027, United States}
\affiliation{Center for Computational Quantum Physics, Flatiron Institute, New York, New York 10010, United States}

%\begin{tocentry}
%
%\includegraphics[width=2in]{Figures/TOC_square.pdf}
%
%\end{tocentry}

\begin{abstract}
We present a stable and systematically improvable quantum Monte Carlo (QMC) approach to calculating excited-state energies, which we implement using our fast randomized iteration method for the full configuration interaction problem (FCI-FRI).
Unlike previous excited-state quantum Monte Carlo methods, our approach, which is based on an asymmetric variant of subspace iteration, avoids the use of dot products of random vectors and instead relies upon trial vectors to maintain orthogonality and estimate eigenvalues.
%The initiator approximation and our previously developed algorithm for sampling excitations are fully compatible with excited-state FCI-FRI.
By leveraging recent advances, we apply our method to calculate ground- and excited-state energies of challenging molecular systems in large active spaces, including the carbon dimer with 8 electrons in 108 orbitals (8e,108o), an oxo-Mn(salen) transition metal complex (28e,28o), ozone (18e,87o), and butadiene (22e,82o).
In the majority of these test cases, our approach yields total excited-state energies that agree with those from state-of-the-art methods---including heat-bath CI, the density matrix renormalization group approach, and FCIQMC---to within sub-milliHartree accuracy. 
%at modest computational cost 
In all cases, estimated excitation energies agree to within about 0.1~eV.
\end{abstract}

\maketitle

\section{Introduction}
\label{sec:Intro}
Excited electronic states of molecules and solid-state systems play a crucial role in determining their chemical properties, including their response to irradition by light~\cite{Mennucci2010simulation, Beard2019comparative, Gonzalez2020quantum, SerranoAndres2005, Navarrete-Miguel2019, Park2017, Subotnik2016understanding, Barbatti2011nonadiabatic} and their behavior at finite temperature~\cite{Harsha2019thermofield, Zhang2021finite}.
%First-principles calculations of low-energy excited states in systems of interacting electrons are used ubiquitously in chemical simulations.
%For example, such calculations can be used to predict spectra, simulate photochemical reactions, and calculate finite-temperature properties~\cite{}.
But calculating excited-state properties from first principles often proves challenging, particularly for systems of strongly correlated electrons.
%calculating any eigenstates of electronic systems can prove challenging in practice.
Rigorous full configuration interaction (FCI) calculations for such systems require the treatment of a number of electronic configurations that scales combinatorially with the number of electrons and the number of orbitals they may occupy~\cite{Fulde2017dealing, Laughlin2000the, Zhang2004}, which has prevented their application to all but the smallest chemical systems.
This has led to the development of active space techniques, such as complete active space configuration interaction (CASCI) approach or its orbital-optimized extension via a self-consistent field (CASSCF), which are limited to systems containing about 20 electrons occupying 20 spatial orbitals~\cite{Vogiatzis2017}.
Larger active spaces can be accurately treated with the density matrix renormalization group (DMRG) and selected CI methods, which can be used to calculate excited states of systems containing about 50 active orbitals~\cite{OlivaresAmaya2015, Sharma2017, Loos2020}.
More detailed active space selection criteria have also enabled the treatment of larger problems using, for example, restricted~\cite{Malmqvist1990}, generalized~\cite{Ma2011}, and localized~\cite{Hermes2020} active spaces.

Quantum Monte Carlo (QMC) methods enable the efficient treatment of electronic correlation in many chemical systems for which conventional methods are too expensive or unreliable~\cite{Foulkes2001, Zhang2004, Wagner2016}.
QMC methods leverage stochastic sampling to iteratively evolve a quantum state toward the minimum of an energy functional in a specified manifold~\cite{Booth2009, Wouters2014projector, Umrigar2015, Schwarz2017, Han2020}.
%Variational Monte Carlo approaches yield an explicit representation of the state at this energy minimum, while in projector Monte Carlo approaches, the state is represented only on average over multiple iterations~\cite{Umrigar2015}.
Because QMC is fundamentally a stochastic minimization procedure, it is used routinely to calculate ground states, which can be construed as solutions to an energy minimization problem~\cite{Foulkes2001}. 
%but calculating excited-state properties is a significant challenge.
In some cases, this procedure can be straightforwardly extended to calculate excited-state properties if symmetry considerations are used to exclude the ground state from the manifold of possible solutions~\cite{Booth2011, Zhang2018}.
The use of nodal and phase constraints in diffusion Monte Carlo (DMC)~\cite{Grimes1986, Wagner2016, Scemama2018excitation} and auxiliary-field QMC (AFQMC)~\cite{Purwanto2009excited}, respectively, can be understood as variations of this strategy, as can the orbital transformations employed in the graphical unitary group approach~\cite{Dobrautz2019}.
But performing accurate, robust calculations of excited states without relying upon such constraints remains a significant, open challenge.
Addressing this challenge in the context of existing QMC methods could extend their applicability to a wider variety of important problems in chemistry and physics.
%Other strategies involve minimizing a different functional with an excited state at its minimum~\cite{Umrigar1988, Booth2012, Shea2017}.

This paper describes a QMC method for calculating multiple excited states with the same symmetry in a discrete CI basis of electron configurations, like in the FCIQMC family of methods~\cite{Booth2009, Guther2020}.
%The fact that we cannot leverage symmetry considerations makes this challenging due to the need to enforce orthogonality among multiple vectors as optimization proceeds.
%Without such orthogonality constraints, all vectors would become increasingly linearly dependent and eventually converge to the ground-state eigenvector.
The fundamental challenge is that, without imposing any orthogonality constraints, the vectors from each iteration become increasingly linearly dependent and eventually converge to the ground-state eigenvector.
Previous discrete-space approaches have addressed this challenge in different ways.
%In the linear method used in variational Monte Carlo, Hamiltonian eigenstates are calculated in the basis of linear perturbations to a trial function's parameters~\cite{Zimmerman2009, Filippi2009, Feldt2020}.
The first QMC methods addressed this challenge by applying Gram-Schmidt orthogonalization to the random vectors as the iteration proceeds~\cite{Ohtsuka2010, Blunt2015}.
Others generated a random sequence of Krylov vectors starting from the ground state and then orthogonalized them using a canonical L\"owdin procedure~\cite{Blunt2015a}.
Both of these approaches rely upon dot products of random vectors, which yield statistical errors that can scale unfavorably with the dimension of the problem~\cite{Blunt2015a, Greene2022}.
For instance, the variance in the dot product of two vectors each constructed by randomly sparsifying the vector $v = (1, ..., 1) d^{-1/2}$ is directly proportional to the dimension $d$.
Thus, while such approaches may yield acceptable accuracy for systems with eigenvectors that have relatively few elements with large magnitudes, they may fail when more elements have significant magnitudes.
Another QMC approach avoids orthogonalization altogether, starting from initial approximate eigenvectors, evolving them all toward the ground state, and considering only energy estimates obtained before they become too linearly dependent~\cite{Blunt2018nonlinear}.
Because this approach is unstable, systematically improving its results may prove challenging.
These various issues can complicate the application of these existing excited-state methods to larger, more challenging chemical systems.

Motivated by these challenges, we recently introduced an alternative, general stochastic approach to calculating matrix eigenvalues, referred to as subspace iteration with repeated random sparsification~\cite{Greene2022}.  
Introducing randomness into standard subspace iteration
presents issues related to maintaining orthogonality among multiple vectors and estimating energy eigenvalues as the iteration proceeds.
These necessitate careful algorithmic choices that are described in detail in ref~\citenum{Greene2022}.
In particular, we choose to randomize a nonstandard, asymmetric subspace iteration.
%While it can be applied in principle to projector Monte Carlo approaches in either a continuous or discrete basis of electronic configurations, we focus here on its application in discrete space.
Unlike previous approaches, our randomized subspace iteration is stable and avoids the use of dot products of random vectors.
We use approximate eigenvectors (i.e. trial vectors) to estimate energies and maintain orthogonality.
These modifications constitute changes to the underlying stochastic dynamics, unlike other excited-state QMC approaches that rely upon symmetry or nodal constraints.
%Consequently, they may also inform the development of new continuous-space QMC methods, e.g. based on DMC or AFQMC, that do not rely upon such constraints to target excited states.
In ref~\citenum{Greene2022}, we applied our approach to the full configuration interaction problem using a simple implementation of our fast randomized iteration approach (FCI-FRI)~\cite{Lim2017,Greene2019,Greene2020} to accelerate matrix-vector multiplications.
%~\cite{Greene2019}, which can be used to calculate eigenvalues of the full configuration interaction or CASCI Hamiltonian matrix.
FCI-FRI is a stochastic implementation of an iterative linear algebra scheme involving sequential matrix-vector multiplication operations.
Stochastic sampling is used to impose sparsity in vectors and matrices, thereby enabling the use of sparsity-based strategies for reducing the computational cost of these operations.
In ref~\citenum{Greene2022}, we approximated the ground- and excited-state energies of three small molecular systems to high accuracy, but the simple flavor of FCI-FRI that we used prevented us from studying larger systems.

In this work, we apply several strategies that enable applications to larger, more challenging molecular systems~\cite{Greene2020}.
%We introduced several strategies for reducing the cost of ground-state FCI-FRI calculations in ref \citenum{Greene2020}.
%One purpose of this work is to evaluate whether incorporating these into our excited-state FCI-FRI method enables its application to larger chemical systems.
In particular, we focus on specific strategies for further reducing the computational cost and statistical error incurred when multiplying sparse vectors by the Hamiltonian matrix.
We apply the initiator approximation, originally developed for the full configuration interaction quantum Monte Carlo (FCIQMC) method~\cite{Cleland2010, Booth2011}, and our ``unnormalized'' modification to the heat-bath Power-Pitzer scheme for factoring the Hamiltonian matrix~\cite{Neufeld2019, Greene2020}.
Additionally, we employ two strategies designed to further reduce the cost and statistical error of this approach for larger systems: we use a state-of-the-art selected configuration interaction method~\cite{Holmes2017} to calculate accurate trial vectors, and we use a basis of spin-coupled functions instead of Slater determinants to reduce the effective dimension of the Hamiltonian matrix, as is commonly done in other methods~\cite{Booth2011, Holmes2016, Holmes2017,Dobrautz2019}.
We apply our method with these extensions to challenging chemical systems of correlated electrons and assess its accuracy through comparisons with energies calculated using state-of-the-art methods.
To our knowledge, these represent the largest calculations to date of excited states in the same symmetry class as the ground state with discrete-space QMC.
In choosing these examples, our main objective is to illustrate the scalability and accuracy of our new excited-state method.
More systematic comparisons to state-of-the-art methods will be left to future publications.
We refer to our excited-state scheme as FCI-FRI, but we emphasize that the general, randomized subspace iteration on which it is based~\cite{Greene2022} can be applied with other QMC techniques, including FCIQMC.

The remainder of this paper is organized as follows.
Section \ref{sec:methods} introduces each of the methodological aspects of our approach, including an overview of our approach to calculating excited-state energies and discussions of each of the methodological extensions described above.
Section \ref{sec:results} presents the numerical results from our applications to chemical systems, and Section \ref{sec:conclusions} summarizes our key findings and discusses possible future directions.

\section{Methods}
\label{sec:methods}
\subsection{The Configuration Interaction Hamiltonian Matrix}
We focus on the calculation of eigenvalues of a matrix representation $\mathbf{H}$ of the Hamiltonian operator for $N$ interacting electrons in a discrete many-particle basis constructed from $M$ single-particle orbitals.
%Although Hartree-Fock orbitals are commonly used, we use different orbitals in this work, as will be discussed in further detail below.
In contrast to our previous FCI-FRI papers, in which we used a discrete basis of Slater determinants, here we instead use spin-coupled functions~\cite{Booth2011, Holmes2016}.
This imposes a block-diagonal structure on $\mathbf{H}$, effectively reducing the dimension of the eigenproblem and thereby reducing the computational cost of our method.
Each block contains only eigenstates with a particular spin parity, i.e. for which the spin $S$ is either even or odd.
The number of spin-coupled functions for a given spin parity, denoted generically as $N_\text{FCI}$, scales as $\mathcal{O}(M \text{ choose } N)$.
The number of nonzero elements in each column of $\mathbf{H}$, which determines the cost of performing sparse matrix-vector multiplication operations, scales as $\mathcal{O}(N^2 M^2)$.
Definitions of spin-coupled functions and formulas for elements of $\mathbf{H}$ in this basis are provided in Appendix \ref{sec:SCFHamil}.
For some systems, we leverage point-group symmetry to impose additional block-diagonal structure and further reduce the effective dimension.
One-particle orbitals and their associated symmetry labels and Hamiltonian matrix elements were calculated using the PySCF software~\cite{Sun2018}.

\subsection{FCI-FRI for Excited States}
\label{sec:subspIter}
This section describes the application of subspace iteration~\cite{Stewart1969, Stewart1975, Saad2011} to calculate the $N_\text{eigen}$ lowest-energy eigenvalues in a single block of $\mathbf{H}$~\cite{Greene2022}.
%It is based on a deterministic algorithm known as subspace iteration~\cite{Stewart1969, Stewart1975, Saad2011}.
Introducing randomness into standard subspace iteration
%as is done here within the FRI framework~\cite{Lim2017}, 
presents issues related to maintaining orthogonality among multiple vectors 
%as the iteration proceeds 
and estimating energy eigenvalues.
These necessitate algorithmic modifications, which are summarized here.
Further details and mathematical analysis are provided in our prior work~\cite{Greene2022}.
These general principles are applicable to any method for stochastically performing matrix-vector multiplication, including FCIQMC.

%We first present the method in its deterministic (non-stochastic) form, without any of the additional features we will use in this work to enable its application to larger chemical systems.
In each iteration $\tau$, we calculate a $N_\text{FCI} \times N_\text{eigen}$ iterate matrix $\mathbf{X}^{(\tau)}$, whose columns each approximate an eigenvector of $\mathbf{H}$.
Starting from an initial matrix $\mathbf{X}^{(0)} = \mathbf{U}$ of orthonormal, sparse trial vectors obtained from an approximate quantum chemistry method, subsequent iterates are constructed via matrix multiplication~\cite{Stewart1969, Stewart1975}:
\begin{equation}
\label{eq:subspIter}
\mathbf{X}^{(\tau + 1)} = \left( \mathbf{1} - \varepsilon \mathbf{H} \right) \mathbf{X}^{(\tau)} \left[ \mathbf{G}^{(\tau)} \right]^{-1}
\end{equation}
where $\varepsilon$ is a small, positive number, $\mathbf{1}$ is the identity matrix, and $\mathbf{G}^{(\tau)}$ is chosen to approximately enforce orthonormality among columns of iterates, as described in Appendix \ref{sec:orthonormalization}.
For $N_\text{eigen}=1$, this is equivalent to power iteration, which underlies many discrete-space QMC methods.
If $\varepsilon$ is sufficiently small in magnitude, the iterates will converge in the absence of statistical error to the space of the $N_\text{eigen}$ lowest-energy eigenvectors as $\tau \to \infty$~\cite{wilkinson1965convergence}.
In order to ensure memory efficiency of this approach, the iterates are represented in a sparse format.
A stochastic procedure that maintains sparsity but introduces randomness is used to calculate the matrix product $\left( \mathbf{1} - \varepsilon \mathbf{H} \right) \mathbf{X}^{(\tau)}$, as will be described below.

In our randomized method, we use averaging to obtain eigenvalue estimates, as the random iterates only represent the lowest-energy eigenvectors on average.
At regular intervals (every 100 iterations in this work, vide infra), we evaluate and store the small $N_\text{eigen} \times N_\text{eigen}$ matrices $\mathbf{U}^\text{T} \mathbf{HX}^{(\tau)}$ and $\mathbf{U}^\text{T} \mathbf{X}^{(\tau)}$.
Denoting averages of these matrices as $\langle \mathbf{U}^\text{T} \mathbf{HX}^{(\tau)} \rangle_\tau$ and $\langle \mathbf{U}^\text{T} \mathbf{X}^{(\tau)} \rangle_\tau$, we solve the generalized eigenvalue problem
\begin{equation}
\label{eq:genEval}
\langle \mathbf{U}^\text{T} \mathbf{HX}^{(\tau)} \rangle_\tau \mathbf{W} = \langle \mathbf{U}^\text{T} \mathbf{X}^{(\tau)} \rangle_\tau \mathbf{W} \mathbf{\Lambda}
\end{equation}
to obtain a diagonal matrix $\mathbf{\Lambda}$ of eigenvalue estimates.
Eq \eqref{eq:genEval} can be understood as a generalization of the projected energy estimator commonly used in other QMC methods~\cite{Foulkes2001, Booth2009} to multiple eigenvalues.
A related eigenvalue estimator is also used in the QMC method first proposed in ref \citenum{Ceperley1988}, although in that method elements of the analogous matrices are evaluated by analyzing correlations within a single trajectory; here, we instead use multiple orthogonal trajectories.
Initial iterations are excluded from the averages in eq \eqref{eq:genEval} in order to ensure sufficient equilibration.
%In analogy to our ground-state FCI-FRI methods, we find it useful to calculate and store (in sparse form) the matrix product $\mathbf{HU}$ at the beginning of each simulation to efficiently calculate the matrices $\mathbf{U}^\text{T} (\mathbf{1} - \varepsilon \mathbf{H} ) \mathbf{X}^{(\tau)}$ and $\mathbf{U}^\text{T} \mathbf{H} \mathbf{X}^{(\tau)}$.
This particular approach to estimating eigenvalues was chosen to mitigate statistical biases arising from nonlinearities in the eigenvalue equation~\cite{Greene2022}.
In particular, eigenvalue estimates are exact in two limiting cases: with infinitely many samples (in which case $\langle \mathbf{X}^{(\tau)} \rangle_\tau$ exactly spans the eigenvectors as $\tau \to \infty$) or for any eigenvector exactly contained in the column span of the matrix $\mathbf{U}$ of trial vectors.
Due to the latter property, we expect better eigenvalue estimates when more accurate trial vectors are used.

Because subsequent iterates are correlated, the standard error in each eigenvalue estimate $\Lambda_{kk}$ is approximated by applying suitable Markov chain Monte Carlo error estimation techniques~\cite{Foreman2013} to the scalar-valued trajectory
\begin{equation}
\label{eq:errTraj}
\mathbf{z}_k^* ( \langle \mathbf{U}^\text{T} \mathbf{H} \mathbf{X}^{(\tau)} \rangle_\tau - {\Lambda}_{kk} \langle \mathbf{U}^\text{T} \mathbf{X}^{(\tau)} \rangle_\tau ) \mathbf{w}_k
\end{equation}
where $\mathbf{z}_k$ and $\mathbf{w}_k$ represent the left and right generalized eigenvectors, respectively, corresponding to $\Lambda_{kk}$.
We used the emcee software package~\cite{Foreman2013} to estimate standard errors and associated autocorrelation times.
For most of the systems considered in this work, these autocorrelation times exceeded 100 iterations, suggesting that evaluating $\mathbf{U}^\text{T} \mathbf{HX}^{(\tau)}$ and $\mathbf{U}^\text{T} \mathbf{X}^{(\tau)}$ more frequently would not significantly change eigenvalue estimates or their associated standard errors.
In our implementation, the computational cost of each evaluation scales with the number of nonzero elements in $\mathbf{U}$ and can be quite significant in practice.
Evaluating these matrices less frequently enables us to afford the increased cost associated with using more accurate trial vectors.
These long autocorrelation times also render it difficult to converge our standard error estimates using the default parameters in the emcee software.
However, given that the range of values in any single trajectory (i.e. the difference between the maximum and minimum) is usually less than $0.5$ m$E_\text{h}$, we believe our error estimates to be sufficiently accurate for the comparisons reported in this work. 

%Regardless of the quality of our estimates of the autocorrelation time, the standard error should be less than the range of the trajectory values (i.e. the difference between the maximum and minimum values).
%For all trajectories considered here, this range is less than 1 m$E_\text{h}$ (substantially so in most cases).
%We additionally found that changing the number of iterations included in trajectory averages changed eigenvalue estimates by $< 0.1$ m$E_\text{h}$.
%Together, these findings suggest

\subsection{Trial Vector Construction}
\label{sec:trials}
Because our goal is to estimate eigenvalues of challenging systems to high accuracy, we seek to obtain more accurate trial vectors than in our previous studies.
To this end, we employ a selected configuration interaction method, as is increasingly done in other QMC methods~\cite{Scemama2018excitation, Caffarel2016using}.
Generically, selected configuration interaction involves constructing a variational subspace of Slater determinants determined to contribute significantly to the eigenvectors of interest, and then calculating Hamiltonian eigenvectors in this subspace.
Here we use a specific subspace construction strategy known as variational heat-bath configuraton interaction (vHCI)~\cite{Holmes2017}, implemented in the Dice software~\cite{dice} and interfaced to PySCF.
%The HCI method, which involves combining vHCI with a semistochastic evaluation of the second-order perturbation theory contribution to the energy, has already been demonstrated to yield eigenenergies of sub-milliHartree accuracy for a variety of challenging chemical systems~\cite{Holmes2017, Eriksen2020, Yao2020almost}. 
Previously, vHCI was demonstrated to yield eigenvectors of sufficient accuracy for orbital optimization~\cite{Smith2017cheap} and evaluation of the second-order perturbation theory contribution to the energy~\cite{Holmes2017, Eriksen2020, Yao2020almost}.
We therefore expect it to yield accurate eigenvectors for our purposes as well.
After calculating the eigenvectors of interest via a vHCI calculation, we project them into a basis of spin-coupled functions according to eq \eqref{eq:evenSCF} or \eqref{eq:oddSCF}.
%SHCI performs particularly well when the exact eigenvectors are sparse.

\begin{table}
\caption{Parameters defining the sizes of the variational subspaces for HCI calculations used to construct trial vectors. $\varepsilon_1$ indicates the parameter in HCI that determines the size of the subspace, as defined in ref \citenum{Sharma2017}, and $N_\text{HCI}$ denotes the number of determinants in the final subspace.}
\begin{tabular}{c  c  c }
System & $\varepsilon_1$ (m$E_\text{h}$) & $N_\text{HCI} / 10^6$ \\ \hline
equilibrium \ce{C2} & 0.10 & 3.50 \\
stretched \ce{C2} & 0.20 & 1.49 \\
oxo-Mn(Salen) & 0.30 & 1.59 \\
ozone (OM) & 0.20 & 2.67 \\
ozone (RM) & 0.15 & 5.62 \\
ozone (TS) & 0.15 & 5.30 \\
butadiene & 0.10 & 5.29
\end{tabular}
\label{tab:SHCIparams}
\end{table}

Here we provide more specific details on how we chose the orbital basis and calculated the matrix $\mathbf{U}$ of trial vectors for each of the chemical systems considered in this work.
Specific definitions of each of these systems and further computational details are presented in Section \ref{sec:results}.
%ox-Mn(salen): UNO
%ozone: MP2
%butadiene: MP2
We begin by performing an inexpensive state-averaged vHCISCF calculation~\cite{Smith2017cheap}, i.e., with a small variational subspace.
The value of $\varepsilon_1$, which determines the size of this subspace in vHCI, was chosen to be 5 milliHartrees (m$E_\text{h}$) for the oxo-Mn(Salen) system and 0.2 m$E_\text{h}$ for ozone and butadiene; for \ce{C2}, we skip the vHCISCF step.
For oxo-Mn(salen), the initial active space orbitals are chosen as the subset of unrestricted Hartree-Fock natural orbitals with occupations that differ from 0 or 2 by more than $10^{-4}$.
For ozone and butadiene, we use second-order M\" oller-Plesset natural orbitals.
We then recalculate the natural orbitals at vHCISCF convergence (or from a one-shot, inexpensive vHCI calculation with $\varepsilon_1=5$~m$E_\text{h}$ in the case of \ce{C2}).
This orbital basis is used for our subsequent vHCI and FCI-FRI calculations, following previous studies~\cite{Sharma2017, Smith2017cheap}.
Since oxo-Mn(salen) was the only system for which we truncated the valence orbital space, we provide the final orbitals we used for this system in ref ~\citenum{Zenodo}.
In order to generate trial vectors, we perform a final vHCI calculation with a larger variational subspace.
The corresponding value of $\varepsilon_1$ and the size of the resulting subspace (denoted $N_\mathrm{HCI}$) are given in Table \ref{tab:SHCIparams} for each system studied.
%applying the heat-bath configuration interaction self-consistent field (HCISCF) method~\cite{Smith2017cheap}, which optimizes the orbitals to minimize the energies of the eigenvectors in the HCI variational subspace.
%%Additional details of these calculations for each system are presented in Section \ref{sec:results}.
%We then performed an additional orbital transformation to make the eigenvectors more sparse:
%starting from the optimized HCISCF orbitals (or the Hartree-Fock orbitals for \ce{C2}), we applied HCI to calculate state-averaged natural orbitals for each system.
%In these calculations, the value of the parameter $\epsilon_1$, which controls the size of the variational subspace in HCI, was chosen to be 5 m$E_\text{h}$ (milliHartrees) for the \ce{C2} and oxo-Mn(Salen) systems, and 0.2 m$E_\text{h}$ for ozone.
%Both HCI and FCI-FRI are known to achieve higher accuracy when eigenvectors are more sparse, so we used these orbitals for all subsequent calculations.
%We then performed a second HCI calculation in this new single-particle basis to calculate trial vectors.
%\red{[TCB: I'm confused about this procedure. Do we truncate in the NO basis?]}
%The value of $\epsilon_1$ and the size of the resulting variational subspace for each system are presented in Table \ref{tab:SHCIparams}.
These subspaces are smaller than those typically used in state-of-the-art HCI calculations because otherwise the cost of evaluating eq \eqref{eq:genEval} is prohibitive.
(Note that the matrix $\mathbf{U}^\text{T} \mathbf{H}$ is too large to store and so we reevaluate its entries on-the-fly, at a cost that scales as $\mathcal{O}(N_\mathrm{HCI} N^2M^2)$.)

\subsection{Stochastic Compression}
Applying the subspace iteration procedure described in Section \ref{sec:subspIter} in its deterministic form to the chemical systems of interest in this work is intractable due to the size of $N_\text{FCI}$ and the associated memory and CPU costs.
The FCI-FRI framework addresses this challenge by using stochastic compression to impose sparsity and thus reduce the cost of matrix multiplication.
Defining the compression operator $\Phi$, a stochastically compressed vector $\Phi(\mathbf{x})$ has elements that equal those of the input vector $\mathbf{x}$ in expectation (i.e. $\text{E}[\Phi(\mathbf{x})] = \mathbf{x}$) and has at most $m$ nonzero elements, where $m$ is a tunable parameter.
%Here, we define the action of $\Phi$ on a matrix $\mathbf{X}$ (with columns $\mathbf{x}_1$, $\mathbf{x}_2$, ...) in terms of independent compression operations applied to each of its columns, i.e. $\Phi(\mathbf{X}) = [\Phi(\mathbf{x}_1) \quad \Phi(\mathbf{x}_2) \quad ...]$.
We use a specific stochastic compression scheme known as pivotal compression~\cite{Greene2022}.
In many applications, this scheme achieves low statistical error, as confirmed by both theoretical analysis and numerical experiments~\cite{Greene2022}.
We provide a brief description of this two-step procedure here and refer the reader to ref \citenum{Greene2022} for further details.
First, a number $d$ of the largest-magnitude elements in the input vector $\mathbf{x}$ are left unchanged in compression, where $d$ is determined by an algorithm that depends both on $m$ and on the relative magnitudes of elements in $\mathbf{x}$~\cite{Greene2019, Greene2020, Greene2022}.
Then, a number $(m - d)$ of the remaining elements are randomly selected to be nonzero in the compressed vector, according to a pivotal resampling scheme that enforces statistical correlations among the elements~\cite{deville1998unequal, chauvet2012characterization, Chauvet2017}.
The probability of selecting each element is proportional to its magnitude.
Elements not selected are zero in $\Phi(\mathbf{x})$.
%Specific details of this compression scheme and its implementation are provided in ref \citenum{Greene2022}.
In general, the statistical error incurred in compression decreases as $m$ is increased, and it is zero if $m$ equals or exceeds the number of nonzero elements in $\mathbf{x}$.

\subsection{Multiplication and Compression Involving the Hamiltonian Matrix}
The most computationally expensive operation in the subspace iteration described in Section \ref{sec:subspIter} is multiplying each iterate $\mathbf{X}^{(\tau)}$ by the matrix $(\mathbf{1} - \varepsilon \mathbf{H})$.
We therefore apply stochastic compression to reduce this cost, thereby enabling the application of subspace iteration to large chemical systems.
The simplest approach to doing so involves stochastically compressing each column of the iterate $\mathbf{X}^{(\tau)}$ and replacing $\mathbf{X}^{(\tau)}$ in eq \eqref{eq:subspIter} with the resulting matrix.
This is not the approach used in this work and instead corresponds to the one described in ref \citenum{Greene2022}.
If $\mathbf{X}^{(\tau)}$ is compressed to $m$ nonzero elements per column and the resulting sparsity structure is leveraged, the memory and CPU cost of multiplying $\mathbf{X}^{(\tau)}$ by $(\mathbf{1} - \varepsilon \mathbf{H})$ scales as $\mathcal{O}(N^2 M^2 m N_\text{eigen})$.
Thus, one can control the cost of FCI-FRI by tuning $m$.
In practice, however, $m$ cannot be chosen to be arbitrarily small.
As has been demonstrated previously in the context of related methods, statistical error can increase very rapidly as $m$ is decreased~\cite{Booth2011, Spencer2012, Kolodrubetz2013, Shepherd2014, Vigor2016, Greene2019}, rendering it impossible to achieve accurate energy estimates, even after averaging over many iterations.
For the large quantum chemistry problems of interest in this work, the values of $m$ required for acceptable accuracy render this approach too computationally expensive.

In order to further reduce the cost of performing these multiplication operations, we employ a factorization strategy~\cite{Greene2020} related to those developed previously in the context of FCIQMC~\cite{Holmes2016, Neufeld2019}.
%As in the simple strategy introduced above, each iterate has $\mathcal{O}(m)$ nonzero elements per column.
In describing this strategy, it will be useful to introduce notation denoting \textit{compositions} of compression operations and matrix multiplications: for example,  $(\mathbf{H} \circ \Phi) \mathbf{x}$ indicates the vector obtained by first stochastically compressing $\mathbf{x}$ and then multiplying the resulting compressed vector by $\mathbf{H}$.
Within this factorization strategy, the $k^\text{th}$ column of the matrix $(\mathbf{1} - \varepsilon \mathbf{H})\mathbf{X}^{(\tau)}$ is approximated by applying a sequence of matrix multiplication and compression operations to the corresponding column of the previous iterate:

\begin{widetext}
\begin{equation}
\label{eq:factorization}
\left[ \left(\mathbf{1} - \varepsilon \mathbf{H} \right) \mathbf{X}^{(\tau)}\right]_{:k} \approx \left( \mathbf{P}_\text{diag} + \mathbf{B}^{(\tau, k)} \circ \Phi \circ \mathbf{Q}^{(5)} \circ \Phi \circ \mathbf{Q}^{(4)} \circ \Phi \circ \mathbf{Q}^{(3)} \circ \Phi \circ \mathbf{Q}^{(2)} \circ \Phi \circ \mathbf{Q}^{(1)} \right) \left[ \Phi \left( \mathbf{X}^{(\tau)}_{:k} \right) \right]
\end{equation}
\end{widetext}

Each of the matrices $\mathbf{Q}^{(1)}, \mathbf{Q}^{(2)}, ..., \mathbf{Q}^{(5)}$ is constructed to have less than $\mathcal{O}(M)$ nonzero elements per column.
Because the vector resulting after each compression operation has at most $m$ nonzero elements, the cost of the multiplication operations involving these matrices is limited to $\mathcal{O}(Mm)$.
The matrix $\mathbf{B}^{(\tau, k)}$, described in more detail below, depends on $\mathbf{X}^{(\tau)}_{:k}$ and has $\mathcal{O}(1)$ nonzero elements per column.
$\mathbf{P}_\text{diag}$ is a diagonal matrix containing the diagonal elements of $(\mathbf{1} - \varepsilon \mathbf{H})$.
The matrix-vector product $\mathbf{P}_\text{diag} \left[ \Phi \left( \mathbf{X}^{(\tau)}_{:k} \right) \right]$ can be formed at $\mathcal{O}(m)$ cost.
We emphasize that none of the matrices used in this factorization are stored explicitly, and that elements are instead evaluated in the course of each multiplication operation, in order to ensure memory efficiency.
%The matrix $\mathbf{U}^\text{T} \mathbf{HX}^{(\tau)}$
The steps involved in the implementation of this strategy are summarized in Table \ref{tab:algorithm}.
With this factorization strategy, the overall CPU and memory cost of performing subspace iteration with stochastic compression scales as $\mathcal{O}(Mm N_\text{eigen})$ or $\mathcal{O}(Nm N_\text{eigen})$.

\begin{table*}
    \centering
    \begin{tabular}{l | l}
     \\ \hline
Step & Cost \\ \hline
For each column $k$ of $\mathbf{X}^{(\tau)}$: & \\
1. Stochastically compress $\mathbf{X}^{(\tau)}_{:k}$ to m nonzero elements. & $\mathcal{O}\left(||\mathbf{X}^{(\tau)}_{:k}||_0\right)^a$ \\
2. Multiply the resulting vector by $\mathbf{Q}^{(1)}$. & $\mathcal{O}(m)$ \\
3. Stochastically compress the resulting vector to $m$ nonzero elements. & $\mathcal{O}(m)$ \\
4. Multiply the resulting vector by $\mathbf{Q}^{(2)}$. & $\mathcal{O}(Nm)$ \\
5. Stochastically compress the resulting vector to $m$ nonzero elements. & $\mathcal{O}(m)$ \\
6. Multiply the resulting vector by $\mathbf{Q}^{(3)}$. & $\mathcal{O}(Mm)$ \\
7. Stochastically compress the resulting vector to $m$ nonzero elements. & $\mathcal{O}(m)$ \\
8. Multiply the resulting vector by $\mathbf{Q}^{(4)}$. & $\mathcal{O}(Mm)$ \\
9. Stochastically compress the resulting vector to $m$ nonzero elements. & $\mathcal{O}(m)$ \\
10. Multiply the resulting vector by $\mathbf{Q}^{(5)}$. & $\mathcal{O}(Mm)$ \\
11. Stochastically compress the resulting vector to $m$ nonzero elements. & $\mathcal{O}(m)$ \\
12. Multiply the resulting vector by $\mathbf{B}^{(\tau, k)}$. & $\mathcal{O}(m)$ \\
13. Multiply the vector obtained in Step 1 by $\mathbf{P}_\text{diag}$, and add this to the & $\mathcal{O}(m)$\\
result from Step 12. & \\~\\
Assemble all of the resulting $N_\text{eigen}$ vectors into a matrix, and right-multiply & $\mathcal{O}(N_\text{eigen}^2 m)$ \\
it by $\left[ \mathbf{G}^{(\tau)}\right]^{-1}$ to obtain the next iterate, $\mathbf{X}^{(\tau + 1)}$. \\ \hline
    \end{tabular}

$^a$ $||\mathbf{X}^{(\tau)}_{:k}||_0$ denotes the number of nonzero elements in the $k^\text{th}$ column of $\mathbf{X}^{(\tau)}$.
    \caption{The steps involved in calculating the matrix product $(\mathbf{1} - \varepsilon \mathbf{H}) \mathbf{X}^{(\tau)} \left[ \mathbf{G}^{(\tau)} \right]^{-1}$. The costs of steps 2, 4, 6, 8, and 10 are determined by the numbers of nonzero elements in each of the corresponding matrices ($\mathbf{Q}^{(1)}$, $\mathbf{Q}^{(2)}$, ..., $\mathbf{Q}^{(5)}$), as presented in Appendix \ref{sec:Hfactor}.}
    \label{tab:algorithm}
\end{table*}

Although it is possible to choose the matrices $\mathbf{B}^{(\tau, k)}$ and $\mathbf{Q}^{(1)}, \mathbf{Q}^{(2)}, ..., \mathbf{Q}^{(5)}$ in eq \eqref{eq:factorization} such that $\mathbf{X}^{(\tau + 1)}$ equals $(\mathbf{1} - \varepsilon \mathbf{H}) \mathbf{X}^{(\tau)}$ in expectation, it is often advantageous to relax this requirement and construct $\mathbf{B}^{(\tau, k)}$ differently, according to an approach known as the initiator approximation, originally developed for FCIQMC~\cite{Cleland2010, Booth2011}.
This approximation introduces a bias in the resulting eigenvalue estimates.
Although it may be possible to reduce this bias, for example by adapting the adaptive-shift techniques described in refs \citenum{Ghanem2019} and \citenum{Ghanem2020}, further investigation is needed before these can reliably be applied to the excited-state procedure described here.
We use the standard initiator approximation for the calculations presented here, as it was previously found to greatly reduce statistical error in ground-state FCIQMC and FCI-FRI calculations, thus enabling the application of these methods to larger chemical systems.
This approach involves constructing $\mathbf{B}^{(\tau, k)}$ such that, in the course of the matrix multiplications in eq \eqref{eq:factorization}, only elements in $\mathbf{X}^{(\tau)}_{:k}$ with magnitudes greater than an initiator threshold are allowed to contribute to elements of $\mathbf{X}^{(\tau + 1)}_{:k}$ that are zero in $\mathbf{X}^{(\tau)}_{:k}$.
Specific formulas for the elements of $\mathbf{B}^{(\tau, k)}$, as well as those of $\mathbf{Q}^{(1)}, \mathbf{Q}^{(2)}, ..., \mathbf{Q}^{(5)}$, are provided in Appendix \ref{sec:Hfactor}.

Applications of the initiator approximation in an FCIQMC context use a fixed, user-specified value for the initiator threshold, $n_a$.
Such an approach presents an issue for our particular FCI-FRI method for multiple eigenvalue calculations.
The column norms of iterates can become very different as the iteration proceeds, in which case there can be very different numbers of elements with magnitudes greater than $n_a$ in each column.
We therefore use a different threshold $t_k$ for each column $k$, scaled by the column norm, to ensure that the initiator approximation is applied uniformly to all columns:
\begin{equation}
\label{eq:iniThresh}
t_k = n_a \left\lVert \mathbf{X}^{(\tau)}_{:k} \right\rVert_1 m^{-1}
\end{equation}
In many implementations of FCIQMC, the number of samples used for stochastic matrix-vector multiplication (analogous to $m$) is approximately equal to the $\ell_1$-norm of the vector being multiplied.
In this case, our implementation of the initiator approximation is equivalent to previous implementations.
In practice, the $\ell_1$-norms of iterate columns in FCI-FRI are less than those of iterates in FCIQMC (and correspondingly less than $m$), since our algorithm does not require an initial ``population growth phase''~\cite{Spencer2012}.
Note that $t_k$ approaches 0 as $m$ is increased, in which case $\mathbf{X}^{(\tau + 1)}$ approaches $(\mathbf{1} - \varepsilon \mathbf{H}) \mathbf{X}^{(\tau)}$ in expectation, and the bias introduced by the initiator approximation approaches 0.
This behavior also parallels that of the initiator approximation as commonly applied to FCIQMC.

\section{Results}
\label{sec:results}
\subsection{The Carbon Dimer}

%\begin{widetext}
%\begin{table*}[t]
\begin{sidewaystable}
\caption{Energies (in Hartrees) of the lowest-energy even-spin A$_\text{g}$ eigenstates of \ce{C2} at ``equilibrium'' and ``stretched'' geometries, obtained by applying FCI-FRI with different values of $m$. The dimension $N_\text{FCI}$ of these problems is approximately $10^{12}$. Results from previous HCI~\cite{Holmes2017}, FCIQMC~\cite{Blunt2015}, and DMRG~\cite{Sharma2015} calculations, as well as the vHCI calculations we used to generate the matrix $\mathbf{U}$ of trial vectors, are included for comparison.}
\begin{tabular}{c c c c c c c c c c}
\multicolumn{10}{c}{Equilibrium \ce{C2} ($r_\text{C-C} = 1.24253$ \AA)} \\ \hline
State & Trial & $m = 1 \times 10^5$ & $m = 1 \times 10^6$ & $m = 2 \times 10^6$ & $m = 4 \times 10^6$ & $m = 8 \times 10^6$ & HCI & FCIQMC & DMRG \\ \hline
1 $^1$A$_\text{g}$ & $-75.80055$ & $-75.80739 $ & $-75.80396$ & $-75.80329$ & $-75.80295$ & $-75.80278$ & $-75.80271$ & $-75.80258$ & $-75.80264$ \\
2 $^1$A$_\text{g}$ & $-75.72211$ & $-75.72871$ & $-75.72486$ & $-75.72439$ & $-75.72435$ & $-75.72415$ & - & - & - \\
3 $^1$A$_\text{g}$ & $-75.71014$ & $-75.71361$ & $-75.71347$ & $-75.71285$ & $-75.71238$ & $-75.71221$ & $-75.71213$ & $-75.71200$ & $-75.71208$ \\
1 $^5$A$_\text{g}$ & $-75.59843$ & $-75.61099$ & $-75.60492$ & $-75.60178$ & $-75.60146$ & $-75.60126$ & - & - & - \\
2 $^5$A$_\text{g}$ & $-75.55597$ & $-75.59254$ & $-75.56045$ & $-75.55976$ & $-75.55941$ & $-75.55916$ & - & - & - \\
4 $^1$A$_\text{g}$ & $-75.54689$ & $-75.56937$ & $-75.55188$ & $-75.55023$ & $-75.55004$ & $-75.54970$ & $-75.54961$ & $-75.54942$ & $-75.54953$ \\  \hline
\multicolumn{9}{c}{Stretched \ce{C2} ($r_\text{C-C} = 2.0$ \AA)} \\ \hline
State & Trial & $m = 1 \times 10^5$ & $m = 1 \times 10^6$ & $m = 2 \times 10^6$ & $m = 4 \times 10^6$ & $m = 8 \times 10^6$ & HCI & FCIQMC & DMRG \\ \hline
1 $^1$A$_\text{g}$ & $-75.64544$ & - & $-75.65152$ & $-75.65140$ & $-75.65115$ & $-75.65110$ & - & - & - \\
2 $^1$A$_\text{g}$ & $-75.64021$ & - & $-75.64621$ & $-75.64613$ & $-75.64588$ & $-75.64565$ & $-75.64565$ & $-75.64548$ & $-75.64552$ \\
3 $^1$A$_\text{g}$ & $-75.60963$ & - & $-75.61509$ & $-75.61504$ & $-75.61483$ & $-75.61481$ & $-75.61486$ & $-75.61470$ & $-75.61469$ \\
1 $^5$A$_\text{g}$ & $-75.55543$ & - & $-75.56083$ & $-75.56096$ & $-75.56088$ & $-75.56085$ & - & - & - \\
4 $^1$A$_\text{g}$ & $-75.48713$ & - & $-75.46921$ & $-75.47473$ & $-75.49317$ & $-75.49313$ & $-75.49316$ & $-75.49297$ & $-75.49290$ \\ \hline
\end{tabular}
\label{tab:c2eq}
\end{sidewaystable}

We first apply our FCI-FRI subspace iteration to calculate eigenenergies of the carbon dimer (\ce{C2}) at two different geometries.
Others have previously used \ce{C2} as a rigorous test case for new quantum chemistry methods due to the significant multireference character of its lowest-energy eigenstates~\cite{Purwanto2009excited, Sharma2015, Holmes2017}. 
In order to facilitate comparisons with previous results, we employ a large cc-pVQZ basis~\cite{Dunning1989} and correlate all valence electrons in all orbitals (core electrons were frozen), resulting in a CI problem of 8 electrons in 108 spatial orbitals (8e,108o).
We focus here on the Hamiltonian block containing states with even-spin (singlet, quintet, etc.) and A$_\text{g}$ symmetry in the D$_\text{2h}$ point group.
Although it is possible to impose additional block-diagonal structure on the Hamiltonian by leveraging the full D$_{\infty \text{h}}$ symmetry of \ce{C2}~\cite{Holmes2017}, we do not employ such an approach here.

Table \ref{tab:c2eq} shows eigenenergy estimates for the six lowest-energy states within this Hamiltonian block for \ce{C2} at its equilibrium geometry, i.e. with an internuclear separation $r_\text{C-C}$ of 1.24253 \AA.
The leftmost estimates are the variational energies associated with the vHCI trial vectors used to perform FCI-FRI calculations.
Estimates in the next five columns were obtained by applying FCI-FRI with an initiator threshold of $n_a = 1$ and five different values of $m$, the number of nonzero elements used in stochastic compression operations for each iterate column.
%Table \ref{tab:c2eq} displays the eigenvalue estimates resulting from each of these calculations.
Standard error estimates (obtained from \eqref{eq:errTraj}) for all calculations with $m \geq 1 \times 10^6$ are less than 0.1 m$E_\text{h}$, while error estimates for $m = 1 \times 10^5$ are 0.66 m$E_\text{h}$ or less.
Therefore, the discrepancies between estimates for different values of $m$ primarily result from the statistical biases associated with our normalization and orthogonalization procedures, as well as the initiator approximation.
The magnitudes of all of these biases are expected to decrease with increasing $m$, as evidenced here by the convergence in the energy for each state as $m$ is increased.

Estimates obtained using $m = 1 \times 10^6$ differ from those associated with the trial vectors by 2.7 to 6.5 m$E_\text{h}$, and from those obtained using $m = 8 \times 10^6$ by 0.7 to 3.7  m$E_\text{h}$.
This indicates that, for this system, applying FCI-FRI with $m = 1 \times 10^6$ yields improved energy estimates relative to the inexpensive vHCI calculations used to generate the trial vectors, but greater values of $m$ are required to achieve convergence.
Applying FCI-FRI with $m = 1 \times 10^5$ yields less accurate energy estimates for four states, as compared to those associated with the trial vectors.
Nevertheless, FCI-FRI estimates obtained with $m = 8 \times 10^6$---the greatest value of $m$ we tested---exhibited sub-milliHartree agreement with those from from three previous state-of-the-art calculations on \ce{C2} in the same cc-pVQZ basis, obtained using HCI~\cite{Holmes2017}, FCIQMC~\cite{Blunt2015}, and DMRG~\cite{Sharma2015}.
These are presented in the rightmost column of Table \ref{tab:c2eq}.
Our total energy estimates differ from these previous calculations by at most 0.3 m$E_\text{h}$, leading to excitation energies that agree to 0.01~eV or better.
%While these previous calculations also used the cc-pVQZ basis and the full CAS(8,108) active space, they used different orbitals than we did. 
%For example, the HCI results in ref \citenum{Holmes2017} were obtained used state-averaged natural orbitals constructed using a variational subspace larger than the one we used.
%However, all results should agree if the CASCI solution is exact, since the active space includes all of the (non-frozen) orbitals and CASCI is invariant to orbital rotations within the active space.
%Nevertheless, the orbital \textit{spaces} in all three cases are the same, so results are comparable.
Because these previous calculations leveraged the full D$_{\infty \text{h}}$ symmetry of \ce{C2}, the effective dimension of the Hamiltonian was smaller than that considered in this work, and some eigenstates of A$_\text{g}$ symmetry in the D$_\text{2h}$ point group were excluded.

\begin{figure}
    \centering
    \includegraphics[width=\linewidth]{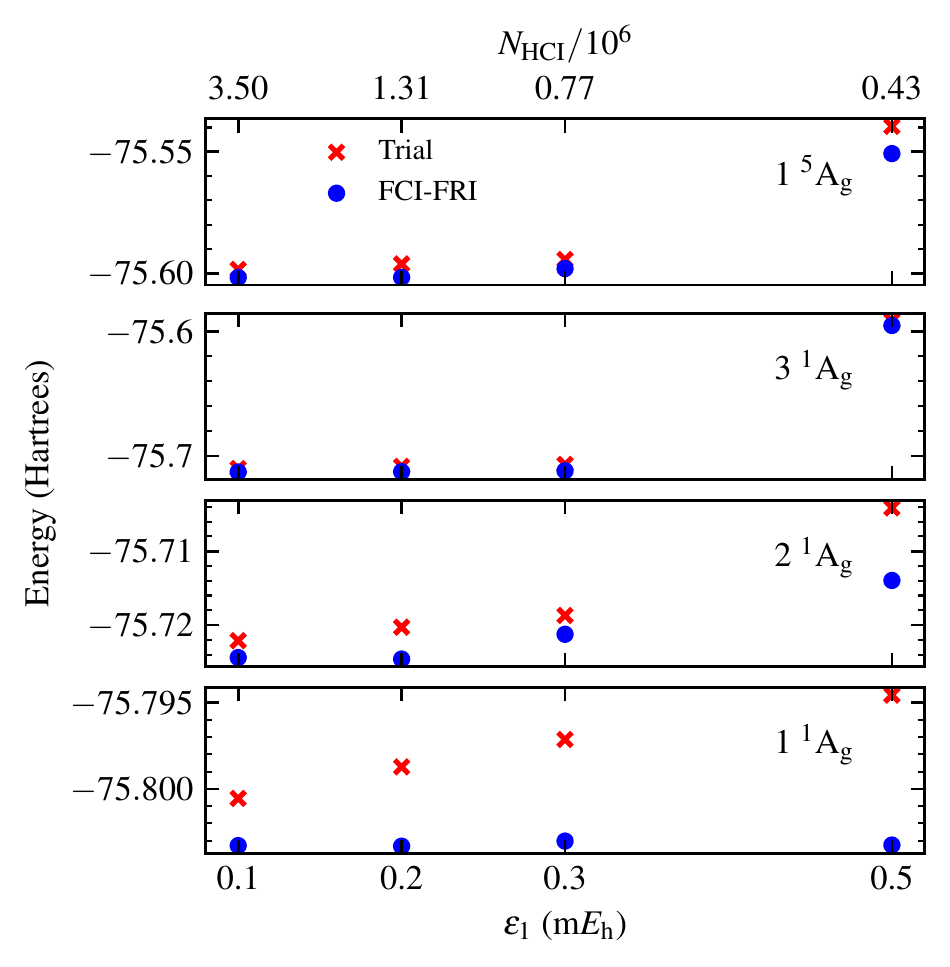}
    \caption{Energy estimates for equilibrium \ce{C2} obtained by applying FCI-FRI with $m = 2 \times 10^6$ with different trial vectors. The parameter $\varepsilon_1$ determines the size $N_\text{HCI}$ of the variational subspace in the vHCI algorithm. Energies of the trial vectors resulting from these vHCI calculations are included for comparison.}
    \label{fig:c2Trials}
\end{figure}

The accuracy of energy estimates from FCI-FRI depends on the quality of the trial vectors.
In order to provide a sense of the strength of this dependence, we performed a series of FCI-FRI calculations using different sets of trial vectors obtained from different variational subspaces.
The value of $\varepsilon_1$ and associated size of each subspace ($N_\text{HCI}$), along with the resulting vHCI and FCI-FRI energies, are shown in Figure \ref{fig:c2Trials}.
Using trial vectors generated using $\varepsilon_1 = 0.3$ m$E_\text{h}$ yielded notably better FCI-FRI energy estimates for equilibrium \ce{C2} than those from a calculation with $\varepsilon = 0.5$ m$E_\text{h}$, except for the ground state, possibly indicating that the ground state for this system is less strongly correlated than the excited states.
Further increasing the size of the variational subspace did not substantially improve the FCI-FRI excited-state energy estimates.
This kind of analysis may enable further reductions in the cost of our FCI-FRI approach, e.g. by using trial vectors from different vHCI calculations for each state.

We additionally consider \ce{C2} in the same cc-pVQZ basis at a nonequilibrium ``stretched'' geometry, with $r_\text{C-C} = 2.0$ \AA. 
Differences between subsequent eigenenergies are smaller at this geometry than at the equilibrium geometry, which makes it more difficult to obtain accurate energy estimates using stochastic methods like FCI-FRI~\cite{Greene2022}. 
This system therefore constitutes an even more rigorous test for our approach than equilibrium \ce{C2}.
Accordingly, we could not obtain energy estimates with $m = 1 \times 10^5$ due to numerical instabilities caused by the statistical error, a phenomenon that we have observed previously~\cite{Greene2022}.
Standard errors for all other estimates are less than 0.1 m$E_\text{h}$.
The trial vectors for this system are further from the exact eigenvectors, as indicated by their associated energies, but we nonetheless obtain converged FCI-FRI energy estimates by increasing $m$ to $8 \times 10^6$.
These estimates differ from those from previous calculations by at most 0.3 m$E_\text{h}$.
%Differences between the trial vector energies and those obtained from FCI-FRI are greater for stretched \ce{C2} than for equilibrium \ce{C2}.
%We observed that HCI eigenvectors for stretched \ce{C2} have more nonzero elements than those for equilibrium \ce{C2} (at constant $\epsilon_1$).
%The fact that FCI-FRI performs similarly for both systems reflects one advantage of stochastic methods like FCI-FRI: they consider all basis functions within configuration space, rather than those only within a variational subspace.

Estimates for the 4 $^1$A$_\text{g}$ state in stretched \ce{C2} exhibit the greatest sensitivity to the value of $m$.
The energy calculated with $m = 1 \times 10^6$ differs from that from the $m = 8 \times 10^6$ calculation by 23.92 m$E_\text{h}$.
We suspect that the increased sensitivity of estimates for this state is caused by the presence of a higher-lying state close in energy.
In the deterministic implementation of our algorithm, each energy estimate $\Lambda_{kk}$ converges at a rate proportional to $|(1 - \varepsilon E_k) / (1 - \varepsilon E_{(N_\text{eigen} + 1)})|^\tau$, where $E_k$ denotes the exact eigenenergy for the $k^\text{th}$ state~\cite{Greene2022}.
Consequently, the rate of convergence is determined by the energy gap between the considered low-energy subspace and the higher-lying eigenenergies.
%smaller energy differences compared to the first eigenvalue \textit{not} considered lead to slower convergence.
Although a similar analysis for the randomized algorithm is more complicated, one can reasonably expect that eigenvalues that converge more slowly in the deterministic algorithm are more susceptible to statistical fluctuations in the randomized algorithm, which are larger at lesser values of $m$.
These statistical fluctuations can give rise to a greater statistical bias, as is observed here.

\subsection{oxo-Mn(salen)}
\label{sec:MnResults}
\begin{figure}
\includegraphics[width=\linewidth]{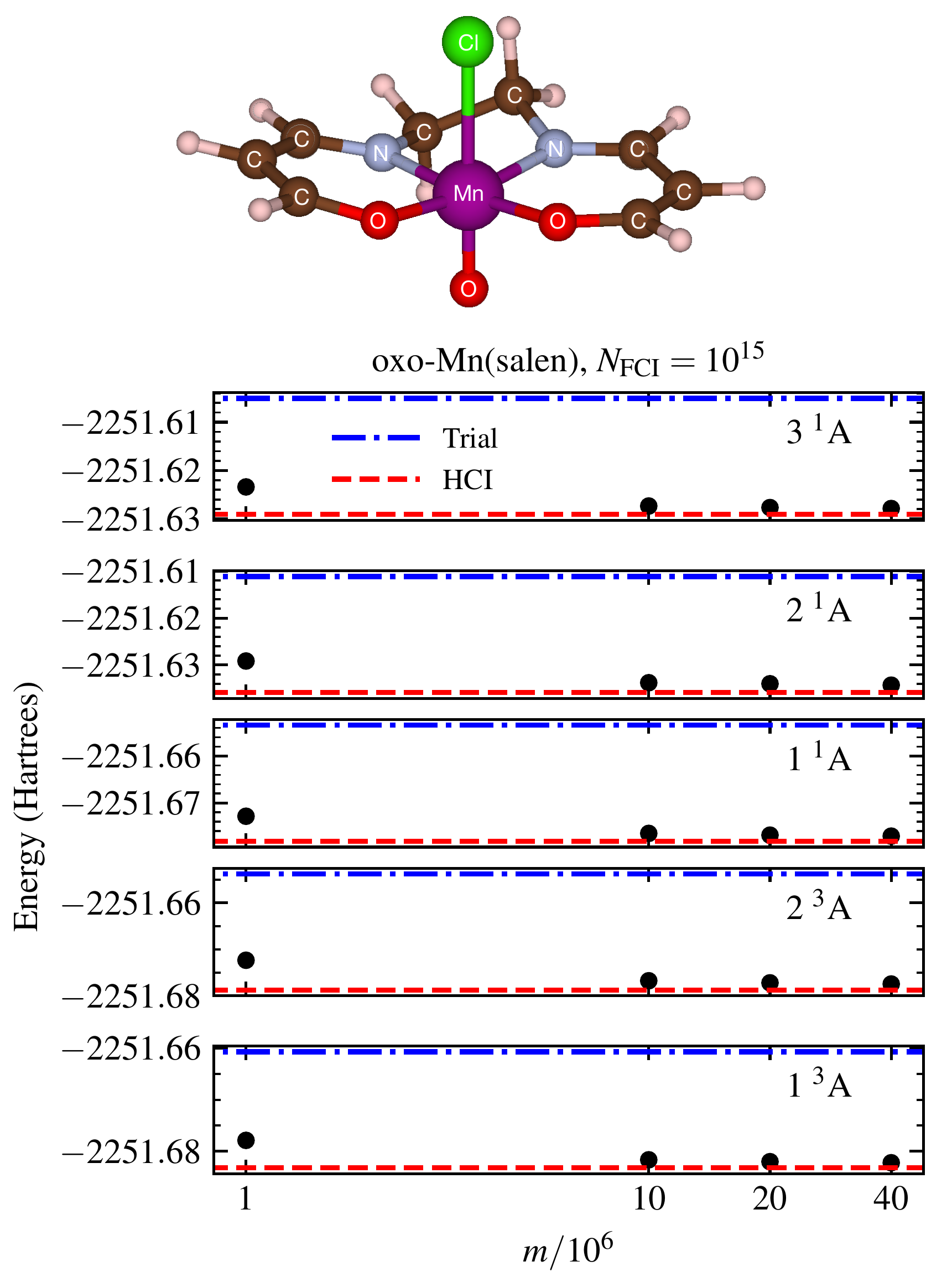}
\caption{(top) The structure of the model oxo-Mn(salen) complex considered in this work. The unlabelled atoms are H atoms. (bottom) Energies of the five lowest-energy eigenstates of this complex, calculated using FCI-FRI with four different values of $m$. 
The energies of the trial vectors used in these calculations are plotted as blue dash-dotted lines. 
Results from our own HCI calculations with perturbative corrections, extrapolated linearly as described in ref \citenum{Holmes2017}, are presented for comparison. Extrapolation was performed using five semistochastic HCI calculations with values of the $\varepsilon_1$ parameter increasing incrementally from 0.1 to 0.5 m$E_\text{h}$. $\varepsilon_2$ was fixed at $10^{-8} E_\text{h}$, and $\varepsilon_2^\text{d}$ was set to $0.1 \varepsilon_1$. (Definitions of these HCI parameters are provided in ref \citenum{Holmes2017}.)}
\label{fig:MnSalen}
\end{figure}

Manganese(salen) complexes are commonly used as catalysts for enantioselective epoxidation of alkenes~\cite{Zhang1990, Irie1990, Jacobsen1991, Katsuki1996, McGarrigle2005}.
The mechanisms of such reactions are not yet fully understood, as various mechanisms have been observed under different reaction conditions and for different alkene reactants~\cite{Srinivasan1986, Fu1991, Norrby1995, Hamada1996, Linker1997, Finney1997}.
A previous theoretical study has suggested that the spin state of the catalyst can play a crucial role in determining the mechanism~\cite{Linde1999}.
Because the catalysts' singlet, triplet, and quintet states are often close in energy, accurately predicting their relative energies at a given geometry is a crucial prerequisite for mechanistic studies~\cite{Ivanic2004}.
Because these states also exhibit strong multireference character, accurately calculating their energies by electronic structure theory is difficult.
This challenge has prompted several theoretical investigations of the electronic structure of these catalysts~\cite{Abashkin2001, Ivanic2004, Sears2006, Ma2011, Stein2016, Dang2021}.
Here, following previous theoretical studies, we calculate the eigenenergies of a model complex (Figure \ref{fig:MnSalen}) with a similar chemical environment around the metal center.
We refer to this model complex as oxo-Mn(salen), and we use the geometry reported in ref \citenum{Ivanic2004}.
%All calculations were performed using the 6-31G* single-particle basis.
Following previous studies~\cite{Ivanic2004, Wouters2014dmrgscf, Sharma2017}, we use the 6-31G* basis and treat only an active subset of the orbitals;
here, we use a (28e,28o) active space.

%Since the complex has $C_1$ symmetry, we did not consider point-group symmetry in these calculations.
%Active space orbitals were determined by HCISCF with $\epsilon_1 = 5$ m$E_\text{h}$, starting from unrestricted natural orbitals from Hartree-Fock. 
%After the resulting orbitals were further transformed according to the procedure outlined in Section \ref{sec:trials}, 
Working in the basis of approximate vHCISCF natural orbitals, as described in Section~\ref{sec:trials}, we perform two sets of FCI-FRI calculations: {one for the two lowest-energy triplets (1~$^3$A and 2~$^3$A) and one for the three lowest-energy singlets (1~$^1$A, 2~$^1$A, and 3~$^1$A)}.
The resulting energy estimates for all five states, obtained from calculations with different values of $m$, are presented in the bottom panel of Figure \ref{fig:MnSalen}.
At all values of $m$ considered, energies obtained using FCI-FRI are approximately 20 m$E_\text{h}$ less than those associated with the trial vectors used for these calculations.
FCI-FRI energy estimates for all five states converge as $m$ is increased from 1 million to 40 million.
Energy estimates obtained with $m = 20$ million differ from those obtained with  $m = 40$ million by less than 0.3 m$E_\text{h}$.
Standard errors for all calculations are less than 0.02 m$E_\text{h}$.
Various other methods have been used to calculate the singlet-triplet gap (i.e. the energy difference between the 1~$^3$A and 1~$^1$A states) for oxo-Mn(salen)~\cite{Ivanic2003, Ivanic2004, Wouters2014dmrgscf, Ma2011}.
These results range from 2.3 to 8.8 kcal/mol, while the corresponding singlet-triplet gap from our $m = 40$ million calculation is 3.26 kcal/mol.
All of these previous calculations used active spaces that differ from ours, which renders a direct comparison of these results difficult.
%For example, ref \citenum{Wouters2014dmrgscf} reports a DMRGSCF calculation with a smaller (28e,22o) active space, 
%orbitals for an active space of 28 electrons in 22 orbitals were optimized using the density matrix renormalization group (DMRG) method with orbital optimization.
%which yielded a singlet-triplet gap (energy difference between the $1^3$A and $1^1$A states) of 5.0 kcal/mol. The, a discrepancy that might be due to our larger active space.

In order to verify the results from our FCI-FRI calculations, we performed our own HCI calculations on this system using the same active space as in our FCI-FRI calculations, including perturbative corrections according to the semistochastic procedure described in ref \citenum{Holmes2017}.
These HCI results are presented for comparison in the bottom panel of Figure \ref{fig:MnSalen}, and computational details are described in Appendix \ref{sec:MnHCI}.
Our HCI results differ from those obtained using FCI-FRI with $m = 40$ million by at most 1.55 m$E_\text{h}$ and predict a singlet-triplet gap of 3.{34~kcal/mol}, in good agreement with our FCI-FRI prediction.
%It should be noted that our HCI calculations were performed with less stringent convergence parameters than those commonly used in state-of-the-art HCI calculations, so this discrepancy may arise from errors remaining in both the FCI-FRI and HCI calculations.
Although we cannot completely rule out that our HCI or FCI-FRI results are unconverged, their mutual agreement is encouraging.

\subsection{Ozone}
Ozone (\ce{O3}) plays important roles in Earth's atmosphere due to its presence in smog~\cite{Zhang2019} and its role in scattering ultraviolet light in the stratosphere~\cite{Bais2019}.
This has prompted a number of computational studies probing the behavior of ozone at a variety of scales~\cite{Crutzen1974, Luecken2019, Zhu2020}.
Electronic structure calculations perhaps represent the most fundamental of these studies.
Ozone has proven particularly challenging for conventional electronic structure methods, in part due to its multireference nature and the large magnitude of its correlation energy relative to that of other energetic properties~\cite{Burton1979}.
Both dynamic and static electron correlation must be carefully considered to achieve quantitative accuracy~\cite{Chien2018}.
Despite these challenges, previous theoretical studies have suggested the presence of a metastable ``ring minimum'' structure of D$_{3h}$ symmetry as an intermediate in the photochemical decomposition of ozone to \ce{O2 + O}~\cite{Lee1990, Xantheas1991, Qu2005, DeVico2008}.
This structure has yet to be observed experimentally, and whether or not it is a stable intermediate depends crucially on its energy relative to the equilibrium structure.
In order to investigate possible formation pathways for this metastable structure, Chien et al.\cite{Chien2018} used HCI to calculate the energies of the two lowest-energy electronic states at the equilibrium geometry (denoted OM), the metastable geometry (RM), and the transition state separating the two (TS).
Here we calculate energies at these same geometries (as reported in ref \citenum{Chien2018}) and compare our results.

\begin{figure}
\includegraphics[width=\linewidth]{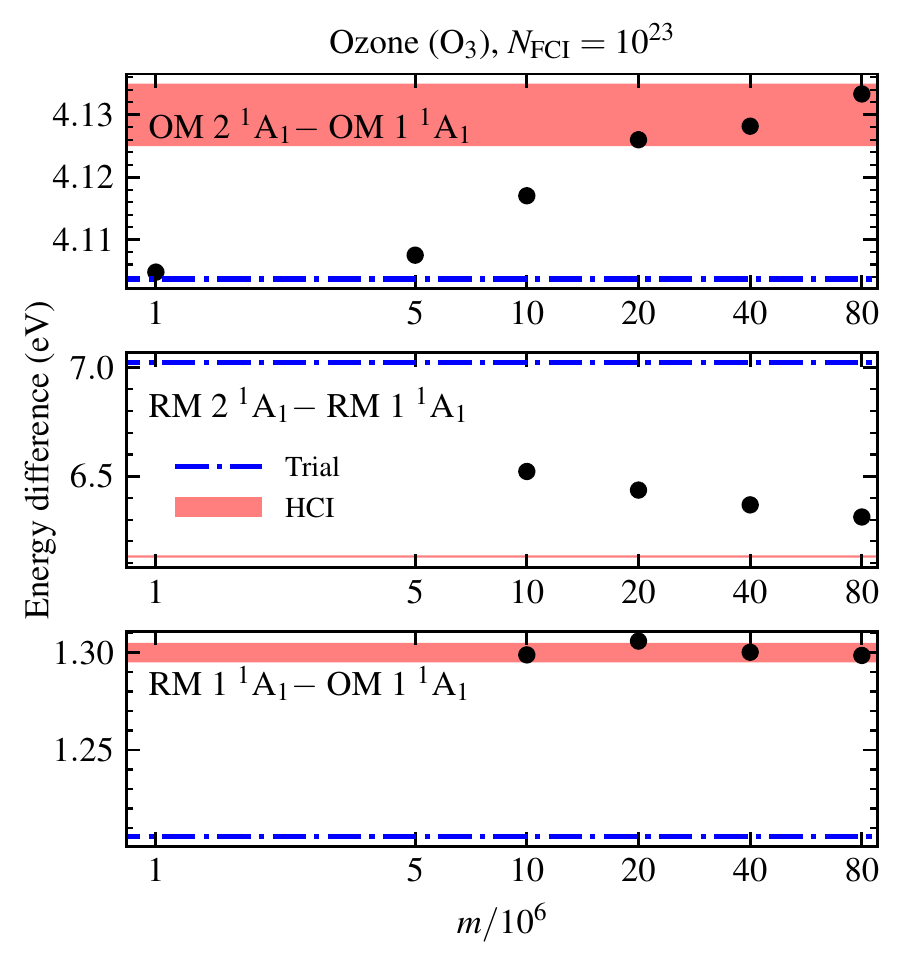}
\caption{Energy differences between the ground and first excited singlet states of ozone at the equilibrium (OM) geometry (top) and metastable (RM) geometry (middle), and between the ground states of ozone at the metastable and equilibrium geometries (bottom). Energies were calculated by performing FCI-FRI calculations at these two geometries with different values of $m$. Energy differences associated with the trial vectors, as well as HCI results from ref \citenum{Chien2018}, are included for comparison. The vertical width of the shaded area represents the reported accuracy of the HCI results (0.01~eV).}
\label{fig:ozoneResults}
\end{figure}

%We focused on the two lowest-energy states of even-spin $A_1$ symmetry at each geometry.
%Analysis of our HCI eigenvectors indicated that both of these states are singlets, in agreement with ref \citenum{Chien2018}.
We consider the singlet ground state and first singlet excited state of $A_1$ symmetry.
{Following ref~\citenum{Chien2018},} we use a cc-pVTZ basis and correlate all valence electrons in all orbitals, yielding a (18e,87o) active space.
%\red{Starting from the natural orbitals from a second-order M{\o}ller-Plesset (MP2) ground-state calculation, we optimized the orbitals by applying HCISCF with $\epsilon_1 = 0.2$ m$E_\text{h}$ and then calculating HCI natural orbitals as described in Section \ref{sec:trials}.}
Using the trial vectors obtained from the subspace specified in Section \ref{sec:trials}, we calculate the energies of the two lowest-energy even-spin eigenstates at each geometry using FCI-FRI.
The resulting estimates are shown in Figure \ref{fig:ozoneResults}.

We report our results in terms of energy differences, rather than absolute energies, to enable direct comparisons with ref \citenum{Chien2018}.
Standard errors for all energy difference estimates from FCI-FRI are less than $6 \times 10^{-5}$ eV.
When $m$ is chosen to be $\geq 10$ million, FCI-FRI offers improved energy estimates relative to those associated with the trial vectors, by as much as 0.5 eV.
Because the HCI energy differences from ref \citenum{Chien2018} were reported to an accuracy of 0.01 eV, we represent these results as shaded regions with widths of 0.01 eV in Figure \ref{fig:ozoneResults}.
At the OM geometry, our estimated energy difference calculated with $m = 80$ million agreed to within 0.01 eV (0.4 m$E_\text{h}$) of the HCI result from ref \citenum{Chien2018}.
Estimates of the difference between the ground-state energies at the OM and RM geometries are relatively insensitive to the value of $m$.
All estimates, except the one at $m = 20$ million, agree with each other to within 0.002 eV and with the HCI result to within 0.01 eV.
Discrepancies between FCI-FRI and HCI estimates are greater at the RM geometry: our best estimate (at $m = 80$ million) differed from the HCI energy difference by 0.18 eV (6.6 m$E_\text{h}$), but FCI-FRI results can be seen to be unconverged with respect to $m$.
%This is likely due to the slower convergence with respect to $m$ observed at this geometry.
Similar convergence issues were also observed in ref \citenum{Chien2018}: HCI energy differences at the RM geometry depended more on the size of the variational subspace than at the other geometries.
These results suggest that the slow convergence we observe is an intrinsic property of the eigenvectors of this system, perhaps associated with the sparsity structure of the RM 2 $^1$A$_1$ eigenvector.

Applying our method to estimate the two lowest-energy eigenvalues at the TS geometry yielded a conjugate pair of complex eigenvalue estimates, as is possible since the matrices $\langle \mathbf{U}^\text{T} \mathbf{HX}^{(\tau)}\rangle_\tau$ and $\langle \mathbf{U}^\text{T} \mathbf{X}^{(\tau)}\rangle_\tau$ in the eigenvalue equation~\eqref{eq:genEval} are real and nonsymmetric.
The appearance of complex eigenvalues is likely due to the fact that the difference of the two lowest-energy eigenvalues for this system (0.01 eV as estimated by HCI~\cite{Chien2018}) is substantially less than that for the OM or RM systems.
In order to better understand how small energy differences can lead to complex eigenvalue estimates, we first recognize that the estimates obtained by solving eq \eqref{eq:genEval} are equivalent to the eigenvalues of the matrix $\langle \mathbf{U}^\text{T} \mathbf{HX}^{(\tau)}\rangle_\tau [\langle \mathbf{U}^\text{T} \mathbf{X}^{(\tau)}\rangle_\tau]^{-1}$. 
Denoting the entries of this $2 \times 2$ matrix as $a_{ij}$,
%$\begin{pmatrix}
%a_{11} & a_{12} \\ a_{21} & a_{22}
%\end{pmatrix}$,
we recall that its eigenvalues are
\begin{equation}
E_\pm = \frac{1}{2} \left\{ a_{11} + a_{22} \pm \left[ \left(a_{11} - a_{22}\right)^2 + 4 a_{21} a_{12}\right]^{1/2} \right\}
\end{equation}
The eigenvalues are real as long as $(a_{11} - a_{22})^2  + 4 a_{21} a_{12} > 0$.
We numerically find that this inequality is violated for the matrices obtained from our calculations with the TS geometry but not with the OM and RM geometries, mainly because the two diagonal elements of $\langle \mathbf{U}^\text{T} \mathbf{HX}^{(\tau)}\rangle_\tau [\langle \mathbf{U}^\text{T} \mathbf{X}^{(\tau)}\rangle_\tau]^{-1}$ are closer in value in our TS calculations than in our OM and RM calculations.
This likely reflects the near-degeneracy of the exact eigenvalues for the TS system.
The FCI-FRI eigenvalue estimates for the TS system will be real for $m$ sufficiently large, since our method becomes exact as $m$ approaches the dimension of the Hamiltonian matrix.
Indeed, we find that as $m$ is increased from 10 million to 40 million, 
%the value of $(a_{11} - a_{22})^2  + 4 a_{21} a_{12}$ increases (i.e. becomes more positive).
the magnitude of the imaginary part of each eigenvalue decreases.
Considering only the real part of our TS eigenvalue estimates at $m = 40$ million and subtracting our corresponding OM 1~$^1$A$_1$ energy estimate, we estimate the difference of the TS and OM ground-state energies to be 2.44 eV. 
This compares favorably to the 2.41 eV difference reported in ref \citenum{Chien2018}, indicating that the slightly complex eigenvalues are not problematic for quantitative predictions.

\subsection{Butadiene}

\begin{figure}
    \centering
    \includegraphics[width=\linewidth]{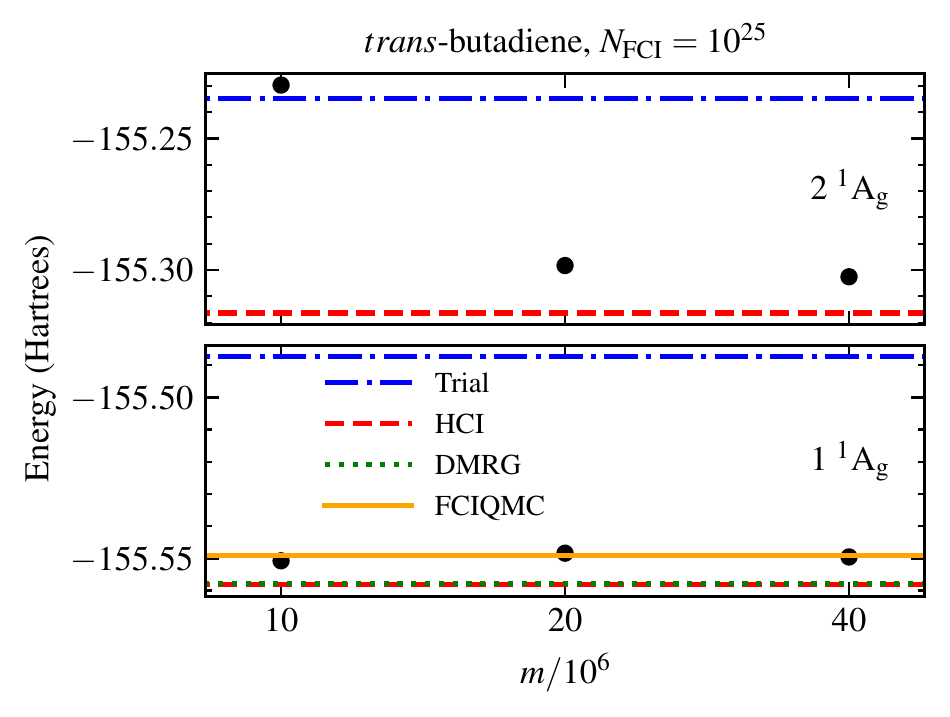}
    \caption{Energies of the two lowest-energy even-spin states of A$_\text{g}$ symmetry for \textit{trans}-butadiene, calculated by applying our FCI-FRI method with varying values of $m$. Trial vector energies, as well as HCI\cite{Chien2018}, DMRG~\cite{OlivaresAmaya2015}, and FCIQMC~\cite{Daday2012} results, are presented for comparison.}
    \label{fig:butadiene}
\end{figure}

Conjugated organic polymers are promising next-generation semiconducting materials due to their tunability and low cost relative to their inorganic counterparts~\cite{Ansari2018, Oka2021}.
The excited-state properties of such materials critically determine their performance in optoelectronic applications.
However, accurately characterizing the excited-state electronic structure of even simple conjugated molecules by theoretical means has proven challenging~\cite{Tavan1987, Watts1996, Starcke2006, Daday2012}.
For these reasons, conjugated organic systems are rigorous tests of electronic structure methods, and their accuracy has important implications for next-generation materials.
Here, we focus on the simplest conjugated organic molecule, \textit{trans}-butadiene, the excited states of which have been studied extensively using a variety of methods~\cite{Watson2012, Daday2012, OlivaresAmaya2015, Chien2018, Loos2020}.
%by HCI~\cite{Chien2018}, DMRG~\cite{OlivaresAmaya2015}, and FCIQMC~\cite{Daday2012}. 
For the sake of comparison, we use the geometry reported in ref \citenum{Daday2012} and the ANO-L-pVDZ basis~\cite{Widmark1990} and correlate all valence electrons in all orbitals, yielding a (22e,82o) active space.
%This same geometry was used for HCI calculations in ref \citenum{Chien2018}, so to enable comparisons to these results, we employ an ANO-L-pVDZ basis~\cite{Widmark1990}.
%Core electrons were frozen and valence electrons were correlated in all possible orbitals, yielding a full CAS(22,82) active space.

%\red{As for ozone, we optimized orbitals by first performing a ground-state MP2 calculation followed by an two-state HCISCF calculation with $\epsilon_1 = 0.2$ m$E_\text{h}$.}
Trial vectors are calculated as described above and used in subsequent FCI-FRI calculations of the two lowest-energy eigenvalues.
Results from these calculations are presented in Figure \ref{fig:butadiene}.
Energy estimates from FCI-FRI are approximately 50 m$E_\text{h}$ less than the trial vector energies, except at $m = 10$ million for the 2~$^1$A$_\text{g}$ state.
Analysis of our FCI-FRI results in comparison to previous results suggests that this is a more difficult system than those discussed above.
Our estimates of the 1 $^1$A$_\text{g}$ ground-state energy are relatively insensitive to the value of $m$, differing by only 0.1 m$E_\text{h}$.
Although our estimate at $m = 40$ million differs from the FCIQMC estimate~\cite{Daday2012} by only 0.4 m$E_\text{h}$, it exhibits greater discrepancies (8 m$E_\text{h}$) with HCI and DMRG estimates~\cite{Chien2018, OlivaresAmaya2015}.
(We compare to the DMRG estimate reported in ref \citenum{Chien2018}, which was obtained by extrapolating the results in ref \citenum{OlivaresAmaya2015}.)
The uncertainties in these estimates are all reported to be less than 0.1 m$E_\text{h}$, so these discrepancies suggest the presence of systematic errors.
The authors of ref \citenum{Chien2018} speculated that the FCIQMC energy is an overestimate due to errors from the initiator approximation and to potential inaccuracies in the reported uncertainty.
Given the similarities between our approach and FCIQMC, as well as the similarities in our estimates, it is likely that our energy is an overestimate for the same reasons.

Estimates of the 2 $^1$A$_\text{g}$ eigenvalue exhibit greater sensitivity to the value of $m$.
Our estimate at $m = 40$ million is 73 m$E_\text{h}$ greater than at $m = 10$ million.
This estimate also differs from the HCI estimate by 14 m$E_\text{h}$, likely due to the reasons discussed above.
Nonetheless, at our largest value of $m$, the FCI-FRI excitation energy is calculated to be 6.72~eV, which is in reasonable agreement with the HCI value of 6.58~eV.

\section{Conclusions}
\label{sec:conclusions}
We presented a general and systematically improvable strategy for calculating excited-state energies of electronic systems in large active spaces within the FCI-FRI framework.
Unlike previous ``replica'' methods for excited-state calculations~\cite{Blunt2015a}, our approach avoids the calculation of dot products of random vectors and instead uses approximate trial vectors to enforce orthogonality and estimate energies.
We expect that this feature will enable the reliable estimation of excited-state energies for large systems.
Applying our method to the carbon dimer (\ce{C2}) in a cc-pVQZ basis at two different geometries yielded energy estimates within 0.3 m$E_\text{h}$ of those from previous calculations.
Estimates for a oxo-Mn(salen) complex differed from our independent HCI calculations by up to 2 m$E_\text{h}$.
Discrepancies for the ozone and butadiene molecules were greater (7 m$E_\text{h}$ and 14 m$E_\text{h}$, respectively).
These total energy discrepancies translate to excitation energy discrepancies of about 0.1~eV or less.
The main sources of error in our calculations are statistical biases associated with the initiator approximation and the operations required to maintain orthonormality of the eigenvectors as the iteration proceeds.
These biases appeared to be greater for the more weakly correlated systems we considered---ozone and butadiene---than for the more strongly correlated \ce{C2} and oxo-Mn(salen) systems.
Previous investigations of the related FCIQMC method found that such errors are not always correlated with the amount of correlation~\cite{Booth2009, Spencer2012}, so our results may or may not be indicative of a more general trend. 
Applying our method to a wider variety of chemical systems could further elucidate general trends in its performance.

Additional developments could enable further reductions in the errors and computational cost of our FCI-FRI method, thereby enabling its application to systems even larger than those considered here.
The largest calculations in this work required 7-14 days of execution time on 448 cores to achieve reliable convergence.
The time required to calculate the matrix products $\mathbf{U}^\text{T} \mathbf{HX}^{(\tau)}$ constituted a significant portion of overall execution time, due to the large number of nonzero elements in both $\mathbf{U}$ and $\mathbf{X}^{(\tau)}$.
Future work could involve developing strategies for reducing this cost, for example by factorizing the matrix $\mathbf{H}$.
It may be possible to obtain more compact forms of the trial vectors without sacrificing accuracy, such as through transformations of the single-particle basis.
An orthogonal improvement, potentially suitable for weakly correlated systems, could involve adding perturbative energy corrections using strategies similar to those developed for FCIQMC methods~\cite{Blunt2018, Blunt2019, Ghanem2019, Ghanem2020}.
Importantly, the excited-state approach described here can be applied in tandem with any of the symmetry-based techniques for targeting excited states introduced in Section \ref{sec:Intro}.
Combining multiple strategies in this way could further extend the applicability of FCI-FRI and other methods to larger, more challenging systems.

More systematic research is needed to better understand how the various parameters used in our calculations affect the accuracy of our estimates.
Such investigation could lead to techniques for automating the selection of certain parameters, following previous FCIQMC studies~\cite{Spencer2016}.
We found that insufficient sampling in our approach leads to an uncontrolled increase in the condition number of the matrix $\mathbf{U}^\text{T} \mathbf{X}^{(\tau)}$~\cite{Greene2022}, so monitoring this condition number could serve as a diagnostic tool.
This could also lead to strategies for choosing different initiator thresholds for each state of interest in a calculation, thus enabling further reductions in the bias introduced by the initiator approximation.

In this paper, we consider only the calculation of energies.
Calculating other observables, such as reduced density matrices, using projector QMC methods like FCI-FRI is more challenging~\cite{Casulleras1995, Motta2017}.
Most existing strategies for doing so in the context of the FCI problem involve ``replica'' methods that require dot products of random vectors~\cite{Overy2014, Thomas2015analytic, Blunt2017density} and are thus affected by the aforementioned statistical errors associated with dot products of random vectors.
The ideas described in this work could potentially be used to inform new strategies for calculating non-energy observables using QMC without relying upon dot products of random vectors.
Such calculations could provide important metrics for assessing the accuracy of our FCI-FRI results and facilitate comparisons with results from other methods.

Notwithstanding these remaining challenges, the results of our calculations in this work suggest the effectiveness of the features of our excited-state FCI-FRI approach for enabling the treatment of large chemical systems.
Given the generality of our approach to excited-state calculations, it could also be implemented within other QMC schemes, such as FCIQMC, auxiliary-field QMC~\cite{Motta2018, Zhang2018}, or diffusion Monte Carlo~\cite{Foulkes2001}.

\appendix
\section{Spin-Coupled Functions and the Hamiltonian Matrix}
\label{sec:SCFHamil}
This section provides formulas for elements of the Hamiltonian matrix in a basis of spin-coupled functions, which are defined as linear combinations of Slater determinants~\cite{Holmes2016}.
Here we use the notation $\ket{\tilde{J}}$ to denote a generic spin-coupled function, constructed from a Slater determinant $\ket{J}$ and possibly $\hat{T}\ket{J}$, where $\hat{T}$ is the time-reversal operator that exchanges spin-up and spin-down electrons.
Spin-coupled functions in the even spin parity block of the Hamiltonian are denoted $\ket{\tilde{J}}_\text{e}$ and have the form
\begin{equation}
\label{eq:evenSCF}
\ket{\tilde{J}}_\text{e} = 
\begin{cases}
\ket{J}, & \hat{T}\ket{J} = \ket{J} \\
2^{-1/2} \left( \ket{J} + \hat{T} \ket{J} \right), & \hat{T} \ket{J} \neq \ket{J}
\end{cases}
\end{equation}
Spin-coupled functions in the odd spin parity block are denoted $\ket{\tilde{J}}_\text{o}$ and have the form
\begin{equation}
\label{eq:oddSCF}
\ket{\tilde{J}}_\text{o} = 2^{-1/2} \left( \ket{J} - \hat{T} \ket{J} \right)
\end{equation}
These can only be comprised of Slater determinants $\ket{J}$ for which $\hat{T}\ket{J} \neq \ket{J}$.

Combining eqs \eqref{eq:evenSCF} and \eqref{eq:oddSCF} with the Slater-Condon rules yields formulas for the elements of the Hamiltonian matrix $\mathbf{H}$.
Diagonal elements are given as
\begin{equation}
\mel{\tilde{J}}{\hat{H}}{\tilde{J}} = \sum_{j \in \ket{J}} h_{jj} + \frac{1}{2} \sum_{i,j \in \ket{J}} \mel{ij}{}{ij} + z \mel{J}{\hat{H} \hat{T}}{J}
\end{equation}
where $h_{jj}$ denotes an element of the one-electron component of the Hamiltonian, $\mel{ij}{}{ij}$ denotes an antisymmetrized two-electron integral, and $j \in \ket{J}$ denotes the constraint that the orbital $j$ is occupied in $\ket{J}$.
The variable $z$ is $+1$ in the even spin parity block and $-1$ in the odd spin parity block.
The last term, $z\mel{J}{\hat{H} \hat{T}}{J}$, is nonzero only if $\ket{J}$ and $\hat{T}\ket{J}$ differ by a double excitation. Denoting the occupied orbitals defining this excitation as $i$ and $j$ and the virtual orbitals $a$ and $b$, and defining $\ket{L} = \hat{T}\ket{J}$, the last term can be evaluated as
\begin{equation}
\label{eq:doubMel}
\mel{J}{\hat{H}\hat{T}}{J} \equiv \mel{J}{\hat{H}}{L} = \gamma_{ia}^J \gamma_{jb}^J \mel{ab}{}{ij}
\end{equation}
where $\gamma_{ia}^J$ is the number of occupied orbitals in between orbitals $i$ and $a$ in $\ket{J}$, as determined by a consistent ordering of orbitals among all Slater determinants~\cite{Holmes2016}.
Off-diagonal elements of $\mathbf{H}$ are given as
\begin{equation}
\label{eq:offdiagMel}
\mel{\tilde{J}}{\hat{H}}{\tilde{K}} = N_{J}^{-1} N_{K}^{-1} \left(\mel{J}{\hat{H}}{K} + z \mel{J}{\hat{T}\hat{H}}{K} \right)
\end{equation}
where
\begin{equation}
N_{J} = 
\begin{cases}
1, & \ket{J} \neq \hat{T}\ket{J} \\
2^{1/2}, & \ket{J} = \hat{T}\ket{J}
\end{cases}
\end{equation}
Eq \eqref{eq:offdiagMel} follows from the observations that $\hat{T}$ and $\hat{H}$ commute and that $\hat{T}^2 \ket{J} = \ket{J}$.
The matrix elements $\mel{J}{\hat{H}}{K}$ and $\mel{J}{\hat{T}\hat{H}}{K}$ can be evaluated according to standard Slater-Condon rules.
The generic matrix element $\mel{J}{\hat{H}}{L}$ is given in eq \eqref{eq:doubMel} if $\ket{J}$ and $\ket{L}$ differ by a double excitation, and as
\begin{equation}
\mel{J}{ \hat{H} }{ L} = \gamma^{J}_{ia} \left(h_{ia} + \sum_{j \in \ket{J}} \mel{ i j}{}{ a j } \right)
\end{equation}
if they differ by a single excitation involving an occupied orbital $i$ and virtual orbital $a$.

\section{Hamiltonian Matrix Factorization}
\label{sec:Hfactor}
This section provides formulas for the elements of the matrices $\mathbf{B}^{(\tau, k)}$ and $\mathbf{Q}^{(1)}, \mathbf{Q}^{(2)}, ..., \mathbf{Q}^{(5)}$ used to generate the iterate $\mathbf{X}^{(\tau + 1)}$ from $\mathbf{X}^{(\tau)}$.
We refer to these matrices as factors of $\mathbf{H}$ because off-diagonal elements of the matrix $\mathbf{B}^{(\tau, k)} \mathbf{Q}^{(5)} \mathbf{Q}^{(4)} \mathbf{Q}^{(3)} \mathbf{Q}^{(2)} \mathbf{Q}^{(1)}$ approximately equal those of $-\varepsilon \mathbf{H}$.
This scheme is based on the modified heat-bath Power-Pitzer factorization scheme introduced in ref \citenum{Greene2020} and includes modifications to enable its application in a basis of spin-coupled functions.

We begin by defining the notation used to index elements of these matrices.
One of the Slater determinants defining each spin-coupled function $\ket{\tilde{J}}$ is arbitrarily chosen as its \textit{representative} Slater determinant and is denoted $\ket{\tilde{J}}_\text{rep}$.
%If $\hat{T} \ket{J} \neq \ket{J}$, then either $\hat{T} \ket{J}$ or $\ket{J}$ is arbitrarily chosen to be $\ket{\tilde{J}}_\text{rep}$.
%If $\hat{T} \ket{J} = \ket{J}$, then $\ket{\tilde{J}}_\text{rep} = \ket{J}$.
Matrices are indexed by excitations from these representative Slater determinants.
For example, $(\tilde{J}, 1, i, a)$ denotes a single excitation involving occupied orbital $i$ and virtual orbital $a$ from $\ket{\tilde{J}}_\text{rep}$, and $(\tilde{J}, 1, i, j, a, b)$ denotes a double excitation involving occupied orbitals $i$ and $j$ and virtual orbitals $a$ and $b$ from $\ket{\tilde{J}}_\text{rep}$.

Elements of the matrices in the factorization are defined in terms of a matrix $\mathbf{D}$ and vectors $\mathbf{S}$ and $\mathbf{Y}$, precomputed at the beginning of each calculation.
The $2M \times 2M$ matrix $\mathbf{D}$ has elements
\begin{equation}
D_{pq} = \left(1 - \delta_{pq} \right) \sum_{r,s \notin \{ p,q \}}  \lvert \langle pq || rs \rangle \rvert
\end{equation}
where $p$ and $q$ represent indices of spin orbitals, $\delta_{pq}$ is a Kronecker delta, and $\lvert \langle pq || rs \rangle \rvert$ is an antisymmetrized two-electron integral.
Due to spin symmetries present in the two-electron integrals, $\mathbf{D}$ has only $M^2 + \begin{pmatrix} M \\ 2 \end{pmatrix}$ unique elements that need to be stored.
The vector $\mathbf{S}$ has elements
\begin{equation}
S_r = \frac{\sum_q D_{rq}}{\sum_{p, q} D_{pq}}
\end{equation}
and $\mathbf{Y}$ has elements
\begin{equation}
Y_i = \sum_a | \langle ia | ai \rangle |^{1/2}
\end{equation}

%The Hamiltonian matrix $\mathbf{H}$ is factorized into a product of six matrices, $\mathbf{H} = \mathbf{B} \mathbf{Q}^{(5)} \mathbf{Q}^{(4)} \mathbf{Q}^{(3)} \mathbf{Q}^{(2)} \mathbf{Q}^{(1)}$.
%The remainder of this section lists formulas for elements of each of these matrices in turn.
For all matrices defined in the remainder of this section, the values of matrices not specified below are 0.
The row space of $\mathbf{Q}^{(1)}$ has dimension $2 N_\text{FCI}$ and contains elements corresponding to either a generic single excitation, $(\tilde{J}, 1)$ or double excitation $(\tilde{J}, 2)$ from the reference determinant for each spin-coupled function $\ket{\tilde{J}}$.
Elements for single excitations are specified as
\begin{equation}
Q^{(1)}_{(\tilde{K}, 1), \tilde{J}} = \frac{n_\text{s}}{n_\text{s} + n_\text{d}} \delta_{KJ}
\end{equation}
and those for double excitations as
\begin{equation}
Q^{(1)}_{(\tilde{K}, 2), \tilde{J}} = \frac{n_\text{d}}{n_\text{s} + n_\text{d}} \delta_{KJ}
\end{equation}
where $n_\text{s}$ and $n_\text{d}$ are the number of symmetry-allowed single and double excitations from the Hartree-Fock determinant, respectively, and $\delta_{KJ}$ is a Kronecker delta.
Here, and in the remainder of this section, we consider only spin and spatial (point-group) symmetries when determining which excitations are allowed.
Each column of $\mathbf{Q}^{(1)}$ contains a maximum of 2 nonzero elements.

Indices for the row space of $\mathbf{Q}^{(2)}$ include an occupied orbital index $i$.
Single-excitation elements are specified as
\begin{equation}
Q^{(2)}_{(\tilde{J}, 1, i), (\tilde{J}, 1)} = \left(n^\text{occ}_{\tilde{J}} \right)^{-1}
\end{equation}
where $n^\text{occ}_J$ is the number of occupied orbitals in $\ket{\tilde{J}}_\text{rep}$ for which there is at least one virtual orbital of the same symmetry.
Double-excitation elements are specified as
\begin{equation}
Q^{(2)}_{(\tilde{J}, 2, i), (\tilde{J}, 2)} = S_i
\end{equation}
where the index $i$ is constrained to be any of the occupied orbitals in $\ket{\tilde{J}}_\text{rep}$ except the first.
Each column of $\mathbf{Q}^{(2)}$ contains $\mathcal{O}(N)$ nonzero elements.

Elements of $\mathbf{Q}^{(3)}$ corresponding to single excitations include a virtual orbital index $a$:
\begin{equation}
Q^{(3)}_{(\tilde{J}, 1, i, a), (\tilde{J}, 1, i)} = \left[ n^\text{virt}_{\tilde{J}} (i) \right]^{-1}
\end{equation}
Here, $n^\text{virt}_{\tilde{J}} (i)$ denotes the number of virtual orbitals in $\ket{\tilde{J}}_\text{rep}$ with the same symmetry as the occupied orbital $i$.
Elements corresponding to double excitations are indexed differently:
\begin{equation}
Q^{(3)}_{(\tilde{J}, 2, i, j), (\tilde{J}, 2, i)} = D_{ij} S_i^{-1}
\end{equation}
Here, the index of the second occupied orbital $j$ in the double excitation is constrained to be less than that of the first $(i)$.
Each column of $\mathbf{Q}^{(3)}$ corresponding to a single excitation contains $\mathcal{O}(M)$ nonzero elements, while each column corresponding to a double excitation contains $\mathcal{O}(N)$ nonzero elements.

All elements in $\mathbf{Q}^{(4)}$ corresponding to single excitations are 1:
\begin{equation}
\label{eq:q4sing}
Q^{(4)}_{(\tilde{J}, 1, i, a), (\tilde{J}, 1, i, a)} = 1
\end{equation}
Double-excitation elements are given as
\begin{equation}
Q^{(4)}_{(\tilde{J}, 2, i, j, a), (\tilde{J}, 2, i, j)} = | \langle ia | ai \rangle |^{1/2} X_i^{-1}
\end{equation}
where $a$ is constrained to be any virtual orbital in $\ket{\tilde{J}}_\text{rep}$ except the first.
Each column of $\mathbf{Q}^{(4)}$ corresponding to a double excitation contains $\mathcal{O}(M)$ nonzero elements.

In analogy to eq \eqref{eq:q4sing}, the values of single-excitation elements in $\mathbf{Q}^{(5)}$ are 1.
Double-excitation elements in $\mathbf{Q}^{(5)}$ are
\begin{equation}
Q^{(5)}_{(2, i, j, a, b), (2, i, j, a)} = | \langle jb | bj \rangle |^{1/2} X_j^{-1}
\end{equation}
where $b$ is constrained to be any virtual orbital in $\ket{\tilde{J}}_\text{rep}$ with an index less than $a$ for which the direct symmetry product $\Gamma_i \otimes \Gamma_j$ is equal to $\Gamma_a \otimes \Gamma_b$.
Further details on computing direct symmetry products can be found in refs \citenum{Greene2019} and \citenum{Booth2014}.
Each column of $\mathbf{Q}^{(5)}$ corresponding to a double excitation contains $\mathcal{O}(M)$ nonzero elements.

Multiplication by the final matrix in the factorization, $\mathbf{B}^{(\tau, k)}$, serves to sum elements corresponding to excitations that map to the same spin-coupled function while enforcing the initiator approximation.
Elements of $\mathbf{B}^{(\tau, k)}$ corresponding to single excitations are specified as
\begin{equation}
B^{(\tau, k)}_{\tilde{K}, (\tilde{J}, 1, i, a)} = 
\begin{cases}
0 & X^{(\tau)}_{\tilde{K}, k} = 0 \text{ and } |X^{(\tau)}_{\tilde{J}, k}| < t_k \\
-\dfrac{\varepsilon H_{\tilde{K}, \tilde{J}}}{Q_{(\tilde{J}, 1, i, a), \tilde{J}}} & \text{otherwise}
%-\varepsilon H_{\tilde{K}, \tilde{J}} \left( Q_{(\tilde{J}, 1, i, a), \tilde{J}} \right)^{-1} & \text{otherwise}
\end{cases}
\end{equation}
for the spin-coupled function $\ket{\tilde{K}}$ connected to $\ket{\tilde{J}}$ by a single excitation involving occupied orbital $i$ and virtual orbital $a$ (i.e. for which $\mel{\tilde{K}}{\hat{c}^\dagger_a \hat{c}_i}{\tilde{J}} \neq 0$).
The variable $t_k$ is defined in eq \eqref{eq:iniThresh}.
The matrix $\mathbf{Q}$ is defined as the product $\mathbf{Q}^{(5)} \mathbf{Q}^{(4)} \mathbf{Q}^{(3)} \mathbf{Q}^{(2)} \mathbf{Q}^{(1)}$.
Its elements can be calculated inexpensively on the fly due to the sparse structure of $\mathbf{Q}^{(1)}, \mathbf{Q}^{(2)}, ..., \mathbf{Q}^{(5)}$.
Elements of $\mathbf{B}^{(\tau, k)}$ corresponding to double excitations are
\begin{equation}
B^{(\tau, k)}_{\tilde{K}, (\tilde{J}, 2, i, j, a, b)} = 
\begin{cases}
0 & X^{(\tau)}_{\tilde{K}, k} = 0 \text{ and } |X^{(\tau)}_{\tilde{J}, k}| < t_k \\
-\dfrac{\varepsilon H_{\tilde{K}, \tilde{J}}}{Q_{(\tilde{J}, 2, i, j, a, b), \tilde{J}}} & \text{otherwise}
%-\varepsilon H_{\tilde{K}, \tilde{J}} \left( Q_{(\tilde{J}, 2, i, j, a, b), \tilde{J}} \right)^{-1} & \text{otherwise}
\end{cases}
\end{equation}
for $\mel{\tilde{K}}{\hat{c}^\dagger_a \hat{c}^\dagger_b \hat{c}_i \hat{c}_j}{\tilde{J}} \neq 0$.
Each column of $\mathbf{B}^{(\tau, k)}$ contains 1 nonzero element.

\begin{figure}
    \centering
    \includegraphics[width=\linewidth]{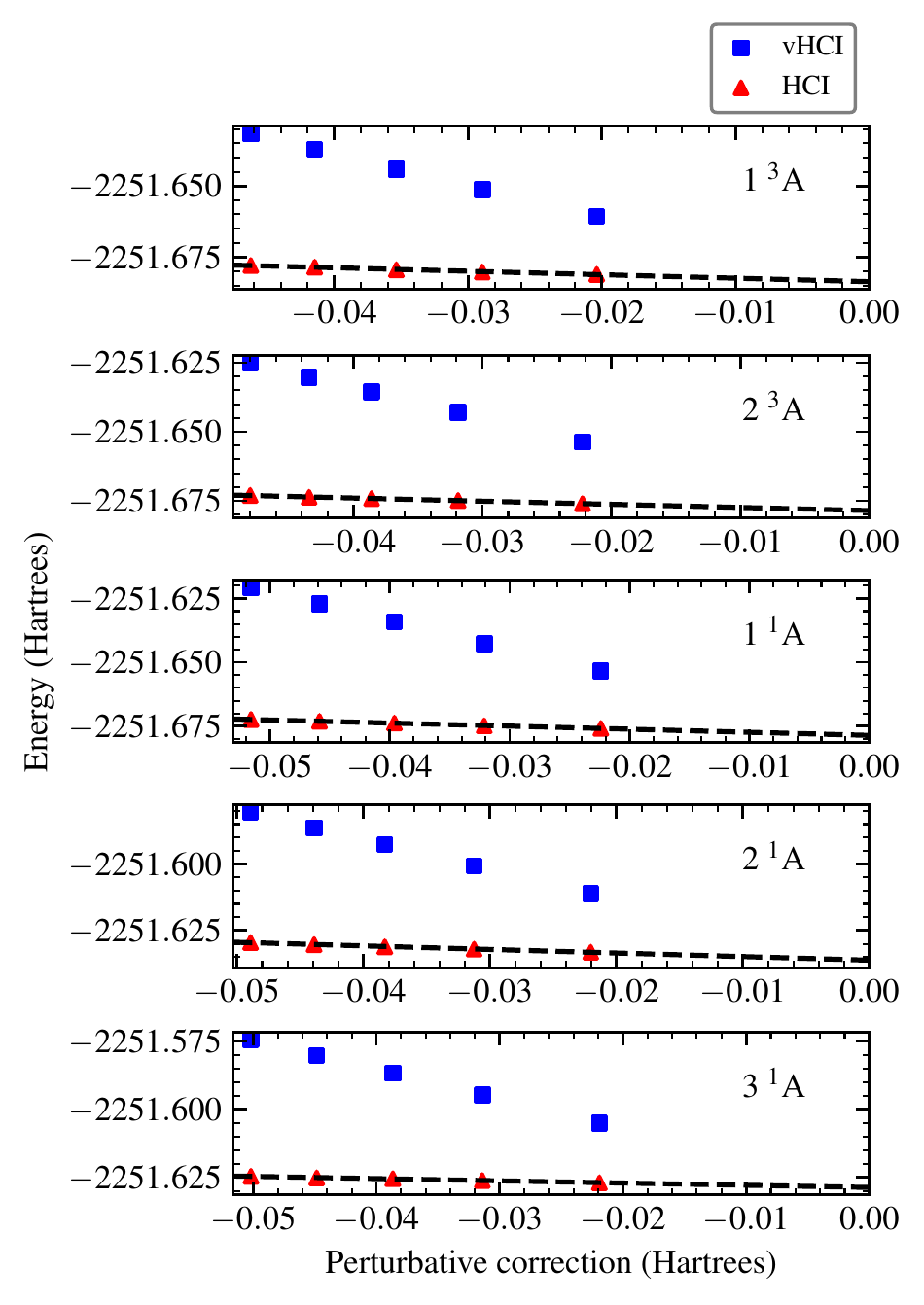}
    \caption{Results for oxo-Mn(salen) obtained by performing vHCI and HCI calculations using ten different values of the $\varepsilon_1$ parameter controlling the size of the variational subspace. The vertical axis represents the absolute energy obtained from each calculation, while the horizontal axis represents the difference between the HCI energies obtained by including semistochastic perturbative corrections and the vHCI energies. Black dashed lines represent the linear model used to extrapolate the HCI energies.}
    \label{fig:MnExtrap}
\end{figure}

\section{HCI Calculations for oxo-Mn(salen)}
\label{sec:MnHCI}
Calculations performed using HCI allow us to evaluate the accuracy of our FCI-FRI results for the oxo-Mn(salen) system introduced in Section \ref{sec:MnResults}.
These HCI calculations use the same active space and single-particle orbitals as those from our FCI-FRI calculations.
Following the extrapolation procedure outlined in ref \citenum{Holmes2017}, we performed five independent HCI calculations using values of the $\varepsilon_1$ parameter ranging from 0.3 m$E_\text{h}$ to 1.5 m$E_\text{h}$.
The parameter $\varepsilon_2$ was fixed at $10^{-7} E_\text{h}$ for all five calculations, and $\varepsilon_2^\text{d}$ was set to 0.1 $\varepsilon_1$.
(The parameters $\varepsilon_2$ and $\varepsilon_2^\text{d}$ control the number of terms included in the semistochastic perturbative corrections and are defined in ref \citenum{Holmes2017}.)

Results from both vHCI and HCI (including perturbative corrections) are presented in Figure \ref{fig:MnExtrap}.
Extrapolation is performed via a linear fit of the HCI energies with respect to the perturbative corrections to the vHCI energies calculated from HCI~\cite{Holmes2017}.
At the smallest value of $\varepsilon_1$ considered in this analysis (i.e. the largest variational subspace), HCI energies are approximately 22 m$E_\text{h}$ less than vHCI energies.
%Extrapolated HCI energies were caclulated as the vertical axis intercept of the black dashed lines.
The uncertainties in our HCI energies, as estimated from the uncertainties in the vertical axis intercepts of our linear fits, are less than $0.05$ m$E_\text{h}$.

\section{Orthonormalization of Iterates}
\label{sec:orthonormalization}
This section describes how we construct the matrix $\mathbf{G}^{(\tau)}$ in eq \eqref{eq:subspIter} to maintain approximate orthonormality of the columns of the iterates $\mathbf{X}^{(\tau)}$.
If $\mathbf{G}^{(\tau)}$ is fixed as the identity matrix, the resulting algorithm will be statistically unbiased, but the norms of the iterate columns will converge to either 0 or $\infty$ as $\tau \to \infty$, and the columns will become increasingly linearly dependent as they all approach the ground-state eigenvector.
These numerical issues would render it impossible to obtain accurate eigenvalue estimates, so a different approach is needed.

In most iterations, $\mathbf{G}^{(\tau)}$ is chosen to be $\mathbf{N}^{(\tau)}$, a diagonal matrix with elements
\begin{equation}
\mathbf{N}^{(\tau)}_{kk} = \left( \frac{\left\lVert \mathbf{X}^{(\tau)}_{:k} \right\rVert_1}{\left\lVert \mathbf{X}^{(\tau-1)}_{:k} \right\rVert_1} \right)^\alpha \left( \textbf{N}^{(\tau-1)}_{kk} \right)^{(1 - \alpha)}
\end{equation}
where $\mathbf{X}^{(\tau)}_{:k}$ denotes the $k^\text{th}$ column of $\mathbf{X}^{(\tau)}$, $|| \cdot ||_1$ denotes the $\ell_1$-norm of a vector (the sum of the magnitudes of its elements), and $\alpha$ is a tunable parameter.
$\mathbf{N}^{(0)}$ is initialized as the identity matrix.
With this choice of $\mathbf{G}^{(\tau)}$, setting $\alpha = 1$ would ensure that the column norms of iterates remain constant as the iteration proceeds.
However, in the randomized implementation of this method, this introduces a statistical bias arising from the nonlinear dependence of $\mathbf{N}^{(\tau)}$ on random variables, i.e. the iterate column norms $\left\lVert \mathbf{X}^{(\tau)}_{:k} \right\rVert_1$.
We therefore choose $\alpha < 1$ so that $\mathbf{N}^{(\tau)}$ depends less strongly on these random variables.
This causes the norms to fluctuate, but still prevents them from tending to 0 or $\infty$ while reducing the magnitude of this bias.
Previous numerical tests~\cite{Greene2022} indicated that $\alpha = 0.5$ is a suitable choice, so it is used for all calculations presented here.
This strategy bears many similarities to the use of a dynamically adjusted energy shift in other QMC methods~\cite{Umrigar1993, Booth2009} but was found to offer better stability for our excited-state calculations.
%We found that for excited-state calculations, the specific construction used here offered better stability than the standard energy shift approach.

At intervals of $\Delta$ iterations, we construct $\mathbf{G}^{(\tau)}$ differently in order to also maintain linear independence of the iterate columns.
In these iterations, $\mathbf{G}^{(\tau)}$ is instead chosen to be $\mathbf{N}^{(\tau)} \mathbf{D}^{(\tau)} \mathbf{R}^{(\tau)}$, where $\mathbf{N}^{(\tau)}$ is defined as above and $\mathbf{R}^{(\tau)}$ is the upper triangular factor of a QR factorization of $\mathbf{U}^\text{T} \mathbf{X}^{(\tau)}$.
This choice of $\mathbf{G}^{(\tau)}$ enforces orthogonality of the iterate columns within the span of the trial vectors $\mathbf{U}$.
Since inclusion of the factor $\mathbf{R}^{(\tau)}$ in $\mathbf{G}^{(\tau)}$ also introduces a normalization constraint, the diagonal matrix $\mathbf{D}^{(\tau)}$ is chosen to remove that constraint and ensure that normalization is controlled only via the matrix $\mathbf{N}^{(\tau)}$.
This reduces the bias associated with orthogonalization.
Elements of $\mathbf{D}^{(\tau)}$ are
\begin{equation}
\label{eq:resetnorms}
\mathbf{D}^{(\tau)}_{kk} = \frac{\left\lVert (\mathbf{X}^{(\tau)} [\mathbf{R}^{(\tau)}]^{-1})_{:k} \right\rVert_1}{\left\lVert \mathbf{X}^{(\tau)}_{:k} \right\rVert_1}
\end{equation}
%Further justification for this approach to orthogonalization is provided in ref \citenum{Greene2021}.
Since elements of $[\mathbf{R}^{(\tau)}]^{-1}$ depend nonlinearly on the random iterates, this orthogonalization procedure also introduces a statistical bias.
This strategy differs slightly from that employed in ref \citenum{Greene2022}, where we instead applied QR factorization to $\mathbf{U}^\text{T} (\mathbf{1} - \varepsilon \mathbf{H} ) \mathbf{X}^{(\tau)}$.
We found that the alternative strategy employed here made little difference to our final results and enabled reductions in the computational cost of our implementation. 
By monitoring the condition number of $\mathbf{U}^\text{T} \mathbf{X}^{(\tau)}$, we found $\Delta=1000$ to be a reasonable choice for the systems discussed here, but our results are relatively unchanged by more frequent orthogonalization.

\begin{acknowledgments}
The authors thank Aaron Dinner, Michael Lindsey, and Verena Neufeld for useful conversations and Benjamin Pritchard for his suggestions for improving the readability and performance of our code.
S.M.G. was supported  by a software fellowship from the Molecular Sciences Software Institute, which is funded by U.S. National Science Foundation grant OAC-1547580.
R.J.W. was supported by New York University's Dean's Dissertation Fellowship and by the National Science Foundation through award DMS-1646339. 
J.W. acknowledges support from the Advanced Scientific Computing Research Program within the DOE Office of Science through award DE-SC0020427. 
The Flatiron Institute is a division of the Simons Foundation.
\end{acknowledgments}

%\section*{References}
\bibliography{Sams_refs,software}

%aipnum4-2.bst 2019-01-14 (MD) hand-edited version of apsrev4-1.bst
\begin{thebibliography}{119}%
\makeatletter
\providecommand \@ifxundefined [1]{%
 \@ifx{#1\undefined}
}%
\providecommand \@ifnum [1]{%
 \ifnum #1\expandafter \@firstoftwo
 \else \expandafter \@secondoftwo
 \fi
}%
\providecommand \@ifx [1]{%
 \ifx #1\expandafter \@firstoftwo
 \else \expandafter \@secondoftwo
 \fi
}%
\providecommand \natexlab [1]{#1}%
\providecommand \enquote  [1]{``#1''}%
\providecommand \bibnamefont  [1]{#1}%
\providecommand \bibfnamefont [1]{#1}%
\providecommand \citenamefont [1]{#1}%
\providecommand \href@noop [0]{\@secondoftwo}%
\providecommand \href [0]{\begingroup \@sanitize@url \@href}%
\providecommand \@href[1]{\@@startlink{#1}\@@href}%
\providecommand \@@href[1]{\endgroup#1\@@endlink}%
\providecommand \@sanitize@url [0]{\catcode `\\12\catcode `\$12\catcode
  `\&12\catcode `\#12\catcode `\^12\catcode `\_12\catcode `\%12\relax}%
\providecommand \@@startlink[1]{}%
\providecommand \@@endlink[0]{}%
\providecommand \url  [0]{\begingroup\@sanitize@url \@url }%
\providecommand \@url [1]{\endgroup\@href {#1}{\urlprefix }}%
\providecommand \urlprefix  [0]{URL }%
\providecommand \Eprint [0]{\href }%
\providecommand \doibase [0]{https://doi.org/}%
\providecommand \selectlanguage [0]{\@gobble}%
\providecommand \bibinfo  [0]{\@secondoftwo}%
\providecommand \bibfield  [0]{\@secondoftwo}%
\providecommand \translation [1]{[#1]}%
\providecommand \BibitemOpen [0]{}%
\providecommand \bibitemStop [0]{}%
\providecommand \bibitemNoStop [0]{.\EOS\space}%
\providecommand \EOS [0]{\spacefactor3000\relax}%
\providecommand \BibitemShut  [1]{\csname bibitem#1\endcsname}%
\let\auto@bib@innerbib\@empty
%</preamble>
\bibitem [{\citenamefont {Mennucci}(2010)}]{Mennucci2010simulation}%
  \BibitemOpen
  \bibfield  {author} {\bibinfo {author} {\bibfnamefont {B.}~\bibnamefont
  {Mennucci}},\ }\enquote {\bibinfo {title} {The simulation of {UV-Vis}
  spectroscopy with computational methods},}\ in\ \href
  {https://doi.org/https://doi.org/10.1002/9783527633272.ch5} {\emph {\bibinfo
  {booktitle} {Computational Spectroscopy}}},\ \bibinfo {editor} {edited by\
  \bibinfo {editor} {\bibfnamefont {J.}~\bibnamefont {Grunenberg}}}\ (\bibinfo
  {publisher} {John Wiley \& Sons, Ltd},\ \bibinfo {year} {2010})\
  Chap.~\bibinfo {chapter} {5}, pp.\ \bibinfo {pages} {151--171}\BibitemShut
  {NoStop}%
\bibitem [{\citenamefont {Beard}\ \emph {et~al.}(2019)\citenamefont {Beard},
  \citenamefont {Sivaraman}, \citenamefont {V\'azquez-Mayagoitia},
  \citenamefont {Vishwanath},\ and\ \citenamefont
  {Cole}}]{Beard2019comparative}%
  \BibitemOpen
  \bibfield  {author} {\bibinfo {author} {\bibfnamefont {E.~J.}\ \bibnamefont
  {Beard}}, \bibinfo {author} {\bibfnamefont {G.}~\bibnamefont {Sivaraman}},
  \bibinfo {author} {\bibfnamefont {A.}~\bibnamefont {V\'azquez-Mayagoitia}},
  \bibinfo {author} {\bibfnamefont {V.}~\bibnamefont {Vishwanath}},\ and\
  \bibinfo {author} {\bibfnamefont {J.~M.}\ \bibnamefont {Cole}},\ }\bibfield
  {title} {\enquote {\bibinfo {title} {Comparative dataset of experimental and
  computational attributes of {UV}/vis absorption spectra},}\ }\href
  {https://doi.org/10.1038/s41597-019-0306-0} {\bibfield  {journal} {\bibinfo
  {journal} {Sci. Data}\ }\textbf {\bibinfo {volume} {6}},\ \bibinfo {pages}
  {307} (\bibinfo {year} {2019})}\BibitemShut {NoStop}%
\bibitem [{\citenamefont {Gonz\'alez}\ and\ \citenamefont
  {Lindh}(2020)}]{Gonzalez2020quantum}%
  \BibitemOpen
  \bibinfo {editor} {\bibfnamefont {L.}~\bibnamefont {Gonz\'alez}}\ and\
  \bibinfo {editor} {\bibfnamefont {R.}~\bibnamefont {Lindh}},\ eds.,\ \href
  {https://doi.org/https://doi.org/10.1002/9781119417774.fmatter} {\emph
  {\bibinfo {title} {Quantum Chemistry and Dynamics of Excited States: Methods
  and Applications}}}\ (\bibinfo  {publisher} {John Wiley \& Sons, Ltd},\
  \bibinfo {address} {Hoboken, {NJ}},\ \bibinfo {year} {2020})\BibitemShut
  {NoStop}%
\bibitem [{\citenamefont {Serrano-Andr\'es}\ and\ \citenamefont
  {Merch\'an}(2005)}]{SerranoAndres2005}%
  \BibitemOpen
  \bibfield  {author} {\bibinfo {author} {\bibfnamefont {L.}~\bibnamefont
  {Serrano-Andr\'es}}\ and\ \bibinfo {author} {\bibfnamefont {M.}~\bibnamefont
  {Merch\'an}},\ }\bibfield  {title} {\enquote {\bibinfo {title} {Quantum
  chemistry of the excited state: 2005 overview},}\ }\href
  {https://doi.org/https://doi.org/10.1016/j.theochem.2005.03.020} {\bibfield
  {journal} {\bibinfo  {journal} {J. Mol. Struct.: THEOCHEM}\ }\textbf
  {\bibinfo {volume} {729}},\ \bibinfo {pages} {99--108} (\bibinfo {year}
  {2005})},\ \bibinfo {note} {proceedings of the 30th International Congress of
  Theoretical Chemists of Latin Expression}\BibitemShut {NoStop}%
\bibitem [{\citenamefont {Navarrete-Miguel}\ \emph {et~al.}(2019)\citenamefont
  {Navarrete-Miguel}, \citenamefont {Segarra-Mart\'i}, \citenamefont
  {Franc\'es-Monerris}, \citenamefont {Giussani}, \citenamefont {Farahani},
  \citenamefont {Ding}, \citenamefont {Monari}, \citenamefont {Liu},\ and\
  \citenamefont {Roca-Sanju\'an}}]{Navarrete-Miguel2019}%
  \BibitemOpen
  \bibfield  {author} {\bibinfo {author} {\bibfnamefont {M.}~\bibnamefont
  {Navarrete-Miguel}}, \bibinfo {author} {\bibfnamefont {J.}~\bibnamefont
  {Segarra-Mart\'i}}, \bibinfo {author} {\bibfnamefont {A.}~\bibnamefont
  {Franc\'es-Monerris}}, \bibinfo {author} {\bibfnamefont {A.}~\bibnamefont
  {Giussani}}, \bibinfo {author} {\bibfnamefont {P.}~\bibnamefont {Farahani}},
  \bibinfo {author} {\bibfnamefont {B.-W.}\ \bibnamefont {Ding}}, \bibinfo
  {author} {\bibfnamefont {A.}~\bibnamefont {Monari}}, \bibinfo {author}
  {\bibfnamefont {Y.-J.}\ \bibnamefont {Liu}},\ and\ \bibinfo {author}
  {\bibfnamefont {D.}~\bibnamefont {Roca-Sanju\'an}},\ }\bibfield  {title}
  {\enquote {\bibinfo {title} {Quantum chemistry of the excited state: recent
  trends in methods developments and applications},}\ }in\ \href
  {https://doi.org/10.1039/9781788013598-00028} {\emph {\bibinfo {booktitle}
  {Photochemistry: Volume 46}}},\ Vol.~\bibinfo {volume} {46}\ (\bibinfo
  {publisher} {The Royal Society of Chemistry},\ \bibinfo {year} {2019})\ pp.\
  \bibinfo {pages} {28--77}\BibitemShut {NoStop}%
\bibitem [{\citenamefont {Park}\ and\ \citenamefont
  {Shiozaki}(2017)}]{Park2017}%
  \BibitemOpen
  \bibfield  {author} {\bibinfo {author} {\bibfnamefont {J.~W.}\ \bibnamefont
  {Park}}\ and\ \bibinfo {author} {\bibfnamefont {T.}~\bibnamefont
  {Shiozaki}},\ }\bibfield  {title} {\enquote {\bibinfo {title} {On-the-fly
  {CASPT2} surface-hopping dynamics},}\ }\bibfield  {booktitle} {\emph
  {\bibinfo {booktitle} {Journal of Chemical Theory and Computation}},\ }\href
  {https://doi.org/10.1021/acs.jctc.7b00559} {\bibfield  {journal} {\bibinfo
  {journal} {J. Chem. Theory Comput.}\ }\textbf {\bibinfo {volume} {13}},\
  \bibinfo {pages} {3676--3683} (\bibinfo {year} {2017})}\BibitemShut {NoStop}%
\bibitem [{\citenamefont {Subotnik}\ \emph {et~al.}(2016)\citenamefont
  {Subotnik}, \citenamefont {Jain}, \citenamefont {Landry}, \citenamefont
  {Petit}, \citenamefont {Ouyang},\ and\ \citenamefont
  {Bellonzi}}]{Subotnik2016understanding}%
  \BibitemOpen
  \bibfield  {author} {\bibinfo {author} {\bibfnamefont {J.~E.}\ \bibnamefont
  {Subotnik}}, \bibinfo {author} {\bibfnamefont {A.}~\bibnamefont {Jain}},
  \bibinfo {author} {\bibfnamefont {B.}~\bibnamefont {Landry}}, \bibinfo
  {author} {\bibfnamefont {A.}~\bibnamefont {Petit}}, \bibinfo {author}
  {\bibfnamefont {W.}~\bibnamefont {Ouyang}},\ and\ \bibinfo {author}
  {\bibfnamefont {N.}~\bibnamefont {Bellonzi}},\ }\bibfield  {title} {\enquote
  {\bibinfo {title} {Understanding the surface hopping view of electronic
  transitions and decoherence},}\ }\href
  {https://doi.org/10.1146/annurev-physchem-040215-112245} {\bibfield
  {journal} {\bibinfo  {journal} {Annu. Rev. Phys. Chem.}\ }\textbf {\bibinfo
  {volume} {67}},\ \bibinfo {pages} {387--417} (\bibinfo {year}
  {2016})}\BibitemShut {NoStop}%
\bibitem [{\citenamefont {Barbatti}(2011)}]{Barbatti2011nonadiabatic}%
  \BibitemOpen
  \bibfield  {author} {\bibinfo {author} {\bibfnamefont {M.}~\bibnamefont
  {Barbatti}},\ }\bibfield  {title} {\enquote {\bibinfo {title} {Nonadiabatic
  dynamics with trajectory surface hopping method},}\ }\href
  {https://doi.org/https://doi.org/10.1002/wcms.64} {\bibfield  {journal}
  {\bibinfo  {journal} {WIREs Comput. Mol. Sci.}\ }\textbf {\bibinfo {volume}
  {1}},\ \bibinfo {pages} {620--633} (\bibinfo {year} {2011})}\BibitemShut
  {NoStop}%
\bibitem [{\citenamefont {Harsha}, \citenamefont {Henderson},\ and\
  \citenamefont {Scuseria}(2019)}]{Harsha2019thermofield}%
  \BibitemOpen
  \bibfield  {author} {\bibinfo {author} {\bibfnamefont {G.}~\bibnamefont
  {Harsha}}, \bibinfo {author} {\bibfnamefont {T.~M.}\ \bibnamefont
  {Henderson}},\ and\ \bibinfo {author} {\bibfnamefont {G.~E.}\ \bibnamefont
  {Scuseria}},\ }\bibfield  {title} {\enquote {\bibinfo {title} {Thermofield
  theory for finite-temperature quantum chemistry},}\ }\href
  {https://doi.org/10.1063/1.5089560} {\bibfield  {journal} {\bibinfo
  {journal} {J. Chem. Phys.}\ }\textbf {\bibinfo {volume} {150}},\ \bibinfo
  {pages} {154109} (\bibinfo {year} {2019})}\BibitemShut {NoStop}%
\bibitem [{\citenamefont {Zhang}\ \emph {et~al.}(2021)\citenamefont {Zhang},
  \citenamefont {Zhang}, \citenamefont {Kang}, \citenamefont {Dai},\ and\
  \citenamefont {Bonitz}}]{Zhang2021finite}%
  \BibitemOpen
  \bibfield  {author} {\bibinfo {author} {\bibfnamefont {H.}~\bibnamefont
  {Zhang}}, \bibinfo {author} {\bibfnamefont {S.}~\bibnamefont {Zhang}},
  \bibinfo {author} {\bibfnamefont {D.}~\bibnamefont {Kang}}, \bibinfo {author}
  {\bibfnamefont {J.}~\bibnamefont {Dai}},\ and\ \bibinfo {author}
  {\bibfnamefont {M.}~\bibnamefont {Bonitz}},\ }\bibfield  {title} {\enquote
  {\bibinfo {title} {Finite-temperature density-functional-theory investigation
  on the nonequilibrium transient warm-dense-matter state created by laser
  excitation},}\ }\href {https://doi.org/10.1103/PhysRevE.103.013210}
  {\bibfield  {journal} {\bibinfo  {journal} {Phys. Rev. E}\ }\textbf {\bibinfo
  {volume} {103}},\ \bibinfo {pages} {013210} (\bibinfo {year}
  {2021})}\BibitemShut {NoStop}%
\bibitem [{\citenamefont {Fulde}\ and\ \citenamefont
  {Stoll}(2017)}]{Fulde2017dealing}%
  \BibitemOpen
  \bibfield  {author} {\bibinfo {author} {\bibfnamefont {P.}~\bibnamefont
  {Fulde}}\ and\ \bibinfo {author} {\bibfnamefont {H.}~\bibnamefont {Stoll}},\
  }\bibfield  {title} {\enquote {\bibinfo {title} {Dealing with the exponential
  wall in electronic structure calculations},}\ }\href
  {https://doi.org/10.1063/1.4983207} {\bibfield  {journal} {\bibinfo
  {journal} {J. Chem. Phys.}\ }\textbf {\bibinfo {volume} {146}},\ \bibinfo
  {pages} {194107} (\bibinfo {year} {2017})}\BibitemShut {NoStop}%
\bibitem [{\citenamefont {Laughlin}\ and\ \citenamefont
  {Pines}(2000)}]{Laughlin2000the}%
  \BibitemOpen
  \bibfield  {author} {\bibinfo {author} {\bibfnamefont {R.~B.}\ \bibnamefont
  {Laughlin}}\ and\ \bibinfo {author} {\bibfnamefont {D.}~\bibnamefont
  {Pines}},\ }\bibfield  {title} {\enquote {\bibinfo {title} {The theory of
  everything},}\ }\href {https://doi.org/10.1073/pnas.97.1.28} {\bibfield
  {journal} {\bibinfo  {journal} {Proc. Natl. Acad. Sci.}\ }\textbf {\bibinfo
  {volume} {97}},\ \bibinfo {pages} {28--31} (\bibinfo {year}
  {2000})}\BibitemShut {NoStop}%
\bibitem [{\citenamefont {Zhang}(2004)}]{Zhang2004}%
  \BibitemOpen
  \bibfield  {author} {\bibinfo {author} {\bibfnamefont {S.}~\bibnamefont
  {Zhang}},\ }\bibfield  {title} {\enquote {\bibinfo {title} {Quantum {M}onte
  {C}arlo methods for strongly correlated electron systems},}\ }in\ \href
  {https://doi.org/10.1007/0-387-21717-7_2} {\emph {\bibinfo {booktitle}
  {Theoretical Methods for Strongly Correlated Electrons}}},\ \bibinfo {series
  and number} {CRM Series in Mathematical Physics},\ \bibinfo {editor} {edited
  by\ \bibinfo {editor} {\bibfnamefont {D.}~\bibnamefont {S\'en\'echal}},
  \bibinfo {editor} {\bibfnamefont {A.-M.}\ \bibnamefont {Tremblay}},\ and\
  \bibinfo {editor} {\bibfnamefont {C.}~\bibnamefont {Bourbonnais}}}\ (\bibinfo
   {publisher} {Springer-Verlag},\ \bibinfo {address} {New York},\ \bibinfo
  {year} {2004})\ pp.\ \bibinfo {pages} {39--74}\BibitemShut {NoStop}%
\bibitem [{\citenamefont {Vogiatzis}\ \emph {et~al.}(2017)\citenamefont
  {Vogiatzis}, \citenamefont {Ma}, \citenamefont {Olsen}, \citenamefont
  {Gagliardi},\ and\ \citenamefont {de~Jong}}]{Vogiatzis2017}%
  \BibitemOpen
  \bibfield  {author} {\bibinfo {author} {\bibfnamefont {K.~D.}\ \bibnamefont
  {Vogiatzis}}, \bibinfo {author} {\bibfnamefont {D.}~\bibnamefont {Ma}},
  \bibinfo {author} {\bibfnamefont {J.}~\bibnamefont {Olsen}}, \bibinfo
  {author} {\bibfnamefont {L.}~\bibnamefont {Gagliardi}},\ and\ \bibinfo
  {author} {\bibfnamefont {W.~A.}\ \bibnamefont {de~Jong}},\ }\bibfield
  {title} {\enquote {\bibinfo {title} {Pushing configuration-interaction to the
  limit: Towards massively parallel {MCSCF} calculations},}\ }\href
  {https://doi.org/10.1063/1.4989858} {\bibfield  {journal} {\bibinfo
  {journal} {J. Chem. Phys.}\ }\textbf {\bibinfo {volume} {147}},\ \bibinfo
  {pages} {184111} (\bibinfo {year} {2017})}\BibitemShut {NoStop}%
\bibitem [{\citenamefont {Olivares-Amaya}\ \emph {et~al.}(2015)\citenamefont
  {Olivares-Amaya}, \citenamefont {Hu}, \citenamefont {Nakatani}, \citenamefont
  {Sharma}, \citenamefont {Yang},\ and\ \citenamefont
  {Chan}}]{OlivaresAmaya2015}%
  \BibitemOpen
  \bibfield  {author} {\bibinfo {author} {\bibfnamefont {R.}~\bibnamefont
  {Olivares-Amaya}}, \bibinfo {author} {\bibfnamefont {W.}~\bibnamefont {Hu}},
  \bibinfo {author} {\bibfnamefont {N.}~\bibnamefont {Nakatani}}, \bibinfo
  {author} {\bibfnamefont {S.}~\bibnamefont {Sharma}}, \bibinfo {author}
  {\bibfnamefont {J.}~\bibnamefont {Yang}},\ and\ \bibinfo {author}
  {\bibfnamefont {G.~K.-L.}\ \bibnamefont {Chan}},\ }\bibfield  {title}
  {\enquote {\bibinfo {title} {The ab-initio density matrix renormalization
  group in practice},}\ }\href {https://doi.org/10.1063/1.4905329} {\bibfield
  {journal} {\bibinfo  {journal} {J. Chem. Phys.}\ }\textbf {\bibinfo {volume}
  {142}},\ \bibinfo {pages} {034102} (\bibinfo {year} {2015})}\BibitemShut
  {NoStop}%
\bibitem [{\citenamefont {Sharma}\ \emph {et~al.}(2017)\citenamefont {Sharma},
  \citenamefont {Holmes}, \citenamefont {Jeanmairet}, \citenamefont {Alavi},\
  and\ \citenamefont {Umrigar}}]{Sharma2017}%
  \BibitemOpen
  \bibfield  {author} {\bibinfo {author} {\bibfnamefont {S.}~\bibnamefont
  {Sharma}}, \bibinfo {author} {\bibfnamefont {A.~A.}\ \bibnamefont {Holmes}},
  \bibinfo {author} {\bibfnamefont {G.}~\bibnamefont {Jeanmairet}}, \bibinfo
  {author} {\bibfnamefont {A.}~\bibnamefont {Alavi}},\ and\ \bibinfo {author}
  {\bibfnamefont {C.~J.}\ \bibnamefont {Umrigar}},\ }\bibfield  {title}
  {\enquote {\bibinfo {title} {Semistochastic heat-bath configuration
  interaction method: Selected configuration interaction with semistochastic
  perturbation theory},}\ }\href {https://doi.org/10.1021/acs.jctc.6b01028}
  {\bibfield  {journal} {\bibinfo  {journal} {J. Chem. Theory Comput.}\
  }\textbf {\bibinfo {volume} {13}},\ \bibinfo {pages} {1595--1604} (\bibinfo
  {year} {2017})}\BibitemShut {NoStop}%
\bibitem [{\citenamefont {Loos}, \citenamefont {Damour},\ and\ \citenamefont
  {Scemama}(2020)}]{Loos2020}%
  \BibitemOpen
  \bibfield  {author} {\bibinfo {author} {\bibfnamefont {P.-F.}\ \bibnamefont
  {Loos}}, \bibinfo {author} {\bibfnamefont {Y.}~\bibnamefont {Damour}},\ and\
  \bibinfo {author} {\bibfnamefont {A.}~\bibnamefont {Scemama}},\ }\bibfield
  {title} {\enquote {\bibinfo {title} {The performance of {CIPSI} on the ground
  state electronic energy of benzene},}\ }\href
  {https://doi.org/10.1063/5.0027617} {\bibfield  {journal} {\bibinfo
  {journal} {J. Chem. Phys.}\ }\textbf {\bibinfo {volume} {153}},\ \bibinfo
  {pages} {176101} (\bibinfo {year} {2020})}\BibitemShut {NoStop}%
\bibitem [{\citenamefont {Malmqvist}, \citenamefont {Rendell},\ and\
  \citenamefont {Roos}(1990)}]{Malmqvist1990}%
  \BibitemOpen
  \bibfield  {author} {\bibinfo {author} {\bibfnamefont {P.~A.}\ \bibnamefont
  {Malmqvist}}, \bibinfo {author} {\bibfnamefont {A.}~\bibnamefont {Rendell}},\
  and\ \bibinfo {author} {\bibfnamefont {B.~O.}\ \bibnamefont {Roos}},\
  }\bibfield  {title} {\enquote {\bibinfo {title} {The restricted active space
  self-consistent-field method, implemented with a split graph unitary group
  approach},}\ }\href {https://doi.org/10.1021/j100377a011} {\ \textbf
  {\bibinfo {volume} {94}},\ \bibinfo {pages} {5477--5482} (\bibinfo {year}
  {1990})}\BibitemShut {NoStop}%
\bibitem [{\citenamefont {Ma}, \citenamefont {Li~Manni},\ and\ \citenamefont
  {Gagliardi}(2011)}]{Ma2011}%
  \BibitemOpen
  \bibfield  {author} {\bibinfo {author} {\bibfnamefont {D.}~\bibnamefont
  {Ma}}, \bibinfo {author} {\bibfnamefont {G.}~\bibnamefont {Li~Manni}},\ and\
  \bibinfo {author} {\bibfnamefont {L.}~\bibnamefont {Gagliardi}},\ }\bibfield
  {title} {\enquote {\bibinfo {title} {The generalized active space concept in
  multiconfigurational self-consistent field methods},}\ }\href
  {https://doi.org/10.1063/1.3611401} {\bibfield  {journal} {\bibinfo
  {journal} {J. Chem. Phys.}\ }\textbf {\bibinfo {volume} {135}},\ \bibinfo
  {pages} {044128} (\bibinfo {year} {2011})}\BibitemShut {NoStop}%
\bibitem [{\citenamefont {Hermes}, \citenamefont {Pandharkar},\ and\
  \citenamefont {Gagliardi}(2020)}]{Hermes2020}%
  \BibitemOpen
  \bibfield  {author} {\bibinfo {author} {\bibfnamefont {M.~R.}\ \bibnamefont
  {Hermes}}, \bibinfo {author} {\bibfnamefont {R.}~\bibnamefont {Pandharkar}},\
  and\ \bibinfo {author} {\bibfnamefont {L.}~\bibnamefont {Gagliardi}},\
  }\bibfield  {title} {\enquote {\bibinfo {title} {Variational localized active
  space self-consistent field method},}\ }\href
  {https://doi.org/10.1021/acs.jctc.0c00222} {\bibfield  {journal} {\bibinfo
  {journal} {J. Chem. Theory Comput.}\ }\textbf {\bibinfo {volume} {16}},\
  \bibinfo {pages} {4923--4937} (\bibinfo {year} {2020})}\BibitemShut {NoStop}%
\bibitem [{\citenamefont {Foulkes}\ \emph {et~al.}(2001)\citenamefont
  {Foulkes}, \citenamefont {Mitas}, \citenamefont {Needs},\ and\ \citenamefont
  {Rajagopal}}]{Foulkes2001}%
  \BibitemOpen
  \bibfield  {author} {\bibinfo {author} {\bibfnamefont {W.~M.~C.}\
  \bibnamefont {Foulkes}}, \bibinfo {author} {\bibfnamefont {L.}~\bibnamefont
  {Mitas}}, \bibinfo {author} {\bibfnamefont {R.~J.}\ \bibnamefont {Needs}},\
  and\ \bibinfo {author} {\bibfnamefont {G.}~\bibnamefont {Rajagopal}},\
  }\bibfield  {title} {\enquote {\bibinfo {title} {Quantum {M}onte {C}arlo
  simulations of solids},}\ }\href {https://doi.org/10.1103/RevModPhys.73.33}
  {\bibfield  {journal} {\bibinfo  {journal} {Rev. Mod. Phys.}\ }\textbf
  {\bibinfo {volume} {73}},\ \bibinfo {pages} {33--83} (\bibinfo {year}
  {2001})}\BibitemShut {NoStop}%
\bibitem [{\citenamefont {Wagner}\ and\ \citenamefont
  {Ceperley}(2016)}]{Wagner2016}%
  \BibitemOpen
  \bibfield  {author} {\bibinfo {author} {\bibfnamefont {L.~K.}\ \bibnamefont
  {Wagner}}\ and\ \bibinfo {author} {\bibfnamefont {D.~M.}\ \bibnamefont
  {Ceperley}},\ }\bibfield  {title} {\enquote {\bibinfo {title} {Discovering
  correlated fermions using quantum {M}onte {C}arlo},}\ }\href
  {https://doi.org/10.1088/0034-4885/79/9/094501} {\bibfield  {journal}
  {\bibinfo  {journal} {Rep. Prog. Phys.}\ }\textbf {\bibinfo {volume} {79}},\
  \bibinfo {pages} {094501} (\bibinfo {year} {2016})}\BibitemShut {NoStop}%
\bibitem [{\citenamefont {Booth}, \citenamefont {Thom},\ and\ \citenamefont
  {Alavi}(2009)}]{Booth2009}%
  \BibitemOpen
  \bibfield  {author} {\bibinfo {author} {\bibfnamefont {G.~H.}\ \bibnamefont
  {Booth}}, \bibinfo {author} {\bibfnamefont {A.~J.~W.}\ \bibnamefont {Thom}},\
  and\ \bibinfo {author} {\bibfnamefont {A.}~\bibnamefont {Alavi}},\ }\bibfield
   {title} {\enquote {\bibinfo {title} {Fermion {M}onte {C}arlo without fixed
  nodes: A game of life, death, and annihilation in {S}later determinant
  space},}\ }\href {https://doi.org/10.1063/1.3193710} {\bibfield  {journal}
  {\bibinfo  {journal} {J. Chem. Phys.}\ }\textbf {\bibinfo {volume} {131}},\
  \bibinfo {pages} {054106} (\bibinfo {year} {2009})}\BibitemShut {NoStop}%
\bibitem [{\citenamefont {Wouters}\ \emph
  {et~al.}(2014{\natexlab{a}})\citenamefont {Wouters}, \citenamefont
  {Verstichel}, \citenamefont {Van~Neck},\ and\ \citenamefont
  {Chan}}]{Wouters2014projector}%
  \BibitemOpen
  \bibfield  {author} {\bibinfo {author} {\bibfnamefont {S.}~\bibnamefont
  {Wouters}}, \bibinfo {author} {\bibfnamefont {B.}~\bibnamefont {Verstichel}},
  \bibinfo {author} {\bibfnamefont {D.}~\bibnamefont {Van~Neck}},\ and\
  \bibinfo {author} {\bibfnamefont {G.~K.-L.}\ \bibnamefont {Chan}},\
  }\bibfield  {title} {\enquote {\bibinfo {title} {Projector quantum {M}onte
  {C}arlo with matrix product states},}\ }\href
  {https://doi.org/10.1103/PhysRevB.90.045104} {\bibfield  {journal} {\bibinfo
  {journal} {Phys. Rev. B}\ }\textbf {\bibinfo {volume} {90}},\ \bibinfo
  {pages} {045104} (\bibinfo {year} {2014}{\natexlab{a}})}\BibitemShut
  {NoStop}%
\bibitem [{\citenamefont {Umrigar}(2015)}]{Umrigar2015}%
  \BibitemOpen
  \bibfield  {author} {\bibinfo {author} {\bibfnamefont {C.~J.}\ \bibnamefont
  {Umrigar}},\ }\bibfield  {title} {\enquote {\bibinfo {title} {Observations on
  variational and projector {M}onte {C}arlo methods},}\ }\href
  {https://doi.org/10.1063/1.4933112} {\bibfield  {journal} {\bibinfo
  {journal} {J. Chem. Phys.}\ }\textbf {\bibinfo {volume} {143}},\ \bibinfo
  {pages} {164105} (\bibinfo {year} {2015})}\BibitemShut {NoStop}%
\bibitem [{\citenamefont {Schwarz}, \citenamefont {Alavi},\ and\ \citenamefont
  {Booth}(2017)}]{Schwarz2017}%
  \BibitemOpen
  \bibfield  {author} {\bibinfo {author} {\bibfnamefont {L.~R.}\ \bibnamefont
  {Schwarz}}, \bibinfo {author} {\bibfnamefont {A.}~\bibnamefont {Alavi}},\
  and\ \bibinfo {author} {\bibfnamefont {G.~H.}\ \bibnamefont {Booth}},\
  }\bibfield  {title} {\enquote {\bibinfo {title} {Projector quantum {M}onte
  {C}arlo method for nonlinear wave functions},}\ }\href
  {https://doi.org/10.1103/PhysRevLett.118.176403} {\bibfield  {journal}
  {\bibinfo  {journal} {Phys. Rev. Lett.}\ }\textbf {\bibinfo {volume} {118}},\
  \bibinfo {pages} {176403} (\bibinfo {year} {2017})}\BibitemShut {NoStop}%
\bibitem [{\citenamefont {Han}, \citenamefont {Lu},\ and\ \citenamefont
  {Zhou}(2020)}]{Han2020}%
  \BibitemOpen
  \bibfield  {author} {\bibinfo {author} {\bibfnamefont {J.}~\bibnamefont
  {Han}}, \bibinfo {author} {\bibfnamefont {J.}~\bibnamefont {Lu}},\ and\
  \bibinfo {author} {\bibfnamefont {M.}~\bibnamefont {Zhou}},\ }\bibfield
  {title} {\enquote {\bibinfo {title} {Solving high-dimensional eigenvalue
  problems using deep neural networks: A diffusion {M}onte {C}arlo like
  approach},}\ }\href
  {https://doi.org/https://doi.org/10.1016/j.jcp.2020.109792} {\bibfield
  {journal} {\bibinfo  {journal} {J. Comput. Phys.}\ }\textbf {\bibinfo
  {volume} {423}},\ \bibinfo {pages} {109792} (\bibinfo {year}
  {2020})}\BibitemShut {NoStop}%
\bibitem [{\citenamefont {Booth}\ \emph {et~al.}(2011)\citenamefont {Booth},
  \citenamefont {Cleland}, \citenamefont {Thom},\ and\ \citenamefont
  {Alavi}}]{Booth2011}%
  \BibitemOpen
  \bibfield  {author} {\bibinfo {author} {\bibfnamefont {G.~H.}\ \bibnamefont
  {Booth}}, \bibinfo {author} {\bibfnamefont {D.}~\bibnamefont {Cleland}},
  \bibinfo {author} {\bibfnamefont {A.~J.~W.}\ \bibnamefont {Thom}},\ and\
  \bibinfo {author} {\bibfnamefont {A.}~\bibnamefont {Alavi}},\ }\bibfield
  {title} {\enquote {\bibinfo {title} {Breaking the carbon dimer: The
  challenges of multiple bond dissociation with full configuration interaction
  quantum {M}onte {C}arlo methods},}\ }\href
  {https://doi.org/10.1063/1.3624383} {\bibfield  {journal} {\bibinfo
  {journal} {J. Chem. Phys.}\ }\textbf {\bibinfo {volume} {135}},\ \bibinfo
  {pages} {084104} (\bibinfo {year} {2011})}\BibitemShut {NoStop}%
\bibitem [{\citenamefont {Zhang}(2018)}]{Zhang2018}%
  \BibitemOpen
  \bibfield  {author} {\bibinfo {author} {\bibfnamefont {S.}~\bibnamefont
  {Zhang}},\ }\enquote {\bibinfo {title} {Ab initio electronic structure
  calculations by auxiliary-field quantum {M}onte {C}arlo},}\ in\ \href
  {https://doi.org/10.1007/978-3-319-42913-7_47-1} {\emph {\bibinfo {booktitle}
  {Handbook of Materials Modeling : Methods: Theory and Modeling}}},\ \bibinfo
  {editor} {edited by\ \bibinfo {editor} {\bibfnamefont {W.}~\bibnamefont
  {Andreoni}}\ and\ \bibinfo {editor} {\bibfnamefont {S.}~\bibnamefont {Yip}}}\
  (\bibinfo  {publisher} {Springer International Publishing},\ \bibinfo {year}
  {2018})\ pp.\ \bibinfo {pages} {1--27}\BibitemShut {NoStop}%
\bibitem [{\citenamefont {Grimes}\ \emph {et~al.}(1986)\citenamefont {Grimes},
  \citenamefont {Hammond}, \citenamefont {Reynolds},\ and\ \citenamefont
  {Lester}}]{Grimes1986}%
  \BibitemOpen
  \bibfield  {author} {\bibinfo {author} {\bibfnamefont {R.~M.}\ \bibnamefont
  {Grimes}}, \bibinfo {author} {\bibfnamefont {B.~L.}\ \bibnamefont {Hammond}},
  \bibinfo {author} {\bibfnamefont {P.~J.}\ \bibnamefont {Reynolds}},\ and\
  \bibinfo {author} {\bibfnamefont {W.~A.}\ \bibnamefont {Lester}},\ }\bibfield
   {title} {\enquote {\bibinfo {title} {Quantum {M}onte {C}arlo approach to
  electronically excited molecules},}\ }\href
  {https://doi.org/10.1063/1.451754} {\bibfield  {journal} {\bibinfo  {journal}
  {J. Chem. Phys.}\ }\textbf {\bibinfo {volume} {85}},\ \bibinfo {pages}
  {4749--4750} (\bibinfo {year} {1986})}\BibitemShut {NoStop}%
\bibitem [{\citenamefont {Scemama}\ \emph {et~al.}(2018)\citenamefont
  {Scemama}, \citenamefont {Benali}, \citenamefont {Jacquemin}, \citenamefont
  {Caffarel},\ and\ \citenamefont {Loos}}]{Scemama2018excitation}%
  \BibitemOpen
  \bibfield  {author} {\bibinfo {author} {\bibfnamefont {A.}~\bibnamefont
  {Scemama}}, \bibinfo {author} {\bibfnamefont {A.}~\bibnamefont {Benali}},
  \bibinfo {author} {\bibfnamefont {D.}~\bibnamefont {Jacquemin}}, \bibinfo
  {author} {\bibfnamefont {M.}~\bibnamefont {Caffarel}},\ and\ \bibinfo
  {author} {\bibfnamefont {P.-F.}\ \bibnamefont {Loos}},\ }\bibfield  {title}
  {\enquote {\bibinfo {title} {Excitation energies from diffusion {M}onte
  {C}arlo using selected configuration interaction nodes},}\ }\href
  {https://doi.org/10.1063/1.5041327} {\bibfield  {journal} {\bibinfo
  {journal} {J. Chem. Phys.}\ }\textbf {\bibinfo {volume} {149}},\ \bibinfo
  {pages} {034108} (\bibinfo {year} {2018})}\BibitemShut {NoStop}%
\bibitem [{\citenamefont {Purwanto}, \citenamefont {Zhang},\ and\ \citenamefont
  {Krakauer}(2009)}]{Purwanto2009excited}%
  \BibitemOpen
  \bibfield  {author} {\bibinfo {author} {\bibfnamefont {W.}~\bibnamefont
  {Purwanto}}, \bibinfo {author} {\bibfnamefont {S.}~\bibnamefont {Zhang}},\
  and\ \bibinfo {author} {\bibfnamefont {H.}~\bibnamefont {Krakauer}},\
  }\bibfield  {title} {\enquote {\bibinfo {title} {Excited state calculations
  using phaseless auxiliary-field quantum {M}onte {C}arlo: Potential energy
  curves of low-lying {C}$_2$ singlet states},}\ }\href
  {https://doi.org/10.1063/1.3077920} {\bibfield  {journal} {\bibinfo
  {journal} {J. Chem. Phys.}\ }\textbf {\bibinfo {volume} {130}},\ \bibinfo
  {pages} {094107} (\bibinfo {year} {2009})}\BibitemShut {NoStop}%
\bibitem [{\citenamefont {Dobrautz}, \citenamefont {Smart},\ and\ \citenamefont
  {Alavi}(2019)}]{Dobrautz2019}%
  \BibitemOpen
  \bibfield  {author} {\bibinfo {author} {\bibfnamefont {W.}~\bibnamefont
  {Dobrautz}}, \bibinfo {author} {\bibfnamefont {S.~D.}\ \bibnamefont
  {Smart}},\ and\ \bibinfo {author} {\bibfnamefont {A.}~\bibnamefont {Alavi}},\
  }\bibfield  {title} {\enquote {\bibinfo {title} {Efficient formulation of
  full configuration interaction quantum {M}onte {C}arlo in a spin eigenbasis
  via the graphical unitary group approach},}\ }\href
  {https://doi.org/10.1063/1.5108908} {\bibfield  {journal} {\bibinfo
  {journal} {J. Chem. Phys.}\ }\textbf {\bibinfo {volume} {151}},\ \bibinfo
  {pages} {094104} (\bibinfo {year} {2019})}\BibitemShut {NoStop}%
\bibitem [{\citenamefont {Guther}\ \emph {et~al.}(2020)\citenamefont {Guther},
  \citenamefont {Anderson}, \citenamefont {Blunt}, \citenamefont {Bogdanov},
  \citenamefont {Cleland}, \citenamefont {Dattani}, \citenamefont {Dobrautz},
  \citenamefont {Ghanem}, \citenamefont {Jeszenski}, \citenamefont
  {Liebermann}, \citenamefont {Manni}, \citenamefont {Lozovoi}, \citenamefont
  {Luo}, \citenamefont {Ma}, \citenamefont {Merz}, \citenamefont {Overy},
  \citenamefont {Rampp}, \citenamefont {Samanta}, \citenamefont {Schwarz},
  \citenamefont {Shepherd}, \citenamefont {Smart}, \citenamefont {Vitale},
  \citenamefont {Weser}, \citenamefont {Booth},\ and\ \citenamefont
  {Alavi}}]{Guther2020}%
  \BibitemOpen
  \bibfield  {author} {\bibinfo {author} {\bibfnamefont {K.}~\bibnamefont
  {Guther}}, \bibinfo {author} {\bibfnamefont {R.~J.}\ \bibnamefont
  {Anderson}}, \bibinfo {author} {\bibfnamefont {N.~S.}\ \bibnamefont {Blunt}},
  \bibinfo {author} {\bibfnamefont {N.~A.}\ \bibnamefont {Bogdanov}}, \bibinfo
  {author} {\bibfnamefont {D.}~\bibnamefont {Cleland}}, \bibinfo {author}
  {\bibfnamefont {N.}~\bibnamefont {Dattani}}, \bibinfo {author} {\bibfnamefont
  {W.}~\bibnamefont {Dobrautz}}, \bibinfo {author} {\bibfnamefont
  {K.}~\bibnamefont {Ghanem}}, \bibinfo {author} {\bibfnamefont
  {P.}~\bibnamefont {Jeszenski}}, \bibinfo {author} {\bibfnamefont
  {N.}~\bibnamefont {Liebermann}}, \bibinfo {author} {\bibfnamefont {G.~L.}\
  \bibnamefont {Manni}}, \bibinfo {author} {\bibfnamefont {A.~Y.}\ \bibnamefont
  {Lozovoi}}, \bibinfo {author} {\bibfnamefont {H.}~\bibnamefont {Luo}},
  \bibinfo {author} {\bibfnamefont {D.}~\bibnamefont {Ma}}, \bibinfo {author}
  {\bibfnamefont {F.}~\bibnamefont {Merz}}, \bibinfo {author} {\bibfnamefont
  {C.}~\bibnamefont {Overy}}, \bibinfo {author} {\bibfnamefont
  {M.}~\bibnamefont {Rampp}}, \bibinfo {author} {\bibfnamefont {P.~K.}\
  \bibnamefont {Samanta}}, \bibinfo {author} {\bibfnamefont {L.~R.}\
  \bibnamefont {Schwarz}}, \bibinfo {author} {\bibfnamefont {J.~J.}\
  \bibnamefont {Shepherd}}, \bibinfo {author} {\bibfnamefont {S.~D.}\
  \bibnamefont {Smart}}, \bibinfo {author} {\bibfnamefont {E.}~\bibnamefont
  {Vitale}}, \bibinfo {author} {\bibfnamefont {O.}~\bibnamefont {Weser}},
  \bibinfo {author} {\bibfnamefont {G.~H.}\ \bibnamefont {Booth}},\ and\
  \bibinfo {author} {\bibfnamefont {A.}~\bibnamefont {Alavi}},\ }\bibfield
  {title} {\enquote {\bibinfo {title} {{NECI}: {N}-electron configuration
  interaction with emphasis on state-of-the-art stochastic methods},}\ }\href
  {https://doi.org/10.1063/5.0005754} {\bibfield  {journal} {\bibinfo
  {journal} {J. Chem. Phys.}\ }\textbf {\bibinfo {volume} {153}},\ \bibinfo
  {pages} {034107} (\bibinfo {year} {2020})}\BibitemShut {NoStop}%
\bibitem [{\citenamefont {Ohtsuka}\ and\ \citenamefont
  {Nagase}(2010)}]{Ohtsuka2010}%
  \BibitemOpen
  \bibfield  {author} {\bibinfo {author} {\bibfnamefont {Y.}~\bibnamefont
  {Ohtsuka}}\ and\ \bibinfo {author} {\bibfnamefont {S.}~\bibnamefont
  {Nagase}},\ }\bibfield  {title} {\enquote {\bibinfo {title} {Projector
  {M}onte {C}arlo method based on {S}later determinants: Test application to
  singlet excited states of {H}$_2${O} and {LiF}},}\ }\href
  {https://doi.org/https://doi.org/10.1016/j.cplett.2009.12.047} {\bibfield
  {journal} {\bibinfo  {journal} {Chem. Phys. Lett.}\ }\textbf {\bibinfo
  {volume} {485}},\ \bibinfo {pages} {367 -- 370} (\bibinfo {year}
  {2010})}\BibitemShut {NoStop}%
\bibitem [{\citenamefont {Blunt}\ \emph {et~al.}(2015)\citenamefont {Blunt},
  \citenamefont {Smart}, \citenamefont {Booth},\ and\ \citenamefont
  {Alavi}}]{Blunt2015}%
  \BibitemOpen
  \bibfield  {author} {\bibinfo {author} {\bibfnamefont {N.~S.}\ \bibnamefont
  {Blunt}}, \bibinfo {author} {\bibfnamefont {S.~D.}\ \bibnamefont {Smart}},
  \bibinfo {author} {\bibfnamefont {G.~H.}\ \bibnamefont {Booth}},\ and\
  \bibinfo {author} {\bibfnamefont {A.}~\bibnamefont {Alavi}},\ }\bibfield
  {title} {\enquote {\bibinfo {title} {An excited-state approach within full
  configuration interaction quantum {M}onte {C}arlo},}\ }\href
  {https://doi.org/10.1063/1.4932595} {\bibfield  {journal} {\bibinfo
  {journal} {J. Chem. Phys.}\ }\textbf {\bibinfo {volume} {143}},\ \bibinfo
  {pages} {134117} (\bibinfo {year} {2015})}\BibitemShut {NoStop}%
\bibitem [{\citenamefont {Blunt}, \citenamefont {Alavi},\ and\ \citenamefont
  {Booth}(2015)}]{Blunt2015a}%
  \BibitemOpen
  \bibfield  {author} {\bibinfo {author} {\bibfnamefont {N.~S.}\ \bibnamefont
  {Blunt}}, \bibinfo {author} {\bibfnamefont {A.}~\bibnamefont {Alavi}},\ and\
  \bibinfo {author} {\bibfnamefont {G.~H.}\ \bibnamefont {Booth}},\ }\bibfield
  {title} {\enquote {\bibinfo {title} {Krylov-projected quantum {M}onte {C}arlo
  method},}\ }\href {https://doi.org/10.1103/PhysRevLett.115.050603} {\bibfield
   {journal} {\bibinfo  {journal} {Phys. Rev. Lett.}\ }\textbf {\bibinfo
  {volume} {115}},\ \bibinfo {pages} {050603} (\bibinfo {year}
  {2015})}\BibitemShut {NoStop}%
\bibitem [{\citenamefont {Greene}\ \emph
  {et~al.}(2022{\natexlab{a}})\citenamefont {Greene}, \citenamefont {Webber},
  \citenamefont {Berkelbach},\ and\ \citenamefont {Weare}}]{Greene2022}%
  \BibitemOpen
  \bibfield  {author} {\bibinfo {author} {\bibfnamefont {S.~M.}\ \bibnamefont
  {Greene}}, \bibinfo {author} {\bibfnamefont {R.~J.}\ \bibnamefont {Webber}},
  \bibinfo {author} {\bibfnamefont {T.~C.}\ \bibnamefont {Berkelbach}},\ and\
  \bibinfo {author} {\bibfnamefont {J.}~\bibnamefont {Weare}},\ }\bibfield
  {title} {\enquote {\bibinfo {title} {Approximating matrix eigenvalues by
  subspace iteration with repeated random sparsification},}\ }\href
  {https://doi.org/10.1137/21M1422513} {\bibfield  {journal} {\bibinfo
  {journal} {SIAM J. Sci. Comput.}\ }\textbf {\bibinfo {volume} {44}},\
  \bibinfo {pages} {A3067--A3097} (\bibinfo {year}
  {2022}{\natexlab{a}})}\BibitemShut {NoStop}%
\bibitem [{\citenamefont {Blunt}, \citenamefont {Alavi},\ and\ \citenamefont
  {Booth}(2018)}]{Blunt2018nonlinear}%
  \BibitemOpen
  \bibfield  {author} {\bibinfo {author} {\bibfnamefont {N.~S.}\ \bibnamefont
  {Blunt}}, \bibinfo {author} {\bibfnamefont {A.}~\bibnamefont {Alavi}},\ and\
  \bibinfo {author} {\bibfnamefont {G.~H.}\ \bibnamefont {Booth}},\ }\bibfield
  {title} {\enquote {\bibinfo {title} {Nonlinear biases, stochastically sampled
  effective {H}amiltonians, and spectral functions in quantum {M}onte {C}arlo
  methods},}\ }\href {https://doi.org/10.1103/PhysRevB.98.085118} {\bibfield
  {journal} {\bibinfo  {journal} {Phys. Rev. B}\ }\textbf {\bibinfo {volume}
  {98}},\ \bibinfo {pages} {085118} (\bibinfo {year} {2018})}\BibitemShut
  {NoStop}%
\bibitem [{\citenamefont {Lim}\ and\ \citenamefont {Weare}(2017)}]{Lim2017}%
  \BibitemOpen
  \bibfield  {author} {\bibinfo {author} {\bibfnamefont {L.-H.}\ \bibnamefont
  {Lim}}\ and\ \bibinfo {author} {\bibfnamefont {J.}~\bibnamefont {Weare}},\
  }\bibfield  {title} {\enquote {\bibinfo {title} {Fast randomized iteration:
  Diffusion {M}onte {C}arlo through the lens of numerical linear algebra},}\
  }\href {https://doi.org/10.1137/15M1040827} {\bibfield  {journal} {\bibinfo
  {journal} {SIAM Rev.}\ }\textbf {\bibinfo {volume} {59}},\ \bibinfo {pages}
  {547--587} (\bibinfo {year} {2017})}\BibitemShut {NoStop}%
\bibitem [{\citenamefont {Greene}\ \emph {et~al.}(2019)\citenamefont {Greene},
  \citenamefont {Webber}, \citenamefont {Weare},\ and\ \citenamefont
  {Berkelbach}}]{Greene2019}%
  \BibitemOpen
  \bibfield  {author} {\bibinfo {author} {\bibfnamefont {S.~M.}\ \bibnamefont
  {Greene}}, \bibinfo {author} {\bibfnamefont {R.~J.}\ \bibnamefont {Webber}},
  \bibinfo {author} {\bibfnamefont {J.}~\bibnamefont {Weare}},\ and\ \bibinfo
  {author} {\bibfnamefont {T.~C.}\ \bibnamefont {Berkelbach}},\ }\bibfield
  {title} {\enquote {\bibinfo {title} {Beyond walkers in stochastic quantum
  chemistry: Reducing error using fast randomized iteration},}\ }\href
  {https://doi.org/10.1021/acs.jctc.9b00422} {\bibfield  {journal} {\bibinfo
  {journal} {J. Chem. Theory Comput.}\ }\textbf {\bibinfo {volume} {15}},\
  \bibinfo {pages} {4834--4850} (\bibinfo {year} {2019})}\BibitemShut {NoStop}%
\bibitem [{\citenamefont {Greene}\ \emph {et~al.}(2020)\citenamefont {Greene},
  \citenamefont {Webber}, \citenamefont {Weare},\ and\ \citenamefont
  {Berkelbach}}]{Greene2020}%
  \BibitemOpen
  \bibfield  {author} {\bibinfo {author} {\bibfnamefont {S.~M.}\ \bibnamefont
  {Greene}}, \bibinfo {author} {\bibfnamefont {R.~J.}\ \bibnamefont {Webber}},
  \bibinfo {author} {\bibfnamefont {J.}~\bibnamefont {Weare}},\ and\ \bibinfo
  {author} {\bibfnamefont {T.~C.}\ \bibnamefont {Berkelbach}},\ }\bibfield
  {title} {\enquote {\bibinfo {title} {Improved fast randomized iteration
  approach to full configuration interaction},}\ }\href
  {https://doi.org/10.1021/acs.jctc.0c00437} {\bibfield  {journal} {\bibinfo
  {journal} {J. Chem. Theory Comput.}\ }\textbf {\bibinfo {volume} {16}},\
  \bibinfo {pages} {5572--5585} (\bibinfo {year} {2020})}\BibitemShut {NoStop}%
\bibitem [{\citenamefont {Cleland}, \citenamefont {Booth},\ and\ \citenamefont
  {Alavi}(2010)}]{Cleland2010}%
  \BibitemOpen
  \bibfield  {author} {\bibinfo {author} {\bibfnamefont {D.}~\bibnamefont
  {Cleland}}, \bibinfo {author} {\bibfnamefont {G.~H.}\ \bibnamefont {Booth}},\
  and\ \bibinfo {author} {\bibfnamefont {A.}~\bibnamefont {Alavi}},\ }\bibfield
   {title} {\enquote {\bibinfo {title} {Communications: Survival of the
  fittest: Accelerating convergence in full configuration-interaction quantum
  {M}onte {C}arlo},}\ }\href@noop {} {\bibfield  {journal} {\bibinfo  {journal}
  {J. Chem. Phys.}\ }\textbf {\bibinfo {volume} {132}} (\bibinfo {year}
  {2010})}\BibitemShut {NoStop}%
\bibitem [{\citenamefont {Neufeld}\ and\ \citenamefont
  {Thom}(2019)}]{Neufeld2019}%
  \BibitemOpen
  \bibfield  {author} {\bibinfo {author} {\bibfnamefont {V.~A.}\ \bibnamefont
  {Neufeld}}\ and\ \bibinfo {author} {\bibfnamefont {A.~J.~W.}\ \bibnamefont
  {Thom}},\ }\bibfield  {title} {\enquote {\bibinfo {title} {Exciting
  determinants in quantum {M}onte {C}arlo: Loading the dice with fast,
  low-memory weights},}\ }\href {https://doi.org/10.1021/acs.jctc.8b00844}
  {\bibfield  {journal} {\bibinfo  {journal} {J. Chem. Theory Comput.}\
  }\textbf {\bibinfo {volume} {15}},\ \bibinfo {pages} {127--140} (\bibinfo
  {year} {2019})}\BibitemShut {NoStop}%
\bibitem [{\citenamefont {Holmes}, \citenamefont {Umrigar},\ and\ \citenamefont
  {Sharma}(2017)}]{Holmes2017}%
  \BibitemOpen
  \bibfield  {author} {\bibinfo {author} {\bibfnamefont {A.~A.}\ \bibnamefont
  {Holmes}}, \bibinfo {author} {\bibfnamefont {C.~J.}\ \bibnamefont
  {Umrigar}},\ and\ \bibinfo {author} {\bibfnamefont {S.}~\bibnamefont
  {Sharma}},\ }\bibfield  {title} {\enquote {\bibinfo {title} {Excited states
  using semistochastic heat-bath configuration interaction},}\ }\href
  {https://doi.org/10.1063/1.4998614} {\bibfield  {journal} {\bibinfo
  {journal} {J. Chem. Phys.}\ }\textbf {\bibinfo {volume} {147}},\ \bibinfo
  {pages} {164111} (\bibinfo {year} {2017})}\BibitemShut {NoStop}%
\bibitem [{\citenamefont {Holmes}, \citenamefont {Changlani},\ and\
  \citenamefont {Umrigar}(2016)}]{Holmes2016}%
  \BibitemOpen
  \bibfield  {author} {\bibinfo {author} {\bibfnamefont {A.~A.}\ \bibnamefont
  {Holmes}}, \bibinfo {author} {\bibfnamefont {H.~J.}\ \bibnamefont
  {Changlani}},\ and\ \bibinfo {author} {\bibfnamefont {C.~J.}\ \bibnamefont
  {Umrigar}},\ }\bibfield  {title} {\enquote {\bibinfo {title} {Efficient
  heat-bath sampling in {F}ock space},}\ }\href
  {https://doi.org/10.1021/acs.jctc.5b01170} {\bibfield  {journal} {\bibinfo
  {journal} {J. Chem. Theory Comput.}\ }\textbf {\bibinfo {volume} {12}},\
  \bibinfo {pages} {1561--1571} (\bibinfo {year} {2016})}\BibitemShut {NoStop}%
\bibitem [{\citenamefont {Sun}\ \emph {et~al.}(2018)\citenamefont {Sun},
  \citenamefont {Berkelbach}, \citenamefont {Blunt}, \citenamefont {Booth},
  \citenamefont {Guo}, \citenamefont {Li}, \citenamefont {Liu}, \citenamefont
  {McClain}, \citenamefont {Sayfutyarova}, \citenamefont {Sharma},
  \citenamefont {Wouters},\ and\ \citenamefont {Chan}}]{Sun2018}%
  \BibitemOpen
  \bibfield  {author} {\bibinfo {author} {\bibfnamefont {Q.}~\bibnamefont
  {Sun}}, \bibinfo {author} {\bibfnamefont {T.~C.}\ \bibnamefont {Berkelbach}},
  \bibinfo {author} {\bibfnamefont {N.~S.}\ \bibnamefont {Blunt}}, \bibinfo
  {author} {\bibfnamefont {G.~H.}\ \bibnamefont {Booth}}, \bibinfo {author}
  {\bibfnamefont {S.}~\bibnamefont {Guo}}, \bibinfo {author} {\bibfnamefont
  {Z.}~\bibnamefont {Li}}, \bibinfo {author} {\bibfnamefont {J.}~\bibnamefont
  {Liu}}, \bibinfo {author} {\bibfnamefont {J.~D.}\ \bibnamefont {McClain}},
  \bibinfo {author} {\bibfnamefont {E.~R.}\ \bibnamefont {Sayfutyarova}},
  \bibinfo {author} {\bibfnamefont {S.}~\bibnamefont {Sharma}}, \bibinfo
  {author} {\bibfnamefont {S.}~\bibnamefont {Wouters}},\ and\ \bibinfo {author}
  {\bibfnamefont {G.~K.-L.}\ \bibnamefont {Chan}},\ }\bibfield  {title}
  {\enquote {\bibinfo {title} {{PySCF}: the {P}ython-based simulations of
  chemistry framework},}\ }\href {https://doi.org/10.1002/wcms.1340} {\bibfield
   {journal} {\bibinfo  {journal} {Wiley Interdiscip. Rev.: Comput. Mol. Sci.}\
  }\textbf {\bibinfo {volume} {8}},\ \bibinfo {pages} {e1340} (\bibinfo {year}
  {2018})}\BibitemShut {NoStop}%
\bibitem [{\citenamefont {Stewart}(1969)}]{Stewart1969}%
  \BibitemOpen
  \bibfield  {author} {\bibinfo {author} {\bibfnamefont {G.~W.}\ \bibnamefont
  {Stewart}},\ }\bibfield  {title} {\enquote {\bibinfo {title} {Accelerating
  the orthogonal iteration for the eigenvectors of a {H}ermitian matrix},}\
  }\href {https://doi.org/10.1007/BF02165413} {\bibfield  {journal} {\bibinfo
  {journal} {Numer. Math.}\ }\textbf {\bibinfo {volume} {13}},\ \bibinfo
  {pages} {362--376} (\bibinfo {year} {1969})}\BibitemShut {NoStop}%
\bibitem [{\citenamefont {Stewart}(1975)}]{Stewart1975}%
  \BibitemOpen
  \bibfield  {author} {\bibinfo {author} {\bibfnamefont {G.}~\bibnamefont
  {Stewart}},\ }\bibfield  {title} {\enquote {\bibinfo {title} {Methods of
  simultaneous iteration for calculating eigenvectors of matrices},}\ }in\
  \href {https://doi.org/https://doi.org/10.1016/B978-0-12-496952-0.50023-4}
  {\emph {\bibinfo {booktitle} {Topics in Numerical Analysis II}}},\ \bibinfo
  {editor} {edited by\ \bibinfo {editor} {\bibfnamefont {J.~J.}\ \bibnamefont
  {Miller}}}\ (\bibinfo  {publisher} {Academic Press},\ \bibinfo {year}
  {1975})\ pp.\ \bibinfo {pages} {185 -- 196}\BibitemShut {NoStop}%
\bibitem [{\citenamefont {Saad}(2011)}]{Saad2011}%
  \BibitemOpen
  \bibfield  {author} {\bibinfo {author} {\bibfnamefont {Y.}~\bibnamefont
  {Saad}},\ }\href {https://books.google.com/books?id=FAkNAQAAIAAJ} {\emph
  {\bibinfo {title} {Numerical Methods for Large Eigenvalue Problems}}},\
  \bibinfo {edition} {2nd}\ ed.\ (\bibinfo  {publisher} {Society for Industrial
  and Applied Mathematics},\ \bibinfo {year} {2011})\BibitemShut {NoStop}%
\bibitem [{\citenamefont {Wilkinson}(1965)}]{wilkinson1965convergence}%
  \BibitemOpen
  \bibfield  {author} {\bibinfo {author} {\bibfnamefont {J.~H.}\ \bibnamefont
  {Wilkinson}},\ }\bibfield  {title} {\enquote {\bibinfo {title} {Convergence
  of the lr, qr, and related algorithms},}\ }\href@noop {} {\bibfield
  {journal} {\bibinfo  {journal} {The Computer Journal}\ }\textbf {\bibinfo
  {volume} {8}},\ \bibinfo {pages} {77--84} (\bibinfo {year}
  {1965})}\BibitemShut {NoStop}%
\bibitem [{\citenamefont {Ceperley}\ and\ \citenamefont
  {Bernu}(1988)}]{Ceperley1988}%
  \BibitemOpen
  \bibfield  {author} {\bibinfo {author} {\bibfnamefont {D.~M.}\ \bibnamefont
  {Ceperley}}\ and\ \bibinfo {author} {\bibfnamefont {B.}~\bibnamefont
  {Bernu}},\ }\bibfield  {title} {\enquote {\bibinfo {title} {The calculation
  of excited state properties with quantum {M}onte {C}arlo},}\ }\href
  {https://doi.org/10.1063/1.455398} {\bibfield  {journal} {\bibinfo  {journal}
  {J. Chem. Phys.}\ }\textbf {\bibinfo {volume} {89}},\ \bibinfo {pages}
  {6316--6328} (\bibinfo {year} {1988})}\BibitemShut {NoStop}%
\bibitem [{\citenamefont {{Foreman-Mackey}}\ \emph {et~al.}(2013)\citenamefont
  {{Foreman-Mackey}}, \citenamefont {{Hogg}}, \citenamefont {{Lang}},\ and\
  \citenamefont {{Goodman}}}]{Foreman2013}%
  \BibitemOpen
  \bibfield  {author} {\bibinfo {author} {\bibfnamefont {D.}~\bibnamefont
  {{Foreman-Mackey}}}, \bibinfo {author} {\bibfnamefont {D.~W.}\ \bibnamefont
  {{Hogg}}}, \bibinfo {author} {\bibfnamefont {D.}~\bibnamefont {{Lang}}},\
  and\ \bibinfo {author} {\bibfnamefont {J.}~\bibnamefont {{Goodman}}},\ }\href
  {https://doi.org/10.1086/670067} {\enquote {\bibinfo {title} {{emcee: The
  MCMC Hammer}},}\ }\bibinfo {howpublished} {Preprint at
  \url{https://arxiv.org/abs/1202.3665}} (\bibinfo {year} {2013}),\ \bibinfo
  {note} {(accessed January 26, 2022)}\BibitemShut {NoStop}%
\bibitem [{\citenamefont {Caffarel}\ \emph {et~al.}(2016)\citenamefont
  {Caffarel}, \citenamefont {Applencourt}, \citenamefont {Giner},\ and\
  \citenamefont {Scemama}}]{Caffarel2016using}%
  \BibitemOpen
  \bibfield  {author} {\bibinfo {author} {\bibfnamefont {M.}~\bibnamefont
  {Caffarel}}, \bibinfo {author} {\bibfnamefont {T.}~\bibnamefont
  {Applencourt}}, \bibinfo {author} {\bibfnamefont {E.}~\bibnamefont {Giner}},\
  and\ \bibinfo {author} {\bibfnamefont {A.}~\bibnamefont {Scemama}},\
  }\bibfield  {title} {\enquote {\bibinfo {title} {Using {CIPSI} nodes in
  diffusion {M}onte {C}arlo},}\ }in\ \href
  {https://doi.org/10.1021/bk-2016-1234.ch002} {\emph {\bibinfo {booktitle}
  {Recent Progress in Quantum {M}onte {C}arlo}}},\ \bibinfo {series} {{ACS}
  Symposium Series}, Vol.\ \bibinfo {volume} {1234}\ (\bibinfo  {publisher}
  {American Chemical Society},\ \bibinfo {year} {2016})\ pp.\ \bibinfo {pages}
  {15--46}\BibitemShut {NoStop}%
\bibitem [{dic(2021)}]{dice}%
  \BibitemOpen
  \href@noop {} {\enquote {\bibinfo {title} {Dice},}\ } (\bibinfo {year}
  {2021}),\ \bibinfo {note} {\url{https://github.com/sanshar/Dice} (accessed
  May 4, 2021)}\BibitemShut {NoStop}%
\bibitem [{\citenamefont {Smith}\ \emph {et~al.}(2017)\citenamefont {Smith},
  \citenamefont {Mussard}, \citenamefont {Holmes},\ and\ \citenamefont
  {Sharma}}]{Smith2017cheap}%
  \BibitemOpen
  \bibfield  {author} {\bibinfo {author} {\bibfnamefont {J.~E.~T.}\
  \bibnamefont {Smith}}, \bibinfo {author} {\bibfnamefont {B.}~\bibnamefont
  {Mussard}}, \bibinfo {author} {\bibfnamefont {A.~A.}\ \bibnamefont
  {Holmes}},\ and\ \bibinfo {author} {\bibfnamefont {S.}~\bibnamefont
  {Sharma}},\ }\bibfield  {title} {\enquote {\bibinfo {title} {Cheap and near
  exact {CASSCF} with large active spaces},}\ }\bibfield  {booktitle} {\emph
  {\bibinfo {booktitle} {Journal of Chemical Theory and Computation}},\ }\href
  {https://doi.org/10.1021/acs.jctc.7b00900} {\bibfield  {journal} {\bibinfo
  {journal} {J. Chem. Theory Comput.}\ }\textbf {\bibinfo {volume} {13}},\
  \bibinfo {pages} {5468--5478} (\bibinfo {year} {2017})}\BibitemShut {NoStop}%
\bibitem [{\citenamefont {Eriksen}\ \emph {et~al.}(2020)\citenamefont
  {Eriksen}, \citenamefont {Anderson}, \citenamefont {Deustua}, \citenamefont
  {Ghanem}, \citenamefont {Hait}, \citenamefont {Hoffmann}, \citenamefont
  {Lee}, \citenamefont {Levine}, \citenamefont {Magoulas}, \citenamefont
  {Shen}, \citenamefont {Tubman}, \citenamefont {Whaley}, \citenamefont {Xu},
  \citenamefont {Yao}, \citenamefont {Zhang}, \citenamefont {Alavi},
  \citenamefont {Chan}, \citenamefont {Head-Gordon}, \citenamefont {Liu},
  \citenamefont {Piecuch}, \citenamefont {Sharma}, \citenamefont {Ten-no},
  \citenamefont {Umrigar},\ and\ \citenamefont {Gauss}}]{Eriksen2020}%
  \BibitemOpen
  \bibfield  {author} {\bibinfo {author} {\bibfnamefont {J.~J.}\ \bibnamefont
  {Eriksen}}, \bibinfo {author} {\bibfnamefont {T.~A.}\ \bibnamefont
  {Anderson}}, \bibinfo {author} {\bibfnamefont {J.~E.}\ \bibnamefont
  {Deustua}}, \bibinfo {author} {\bibfnamefont {K.}~\bibnamefont {Ghanem}},
  \bibinfo {author} {\bibfnamefont {D.}~\bibnamefont {Hait}}, \bibinfo {author}
  {\bibfnamefont {M.~R.}\ \bibnamefont {Hoffmann}}, \bibinfo {author}
  {\bibfnamefont {S.}~\bibnamefont {Lee}}, \bibinfo {author} {\bibfnamefont
  {D.~S.}\ \bibnamefont {Levine}}, \bibinfo {author} {\bibfnamefont
  {I.}~\bibnamefont {Magoulas}}, \bibinfo {author} {\bibfnamefont
  {J.}~\bibnamefont {Shen}}, \bibinfo {author} {\bibfnamefont {N.~M.}\
  \bibnamefont {Tubman}}, \bibinfo {author} {\bibfnamefont {K.~B.}\
  \bibnamefont {Whaley}}, \bibinfo {author} {\bibfnamefont {E.}~\bibnamefont
  {Xu}}, \bibinfo {author} {\bibfnamefont {Y.}~\bibnamefont {Yao}}, \bibinfo
  {author} {\bibfnamefont {N.}~\bibnamefont {Zhang}}, \bibinfo {author}
  {\bibfnamefont {A.}~\bibnamefont {Alavi}}, \bibinfo {author} {\bibfnamefont
  {G.~K.-L.}\ \bibnamefont {Chan}}, \bibinfo {author} {\bibfnamefont
  {M.}~\bibnamefont {Head-Gordon}}, \bibinfo {author} {\bibfnamefont
  {W.}~\bibnamefont {Liu}}, \bibinfo {author} {\bibfnamefont {P.}~\bibnamefont
  {Piecuch}}, \bibinfo {author} {\bibfnamefont {S.}~\bibnamefont {Sharma}},
  \bibinfo {author} {\bibfnamefont {S.~L.}\ \bibnamefont {Ten-no}}, \bibinfo
  {author} {\bibfnamefont {C.~J.}\ \bibnamefont {Umrigar}},\ and\ \bibinfo
  {author} {\bibfnamefont {J.}~\bibnamefont {Gauss}},\ }\bibfield  {title}
  {\enquote {\bibinfo {title} {The ground state electronic energy of
  benzene},}\ }\bibfield  {booktitle} {\emph {\bibinfo {booktitle} {The Journal
  of Physical Chemistry Letters}},\ }\href
  {https://doi.org/10.1021/acs.jpclett.0c02621} {\bibfield  {journal} {\bibinfo
   {journal} {J. Phys. Chem. Lett.}\ }\textbf {\bibinfo {volume} {11}},\
  \bibinfo {pages} {8922--8929} (\bibinfo {year} {2020})}\BibitemShut {NoStop}%
\bibitem [{\citenamefont {Yao}\ \emph {et~al.}(2020)\citenamefont {Yao},
  \citenamefont {Giner}, \citenamefont {Li}, \citenamefont {Toulouse},\ and\
  \citenamefont {Umrigar}}]{Yao2020almost}%
  \BibitemOpen
  \bibfield  {author} {\bibinfo {author} {\bibfnamefont {Y.}~\bibnamefont
  {Yao}}, \bibinfo {author} {\bibfnamefont {E.}~\bibnamefont {Giner}}, \bibinfo
  {author} {\bibfnamefont {J.}~\bibnamefont {Li}}, \bibinfo {author}
  {\bibfnamefont {J.}~\bibnamefont {Toulouse}},\ and\ \bibinfo {author}
  {\bibfnamefont {C.~J.}\ \bibnamefont {Umrigar}},\ }\bibfield  {title}
  {\enquote {\bibinfo {title} {Almost exact energies for the {G}aussian-2 set
  with the semistochastic heat-bath configuration interaction method},}\ }\href
  {https://doi.org/10.1063/5.0018577} {\bibfield  {journal} {\bibinfo
  {journal} {J. Chem. Phys.}\ }\textbf {\bibinfo {volume} {153}},\ \bibinfo
  {pages} {124117} (\bibinfo {year} {2020})}\BibitemShut {NoStop}%
\bibitem [{\citenamefont {Greene}\ \emph
  {et~al.}(2022{\natexlab{b}})\citenamefont {Greene}, \citenamefont {Webber},
  \citenamefont {Smith}, \citenamefont {Weare},\ and\ \citenamefont
  {Berkelbach}}]{Zenodo}%
  \BibitemOpen
  \bibfield  {author} {\bibinfo {author} {\bibfnamefont {S.~M.}\ \bibnamefont
  {Greene}}, \bibinfo {author} {\bibfnamefont {R.~J.}\ \bibnamefont {Webber}},
  \bibinfo {author} {\bibfnamefont {J.~E.~T.}\ \bibnamefont {Smith}}, \bibinfo
  {author} {\bibfnamefont {J.}~\bibnamefont {Weare}},\ and\ \bibinfo {author}
  {\bibfnamefont {T.~C.}\ \bibnamefont {Berkelbach}},\ }\href
  {http://dx.doi.org/10.5281/zenodo.6638519} {\enquote {\bibinfo {title} {Full
  configuration interaction excited-state energies in large active spaces from
  subspace iteration with repeated random sparsification: Active space orbitals
  [data set]},}\ } (\bibinfo {year} {2022}{\natexlab{b}})\BibitemShut {NoStop}%
\bibitem [{\citenamefont {Deville}\ and\ \citenamefont
  {Till\'e}(1998)}]{deville1998unequal}%
  \BibitemOpen
  \bibfield  {author} {\bibinfo {author} {\bibfnamefont {J.-C.}\ \bibnamefont
  {Deville}}\ and\ \bibinfo {author} {\bibfnamefont {Y.}~\bibnamefont
  {Till\'e}},\ }\bibfield  {title} {\enquote {\bibinfo {title} {Unequal
  probability sampling without replacement through a splitting method},}\
  }\href {https://doi.org/10.1093/biomet/85.1.89} {\bibfield  {journal}
  {\bibinfo  {journal} {Biometrika}\ }\textbf {\bibinfo {volume} {85}},\
  \bibinfo {pages} {89--101} (\bibinfo {year} {1998})}\BibitemShut {NoStop}%
\bibitem [{\citenamefont {Chauvet}(2012)}]{chauvet2012characterization}%
  \BibitemOpen
  \bibfield  {author} {\bibinfo {author} {\bibfnamefont {G.}~\bibnamefont
  {Chauvet}},\ }\bibfield  {title} {\enquote {\bibinfo {title} {On a
  characterization of ordered pivotal sampling},}\ }\href
  {http://www.jstor.org/stable/41714093} {\bibfield  {journal} {\bibinfo
  {journal} {Bernoulli}\ }\textbf {\bibinfo {volume} {18}},\ \bibinfo {pages}
  {1320--1340} (\bibinfo {year} {2012})}\BibitemShut {NoStop}%
\bibitem [{\citenamefont {Chauvet}(2017)}]{Chauvet2017}%
  \BibitemOpen
  \bibfield  {author} {\bibinfo {author} {\bibfnamefont {G.}~\bibnamefont
  {Chauvet}},\ }\bibfield  {title} {\enquote {\bibinfo {title} {A comparison of
  pivotal sampling and unequal probability sampling with replacement},}\ }\href
  {https://doi.org/https://doi.org/10.1016/j.spl.2016.09.027} {\bibfield
  {journal} {\bibinfo  {journal} {Statistics and Probability Letters}\ }\textbf
  {\bibinfo {volume} {121}},\ \bibinfo {pages} {1--5} (\bibinfo {year}
  {2017})}\BibitemShut {NoStop}%
\bibitem [{\citenamefont {Spencer}, \citenamefont {Blunt},\ and\ \citenamefont
  {Foulkes}(2012)}]{Spencer2012}%
  \BibitemOpen
  \bibfield  {author} {\bibinfo {author} {\bibfnamefont {J.~S.}\ \bibnamefont
  {Spencer}}, \bibinfo {author} {\bibfnamefont {N.~S.}\ \bibnamefont {Blunt}},\
  and\ \bibinfo {author} {\bibfnamefont {W.~M.}\ \bibnamefont {Foulkes}},\
  }\bibfield  {title} {\enquote {\bibinfo {title} {The sign problem and
  population dynamics in the full configuration interaction quantum {M}onte
  {C}arlo method},}\ }\href {https://doi.org/10.1063/1.3681396} {\bibfield
  {journal} {\bibinfo  {journal} {J. Chem. Phys.}\ }\textbf {\bibinfo {volume}
  {136}},\ \bibinfo {pages} {054110} (\bibinfo {year} {2012})}\BibitemShut
  {NoStop}%
\bibitem [{\citenamefont {Kolodrubetz}\ \emph {et~al.}(2013)\citenamefont
  {Kolodrubetz}, \citenamefont {Spencer}, \citenamefont {Clark},\ and\
  \citenamefont {Foulkes}}]{Kolodrubetz2013}%
  \BibitemOpen
  \bibfield  {author} {\bibinfo {author} {\bibfnamefont {M.~H.}\ \bibnamefont
  {Kolodrubetz}}, \bibinfo {author} {\bibfnamefont {J.~S.}\ \bibnamefont
  {Spencer}}, \bibinfo {author} {\bibfnamefont {B.~K.}\ \bibnamefont {Clark}},\
  and\ \bibinfo {author} {\bibfnamefont {W.~M.~C.}\ \bibnamefont {Foulkes}},\
  }\bibfield  {title} {\enquote {\bibinfo {title} {The effect of quantization
  on the full configuration interaction quantum {M}onte {C}arlo sign
  problem},}\ }\href@noop {} {\bibfield  {journal} {\bibinfo  {journal} {J.
  Chem. Phys.}\ }\textbf {\bibinfo {volume} {138}},\ \bibinfo {pages} {024110}
  (\bibinfo {year} {2013})}\BibitemShut {NoStop}%
\bibitem [{\citenamefont {Shepherd}, \citenamefont {Scuseria},\ and\
  \citenamefont {Spencer}(2014)}]{Shepherd2014}%
  \BibitemOpen
  \bibfield  {author} {\bibinfo {author} {\bibfnamefont {J.~J.}\ \bibnamefont
  {Shepherd}}, \bibinfo {author} {\bibfnamefont {G.~E.}\ \bibnamefont
  {Scuseria}},\ and\ \bibinfo {author} {\bibfnamefont {J.~S.}\ \bibnamefont
  {Spencer}},\ }\bibfield  {title} {\enquote {\bibinfo {title} {Sign problem in
  full configuration interaction quantum {M}onte {C}arlo: Linear and sublinear
  representation regimes for the exact wave function},}\ }\href
  {https://doi.org/10.1103/PhysRevB.90.155130} {\bibfield  {journal} {\bibinfo
  {journal} {Phys. Rev. B}\ }\textbf {\bibinfo {volume} {90}},\ \bibinfo
  {pages} {155130} (\bibinfo {year} {2014})}\BibitemShut {NoStop}%
\bibitem [{\citenamefont {Vigor}\ \emph {et~al.}(2016)\citenamefont {Vigor},
  \citenamefont {Spencer}, \citenamefont {Bearpark},\ and\ \citenamefont
  {Thom}}]{Vigor2016}%
  \BibitemOpen
  \bibfield  {author} {\bibinfo {author} {\bibfnamefont {W.~A.}\ \bibnamefont
  {Vigor}}, \bibinfo {author} {\bibfnamefont {J.~S.}\ \bibnamefont {Spencer}},
  \bibinfo {author} {\bibfnamefont {M.~J.}\ \bibnamefont {Bearpark}},\ and\
  \bibinfo {author} {\bibfnamefont {A.~J.~W.}\ \bibnamefont {Thom}},\
  }\bibfield  {title} {\enquote {\bibinfo {title} {Understanding and improving
  the efficiency of full configuration interaction quantum {M}onte {C}arlo},}\
  }\href {https://doi.org/10.1063/1.4943113} {\bibfield  {journal} {\bibinfo
  {journal} {J. Chem. Phys.}\ }\textbf {\bibinfo {volume} {144}},\ \bibinfo
  {pages} {094110} (\bibinfo {year} {2016})}\BibitemShut {NoStop}%
\bibitem [{\citenamefont {Ghanem}, \citenamefont {Lozovoi},\ and\ \citenamefont
  {Alavi}(2019)}]{Ghanem2019}%
  \BibitemOpen
  \bibfield  {author} {\bibinfo {author} {\bibfnamefont {K.}~\bibnamefont
  {Ghanem}}, \bibinfo {author} {\bibfnamefont {A.~Y.}\ \bibnamefont
  {Lozovoi}},\ and\ \bibinfo {author} {\bibfnamefont {A.}~\bibnamefont
  {Alavi}},\ }\bibfield  {title} {\enquote {\bibinfo {title} {Unbiasing the
  initiator approximation in full configuration interaction quantum {M}onte
  {C}arlo},}\ }\href {https://doi.org/10.1063/1.5134006} {\bibfield  {journal}
  {\bibinfo  {journal} {J. Chem. Phys.}\ ,\ \bibinfo {pages} {224108}}
  (\bibinfo {year} {2019})}\BibitemShut {NoStop}%
\bibitem [{\citenamefont {Ghanem}, \citenamefont {Guther},\ and\ \citenamefont
  {Alavi}(2020)}]{Ghanem2020}%
  \BibitemOpen
  \bibfield  {author} {\bibinfo {author} {\bibfnamefont {K.}~\bibnamefont
  {Ghanem}}, \bibinfo {author} {\bibfnamefont {K.}~\bibnamefont {Guther}},\
  and\ \bibinfo {author} {\bibfnamefont {A.}~\bibnamefont {Alavi}},\ }\bibfield
   {title} {\enquote {\bibinfo {title} {The adaptive shift method in full
  configuration interaction quantum monte carlo: Development and
  applications},}\ }\href {https://doi.org/10.1063/5.0032617} {\bibfield
  {journal} {\bibinfo  {journal} {J. Chem. Phys.}\ }\textbf {\bibinfo {volume}
  {153}},\ \bibinfo {pages} {224115} (\bibinfo {year} {2020})}\BibitemShut
  {NoStop}%
\bibitem [{\citenamefont {Sharma}(2015)}]{Sharma2015}%
  \BibitemOpen
  \bibfield  {author} {\bibinfo {author} {\bibfnamefont {S.}~\bibnamefont
  {Sharma}},\ }\bibfield  {title} {\enquote {\bibinfo {title} {A general
  non-{A}belian density matrix renormalization group algorithm with application
  to the {C}$_2$ dimer},}\ }\href {https://doi.org/10.1063/1.4905237}
  {\bibfield  {journal} {\bibinfo  {journal} {J. Chem. Phys.}\ }\textbf
  {\bibinfo {volume} {142}},\ \bibinfo {pages} {024107} (\bibinfo {year}
  {2015})}\BibitemShut {NoStop}%
\bibitem [{\citenamefont {Dunning}(1989)}]{Dunning1989}%
  \BibitemOpen
  \bibfield  {author} {\bibinfo {author} {\bibfnamefont {T.~H.}\ \bibnamefont
  {Dunning}},\ }\bibfield  {title} {\enquote {\bibinfo {title} {Gaussian basis
  sets for use in correlated molecular calculations. {I}. the atoms boron
  through neon and hydrogen},}\ }\href {https://doi.org/10.1063/1.456153}
  {\bibfield  {journal} {\bibinfo  {journal} {J. Chem. Phys.}\ }\textbf
  {\bibinfo {volume} {90}},\ \bibinfo {pages} {1007--1023} (\bibinfo {year}
  {1989})}\BibitemShut {NoStop}%
\bibitem [{\citenamefont {Zhang}\ \emph {et~al.}(1990)\citenamefont {Zhang},
  \citenamefont {Loebach}, \citenamefont {Wilson},\ and\ \citenamefont
  {Jacobsen}}]{Zhang1990}%
  \BibitemOpen
  \bibfield  {author} {\bibinfo {author} {\bibfnamefont {W.}~\bibnamefont
  {Zhang}}, \bibinfo {author} {\bibfnamefont {J.~L.}\ \bibnamefont {Loebach}},
  \bibinfo {author} {\bibfnamefont {S.~R.}\ \bibnamefont {Wilson}},\ and\
  \bibinfo {author} {\bibfnamefont {E.~N.}\ \bibnamefont {Jacobsen}},\
  }\bibfield  {title} {\enquote {\bibinfo {title} {Enantioselective epoxidation
  of unfunctionalized olefins catalyzed by salen manganese complexes},}\ }\href
  {https://doi.org/10.1021/ja00163a052} {\bibfield  {journal} {\bibinfo
  {journal} {J .Am. Chem. Soc.}\ }\textbf {\bibinfo {volume} {112}},\ \bibinfo
  {pages} {2801--2803} (\bibinfo {year} {1990})}\BibitemShut {NoStop}%
\bibitem [{\citenamefont {Irie}\ \emph {et~al.}(1990)\citenamefont {Irie},
  \citenamefont {Noda}, \citenamefont {Ito}, \citenamefont {Matsumoto},\ and\
  \citenamefont {Katsuki}}]{Irie1990}%
  \BibitemOpen
  \bibfield  {author} {\bibinfo {author} {\bibfnamefont {R.}~\bibnamefont
  {Irie}}, \bibinfo {author} {\bibfnamefont {K.}~\bibnamefont {Noda}}, \bibinfo
  {author} {\bibfnamefont {Y.}~\bibnamefont {Ito}}, \bibinfo {author}
  {\bibfnamefont {N.}~\bibnamefont {Matsumoto}},\ and\ \bibinfo {author}
  {\bibfnamefont {T.}~\bibnamefont {Katsuki}},\ }\bibfield  {title} {\enquote
  {\bibinfo {title} {Catalytic asymmetric epoxidation of unfunctionalized
  olefins},}\ }\href
  {https://doi.org/https://doi.org/10.1016/S0040-4039(00)88562-7} {\bibfield
  {journal} {\bibinfo  {journal} {Tetrahedron Lett.}\ }\textbf {\bibinfo
  {volume} {31}},\ \bibinfo {pages} {7345--7348} (\bibinfo {year}
  {1990})}\BibitemShut {NoStop}%
\bibitem [{\citenamefont {Jacobsen}\ \emph {et~al.}(1991)\citenamefont
  {Jacobsen}, \citenamefont {Zhang}, \citenamefont {Muci}, \citenamefont
  {Ecker},\ and\ \citenamefont {Deng}}]{Jacobsen1991}%
  \BibitemOpen
  \bibfield  {author} {\bibinfo {author} {\bibfnamefont {E.~N.}\ \bibnamefont
  {Jacobsen}}, \bibinfo {author} {\bibfnamefont {W.}~\bibnamefont {Zhang}},
  \bibinfo {author} {\bibfnamefont {A.~R.}\ \bibnamefont {Muci}}, \bibinfo
  {author} {\bibfnamefont {J.~R.}\ \bibnamefont {Ecker}},\ and\ \bibinfo
  {author} {\bibfnamefont {L.}~\bibnamefont {Deng}},\ }\bibfield  {title}
  {\enquote {\bibinfo {title} {Highly enantioselective epoxidation catalysts
  derived from 1,2-diaminocyclohexane},}\ }\href
  {https://doi.org/10.1021/ja00018a068} {\bibfield  {journal} {\bibinfo
  {journal} {J. Am. Chem. Soc.}\ }\textbf {\bibinfo {volume} {113}},\ \bibinfo
  {pages} {7063--7064} (\bibinfo {year} {1991})}\BibitemShut {NoStop}%
\bibitem [{\citenamefont {Katsuki}(1996)}]{Katsuki1996}%
  \BibitemOpen
  \bibfield  {author} {\bibinfo {author} {\bibfnamefont {T.}~\bibnamefont
  {Katsuki}},\ }\bibfield  {title} {\enquote {\bibinfo {title} {Mn-salen
  catalyst, competitor of enzymes, for asymmetric epoxidation},}\ }\href
  {https://doi.org/https://doi.org/10.1016/S1381-1169(96)00106-9} {\bibfield
  {journal} {\bibinfo  {journal} {J. Mol. Catal. A: Chem.}\ }\textbf {\bibinfo
  {volume} {113}},\ \bibinfo {pages} {87--107} (\bibinfo {year} {1996})},\
  \bibinfo {note} {recent Developments in Biomimetic Oxidation
  Catalysis}\BibitemShut {NoStop}%
\bibitem [{\citenamefont {McGarrigle}\ and\ \citenamefont
  {Gilheany}(2005)}]{McGarrigle2005}%
  \BibitemOpen
  \bibfield  {author} {\bibinfo {author} {\bibfnamefont {E.~M.}\ \bibnamefont
  {McGarrigle}}\ and\ \bibinfo {author} {\bibfnamefont {D.~G.}\ \bibnamefont
  {Gilheany}},\ }\bibfield  {title} {\enquote {\bibinfo {title} {Chromium- and
  manganese-salen promoted epoxidation of alkenes},}\ }\href
  {https://doi.org/10.1021/cr0306945} {\bibfield  {journal} {\bibinfo
  {journal} {Chem. Rev.}\ }\textbf {\bibinfo {volume} {105}},\ \bibinfo {pages}
  {1563--1602} (\bibinfo {year} {2005})}\BibitemShut {NoStop}%
\bibitem [{\citenamefont {Srinivasan}, \citenamefont {Michaud},\ and\
  \citenamefont {Kochi}(1986)}]{Srinivasan1986}%
  \BibitemOpen
  \bibfield  {author} {\bibinfo {author} {\bibfnamefont {K.}~\bibnamefont
  {Srinivasan}}, \bibinfo {author} {\bibfnamefont {P.}~\bibnamefont
  {Michaud}},\ and\ \bibinfo {author} {\bibfnamefont {J.~K.}\ \bibnamefont
  {Kochi}},\ }\bibfield  {title} {\enquote {\bibinfo {title} {Epoxidation of
  olefins with cationic (salen)manganese(iii) complexes. the modulation of
  catalytic activity by substituents},}\ }\href
  {https://doi.org/10.1021/ja00269a029} {\bibfield  {journal} {\bibinfo
  {journal} {J. Am. Chem. Soc.}\ }\textbf {\bibinfo {volume} {108}},\ \bibinfo
  {pages} {2309--2320} (\bibinfo {year} {1986})}\BibitemShut {NoStop}%
\bibitem [{\citenamefont {Fu}\ \emph {et~al.}(1991)\citenamefont {Fu},
  \citenamefont {Look}, \citenamefont {Zhang}, \citenamefont {Jacobsen},\ and\
  \citenamefont {Wong}}]{Fu1991}%
  \BibitemOpen
  \bibfield  {author} {\bibinfo {author} {\bibfnamefont {H.}~\bibnamefont
  {Fu}}, \bibinfo {author} {\bibfnamefont {G.~C.}\ \bibnamefont {Look}},
  \bibinfo {author} {\bibfnamefont {W.}~\bibnamefont {Zhang}}, \bibinfo
  {author} {\bibfnamefont {E.~N.}\ \bibnamefont {Jacobsen}},\ and\ \bibinfo
  {author} {\bibfnamefont {C.~H.}\ \bibnamefont {Wong}},\ }\bibfield  {title}
  {\enquote {\bibinfo {title} {Mechanistic study of a synthetically useful
  monooxygenase model using the hypersensitive probe
  trans-2-phenyl-1-vinylcyclopropane},}\ }\href
  {https://doi.org/10.1021/jo00023a008} {\bibfield  {journal} {\bibinfo
  {journal} {J. Org. Chem.}\ }\textbf {\bibinfo {volume} {56}},\ \bibinfo
  {pages} {6497--6500} (\bibinfo {year} {1991})}\BibitemShut {NoStop}%
\bibitem [{\citenamefont {Norrby}, \citenamefont {Linde},\ and\ \citenamefont
  {{\AA}kermark}(1995)}]{Norrby1995}%
  \BibitemOpen
  \bibfield  {author} {\bibinfo {author} {\bibfnamefont {P.-O.}\ \bibnamefont
  {Norrby}}, \bibinfo {author} {\bibfnamefont {C.}~\bibnamefont {Linde}},\ and\
  \bibinfo {author} {\bibfnamefont {B.}~\bibnamefont {{\AA}kermark}},\
  }\bibfield  {title} {\enquote {\bibinfo {title} {On the chirality transfer in
  the epoxidation of alkenes catalyzed by {M}n(salen) complexes},}\ }\href
  {https://doi.org/10.1021/ja00149a038} {\bibfield  {journal} {\bibinfo
  {journal} {J. Am. Chem. Soc.}\ }\textbf {\bibinfo {volume} {117}},\ \bibinfo
  {pages} {11035--11036} (\bibinfo {year} {1995})}\BibitemShut {NoStop}%
\bibitem [{\citenamefont {Hamada}\ \emph {et~al.}(1996)\citenamefont {Hamada},
  \citenamefont {Fukuda}, \citenamefont {Imanishi},\ and\ \citenamefont
  {Katsuki}}]{Hamada1996}%
  \BibitemOpen
  \bibfield  {author} {\bibinfo {author} {\bibfnamefont {T.}~\bibnamefont
  {Hamada}}, \bibinfo {author} {\bibfnamefont {T.}~\bibnamefont {Fukuda}},
  \bibinfo {author} {\bibfnamefont {H.}~\bibnamefont {Imanishi}},\ and\
  \bibinfo {author} {\bibfnamefont {T.}~\bibnamefont {Katsuki}},\ }\bibfield
  {title} {\enquote {\bibinfo {title} {Mechanism of one oxygen atom transfer
  from oxo (salen) manganese({V}) complex to olefins},}\ }\href
  {https://doi.org/https://doi.org/10.1016/0040-4020(95)00904-3} {\bibfield
  {journal} {\bibinfo  {journal} {Tetrahedron}\ }\textbf {\bibinfo {volume}
  {52}},\ \bibinfo {pages} {515--530} (\bibinfo {year} {1996})}\BibitemShut
  {NoStop}%
\bibitem [{\citenamefont {Linker}(1997)}]{Linker1997}%
  \BibitemOpen
  \bibfield  {author} {\bibinfo {author} {\bibfnamefont {T.}~\bibnamefont
  {Linker}},\ }\bibfield  {title} {\enquote {\bibinfo {title} {The
  jacobsen-katsuki epoxidation and its controversial mechanism},}\ }\href
  {https://doi.org/https://doi.org/10.1002/anie.199720601} {\bibfield
  {journal} {\bibinfo  {journal} {Angew. Chem., Int. Ed. Engl.}\ }\textbf
  {\bibinfo {volume} {36}},\ \bibinfo {pages} {2060--2062} (\bibinfo {year}
  {1997})}\BibitemShut {NoStop}%
\bibitem [{\citenamefont {Finney}\ \emph {et~al.}(1997)\citenamefont {Finney},
  \citenamefont {Pospisil}, \citenamefont {Chang}, \citenamefont {Palucki},
  \citenamefont {Konsler}, \citenamefont {Hansen},\ and\ \citenamefont
  {Jacobsen}}]{Finney1997}%
  \BibitemOpen
  \bibfield  {author} {\bibinfo {author} {\bibfnamefont {N.~S.}\ \bibnamefont
  {Finney}}, \bibinfo {author} {\bibfnamefont {P.~J.}\ \bibnamefont
  {Pospisil}}, \bibinfo {author} {\bibfnamefont {S.}~\bibnamefont {Chang}},
  \bibinfo {author} {\bibfnamefont {M.}~\bibnamefont {Palucki}}, \bibinfo
  {author} {\bibfnamefont {R.~G.}\ \bibnamefont {Konsler}}, \bibinfo {author}
  {\bibfnamefont {K.~B.}\ \bibnamefont {Hansen}},\ and\ \bibinfo {author}
  {\bibfnamefont {E.~N.}\ \bibnamefont {Jacobsen}},\ }\bibfield  {title}
  {\enquote {\bibinfo {title} {On the viability of oxametallacyclic
  intermediates in the (salen)mn-catalyzed asymmetric epoxidation},}\ }\href
  {https://doi.org/https://doi.org/10.1002/anie.199717201} {\bibfield
  {journal} {\bibinfo  {journal} {Angew. Chem., Int. Ed.}\ }\textbf {\bibinfo
  {volume} {36}},\ \bibinfo {pages} {1720--1723} (\bibinfo {year}
  {1997})}\BibitemShut {NoStop}%
\bibitem [{\citenamefont {Linde}\ \emph {et~al.}(1999)\citenamefont {Linde},
  \citenamefont {{\AA}akermark}, \citenamefont {Norrby},\ and\ \citenamefont
  {Svensson}}]{Linde1999}%
  \BibitemOpen
  \bibfield  {author} {\bibinfo {author} {\bibfnamefont {C.}~\bibnamefont
  {Linde}}, \bibinfo {author} {\bibfnamefont {B.}~\bibnamefont
  {{\AA}akermark}}, \bibinfo {author} {\bibfnamefont {P.-O.}\ \bibnamefont
  {Norrby}},\ and\ \bibinfo {author} {\bibfnamefont {M.}~\bibnamefont
  {Svensson}},\ }\bibfield  {title} {\enquote {\bibinfo {title} {Timing is
  critical:: Effect of spin changes on the diastereoselectivity in
  mn(salen)-catalyzed epoxidation},}\ }\href
  {https://doi.org/10.1021/ja9809915} {\bibfield  {journal} {\bibinfo
  {journal} {J. Am. Chem. Soc.}\ }\textbf {\bibinfo {volume} {121}},\ \bibinfo
  {pages} {5083--5084} (\bibinfo {year} {1999})}\BibitemShut {NoStop}%
\bibitem [{\citenamefont {Ivanic}, \citenamefont {Collins},\ and\ \citenamefont
  {Burt}(2004)}]{Ivanic2004}%
  \BibitemOpen
  \bibfield  {author} {\bibinfo {author} {\bibfnamefont {J.}~\bibnamefont
  {Ivanic}}, \bibinfo {author} {\bibfnamefont {J.~R.}\ \bibnamefont
  {Collins}},\ and\ \bibinfo {author} {\bibfnamefont {S.~K.}\ \bibnamefont
  {Burt}},\ }\bibfield  {title} {\enquote {\bibinfo {title} {Theoretical study
  of the low lying electronic states of oxox(salen) (x = mn, mn-, fe, and cr-)
  complexes},}\ }\href {https://doi.org/10.1021/jp031214g} {\bibfield
  {journal} {\bibinfo  {journal} {J. Phys. Chem. A}\ }\textbf {\bibinfo
  {volume} {108}},\ \bibinfo {pages} {2314--2323} (\bibinfo {year}
  {2004})}\BibitemShut {NoStop}%
\bibitem [{\citenamefont {Abashkin}, \citenamefont {Collins},\ and\
  \citenamefont {Burt}(2001)}]{Abashkin2001}%
  \BibitemOpen
  \bibfield  {author} {\bibinfo {author} {\bibfnamefont {Y.~G.}\ \bibnamefont
  {Abashkin}}, \bibinfo {author} {\bibfnamefont {J.~R.}\ \bibnamefont
  {Collins}},\ and\ \bibinfo {author} {\bibfnamefont {S.~K.}\ \bibnamefont
  {Burt}},\ }\bibfield  {title} {\enquote {\bibinfo {title}
  {(salen){M}n({III})-catalyzed epoxidation reaction as a multichannel process
  with different spin states. electronic tuning of asymmetric catalysis: A
  theoretical study},}\ }\href {https://doi.org/10.1021/ic0012221} {\bibfield
  {journal} {\bibinfo  {journal} {Inorg. Chem.}\ }\textbf {\bibinfo {volume}
  {40}},\ \bibinfo {pages} {4040--4048} (\bibinfo {year} {2001})}\BibitemShut
  {NoStop}%
\bibitem [{\citenamefont {Sears}\ and\ \citenamefont
  {Sherrill}(2006)}]{Sears2006}%
  \BibitemOpen
  \bibfield  {author} {\bibinfo {author} {\bibfnamefont {J.~S.}\ \bibnamefont
  {Sears}}\ and\ \bibinfo {author} {\bibfnamefont {C.~D.}\ \bibnamefont
  {Sherrill}},\ }\bibfield  {title} {\enquote {\bibinfo {title} {The electronic
  structure of oxo-mn(salen): Single-reference and multireference
  approaches},}\ }\href {https://doi.org/10.1063/1.2187974} {\bibfield
  {journal} {\bibinfo  {journal} {J. Chem. Phys.}\ }\textbf {\bibinfo {volume}
  {124}},\ \bibinfo {pages} {144314} (\bibinfo {year} {2006})}\BibitemShut
  {NoStop}%
\bibitem [{\citenamefont {Stein}\ and\ \citenamefont
  {Reiher}(2016)}]{Stein2016}%
  \BibitemOpen
  \bibfield  {author} {\bibinfo {author} {\bibfnamefont {C.~J.}\ \bibnamefont
  {Stein}}\ and\ \bibinfo {author} {\bibfnamefont {M.}~\bibnamefont {Reiher}},\
  }\bibfield  {title} {\enquote {\bibinfo {title} {Automated selection of
  active orbital spaces},}\ }\href {https://doi.org/10.1021/acs.jctc.6b00156}
  {\bibfield  {journal} {\bibinfo  {journal} {J. Chem. Theory Comput.}\
  }\textbf {\bibinfo {volume} {12}},\ \bibinfo {pages} {1760--1771} (\bibinfo
  {year} {2016})}\BibitemShut {NoStop}%
\bibitem [{\citenamefont {Dang}\ and\ \citenamefont
  {Zimmerman}(2021)}]{Dang2021}%
  \BibitemOpen
  \bibfield  {author} {\bibinfo {author} {\bibfnamefont {D.-K.}\ \bibnamefont
  {Dang}}\ and\ \bibinfo {author} {\bibfnamefont {P.~M.}\ \bibnamefont
  {Zimmerman}},\ }\bibfield  {title} {\enquote {\bibinfo {title} {Fully
  variational incremental {CASSCF}},}\ }\href
  {https://doi.org/10.1063/5.0031208} {\bibfield  {journal} {\bibinfo
  {journal} {J. Chem. Phys.}\ }\textbf {\bibinfo {volume} {154}},\ \bibinfo
  {pages} {014105} (\bibinfo {year} {2021})}\BibitemShut {NoStop}%
\bibitem [{\citenamefont {Wouters}\ \emph
  {et~al.}(2014{\natexlab{b}})\citenamefont {Wouters}, \citenamefont
  {Bogaerts}, \citenamefont {Van Der~Voort}, \citenamefont {Van~Speybroeck},\
  and\ \citenamefont {Van~Neck}}]{Wouters2014dmrgscf}%
  \BibitemOpen
  \bibfield  {author} {\bibinfo {author} {\bibfnamefont {S.}~\bibnamefont
  {Wouters}}, \bibinfo {author} {\bibfnamefont {T.}~\bibnamefont {Bogaerts}},
  \bibinfo {author} {\bibfnamefont {P.}~\bibnamefont {Van Der~Voort}}, \bibinfo
  {author} {\bibfnamefont {V.}~\bibnamefont {Van~Speybroeck}},\ and\ \bibinfo
  {author} {\bibfnamefont {D.}~\bibnamefont {Van~Neck}},\ }\bibfield  {title}
  {\enquote {\bibinfo {title} {Communication: {DMRG-SCF} study of the singlet,
  triplet, and quintet states of {oxo-Mn(Salen)}},}\ }\href
  {https://doi.org/10.1063/1.4885815} {\bibfield  {journal} {\bibinfo
  {journal} {J. Chem. Phys.}\ }\textbf {\bibinfo {volume} {140}},\ \bibinfo
  {pages} {241103} (\bibinfo {year} {2014}{\natexlab{b}})}\BibitemShut
  {NoStop}%
\bibitem [{\citenamefont {Ivanic}(2003)}]{Ivanic2003}%
  \BibitemOpen
  \bibfield  {author} {\bibinfo {author} {\bibfnamefont {J.}~\bibnamefont
  {Ivanic}},\ }\bibfield  {title} {\enquote {\bibinfo {title} {Direct
  configuration interaction and multiconfigurational self-consistent-field
  method for multiple active spaces with variable occupations. ii. application
  to oxomn(salen) and n$_2$o$_4$},}\ }\href {https://doi.org/10.1063/1.1615955}
  {\bibfield  {journal} {\bibinfo  {journal} {J. Chem. Phys.}\ }\textbf
  {\bibinfo {volume} {119}},\ \bibinfo {pages} {9377--9385} (\bibinfo {year}
  {2003})}\BibitemShut {NoStop}%
\bibitem [{\citenamefont {Zhang}, \citenamefont {Wei},\ and\ \citenamefont
  {Fang}(2019)}]{Zhang2019}%
  \BibitemOpen
  \bibfield  {author} {\bibinfo {author} {\bibfnamefont {J.~J.}\ \bibnamefont
  {Zhang}}, \bibinfo {author} {\bibfnamefont {Y.}~\bibnamefont {Wei}},\ and\
  \bibinfo {author} {\bibfnamefont {Z.}~\bibnamefont {Fang}},\ }\bibfield
  {title} {\enquote {\bibinfo {title} {Ozone pollution: A major health hazard
  worldwide},}\ }\href {https://doi.org/10.3389/fimmu.2019.02518} {\bibfield
  {journal} {\bibinfo  {journal} {Front. Immunol.}\ }\textbf {\bibinfo {volume}
  {10}},\ \bibinfo {pages} {2518} (\bibinfo {year} {2019})}\BibitemShut
  {NoStop}%
\bibitem [{\citenamefont {Bais}\ \emph {et~al.}(2019)\citenamefont {Bais},
  \citenamefont {Bernhard}, \citenamefont {McKenzie}, \citenamefont {Aucamp},
  \citenamefont {Young}, \citenamefont {Ilyas}, \citenamefont {J\"ockel},\ and\
  \citenamefont {Deushi}}]{Bais2019}%
  \BibitemOpen
  \bibfield  {author} {\bibinfo {author} {\bibfnamefont {A.~F.}\ \bibnamefont
  {Bais}}, \bibinfo {author} {\bibfnamefont {G.}~\bibnamefont {Bernhard}},
  \bibinfo {author} {\bibfnamefont {R.~L.}\ \bibnamefont {McKenzie}}, \bibinfo
  {author} {\bibfnamefont {P.~J.}\ \bibnamefont {Aucamp}}, \bibinfo {author}
  {\bibfnamefont {P.~J.}\ \bibnamefont {Young}}, \bibinfo {author}
  {\bibfnamefont {M.}~\bibnamefont {Ilyas}}, \bibinfo {author} {\bibfnamefont
  {P.}~\bibnamefont {J\"ockel}},\ and\ \bibinfo {author} {\bibfnamefont
  {M.}~\bibnamefont {Deushi}},\ }\bibfield  {title} {\enquote {\bibinfo {title}
  {Ozone-climate interactions and effects on solar ultraviolet radiation},}\
  }\href {https://doi.org/10.1039/C8PP90059K} {\bibfield  {journal} {\bibinfo
  {journal} {Photochem. Photobiol. Sci.}\ }\textbf {\bibinfo {volume} {18}},\
  \bibinfo {pages} {602--640} (\bibinfo {year} {2019})}\BibitemShut {NoStop}%
\bibitem [{\citenamefont {Crutzen}(1974)}]{Crutzen1974}%
  \BibitemOpen
  \bibfield  {author} {\bibinfo {author} {\bibfnamefont {P.}~\bibnamefont
  {Crutzen}},\ }\bibfield  {title} {\enquote {\bibinfo {title} {A review of
  upper atmospheric photochemistry},}\ }\href {https://doi.org/10.1139/v74-229}
  {\bibfield  {journal} {\bibinfo  {journal} {Can. J. Chem.}\ }\textbf
  {\bibinfo {volume} {52}},\ \bibinfo {pages} {1569--1581} (\bibinfo {year}
  {1974})}\BibitemShut {NoStop}%
\bibitem [{\citenamefont {Luecken}, \citenamefont {Yarwood},\ and\
  \citenamefont {Hutzell}(2019)}]{Luecken2019}%
  \BibitemOpen
  \bibfield  {author} {\bibinfo {author} {\bibfnamefont {D.}~\bibnamefont
  {Luecken}}, \bibinfo {author} {\bibfnamefont {G.}~\bibnamefont {Yarwood}},\
  and\ \bibinfo {author} {\bibfnamefont {W.}~\bibnamefont {Hutzell}},\
  }\bibfield  {title} {\enquote {\bibinfo {title} {Multipollutant modeling of
  ozone, reactive nitrogen and haps across the continental us with cmaq-cb6},}\
  }\href {https://doi.org/https://doi.org/10.1016/j.atmosenv.2018.11.060}
  {\bibfield  {journal} {\bibinfo  {journal} {Atmos. Environ.}\ }\textbf
  {\bibinfo {volume} {201}},\ \bibinfo {pages} {62--72} (\bibinfo {year}
  {2019})}\BibitemShut {NoStop}%
\bibitem [{\citenamefont {Zhu}\ \emph {et~al.}(2020)\citenamefont {Zhu},
  \citenamefont {Wang}, \citenamefont {Wang}, \citenamefont {Jing},
  \citenamefont {Lou}, \citenamefont {Saiz-Lopez},\ and\ \citenamefont
  {Zhou}}]{Zhu2020}%
  \BibitemOpen
  \bibfield  {author} {\bibinfo {author} {\bibfnamefont {J.}~\bibnamefont
  {Zhu}}, \bibinfo {author} {\bibfnamefont {S.}~\bibnamefont {Wang}}, \bibinfo
  {author} {\bibfnamefont {H.}~\bibnamefont {Wang}}, \bibinfo {author}
  {\bibfnamefont {S.}~\bibnamefont {Jing}}, \bibinfo {author} {\bibfnamefont
  {S.}~\bibnamefont {Lou}}, \bibinfo {author} {\bibfnamefont {A.}~\bibnamefont
  {Saiz-Lopez}},\ and\ \bibinfo {author} {\bibfnamefont {B.}~\bibnamefont
  {Zhou}},\ }\bibfield  {title} {\enquote {\bibinfo {title} {Observationally
  constrained modeling of atmospheric oxidation capacity and photochemical
  reactivity in shanghai, china},}\ }\href
  {https://doi.org/10.5194/acp-20-1217-2020} {\bibfield  {journal} {\bibinfo
  {journal} {Atmos. Chem. Phys.}\ }\textbf {\bibinfo {volume} {20}},\ \bibinfo
  {pages} {1217--1232} (\bibinfo {year} {2020})}\BibitemShut {NoStop}%
\bibitem [{\citenamefont {Burton}(1979)}]{Burton1979}%
  \BibitemOpen
  \bibfield  {author} {\bibinfo {author} {\bibfnamefont {P.~G.}\ \bibnamefont
  {Burton}},\ }\bibfield  {title} {\enquote {\bibinfo {title} {The cyclic ozone
  isomer},}\ }\href {https://doi.org/10.1063/1.438387} {\bibfield  {journal}
  {\bibinfo  {journal} {J. Chem. Phys.}\ }\textbf {\bibinfo {volume} {71}},\
  \bibinfo {pages} {961--972} (\bibinfo {year} {1979})}\BibitemShut {NoStop}%
\bibitem [{\citenamefont {Chien}\ \emph {et~al.}(2018)\citenamefont {Chien},
  \citenamefont {Holmes}, \citenamefont {Otten}, \citenamefont {Umrigar},
  \citenamefont {Sharma},\ and\ \citenamefont {Zimmerman}}]{Chien2018}%
  \BibitemOpen
  \bibfield  {author} {\bibinfo {author} {\bibfnamefont {A.~D.}\ \bibnamefont
  {Chien}}, \bibinfo {author} {\bibfnamefont {A.~A.}\ \bibnamefont {Holmes}},
  \bibinfo {author} {\bibfnamefont {M.}~\bibnamefont {Otten}}, \bibinfo
  {author} {\bibfnamefont {C.~J.}\ \bibnamefont {Umrigar}}, \bibinfo {author}
  {\bibfnamefont {S.}~\bibnamefont {Sharma}},\ and\ \bibinfo {author}
  {\bibfnamefont {P.~M.}\ \bibnamefont {Zimmerman}},\ }\bibfield  {title}
  {\enquote {\bibinfo {title} {Excited states of methylene, polyenes, and ozone
  from heat-bath configuration interaction},}\ }\href
  {https://doi.org/10.1021/acs.jpca.8b01554} {\bibfield  {journal} {\bibinfo
  {journal} {J. Phys. Chem. A}\ }\textbf {\bibinfo {volume} {122}},\ \bibinfo
  {pages} {2714--2722} (\bibinfo {year} {2018})}\BibitemShut {NoStop}%
\bibitem [{\citenamefont {Lee}(1990)}]{Lee1990}%
  \BibitemOpen
  \bibfield  {author} {\bibinfo {author} {\bibfnamefont {T.~J.}\ \bibnamefont
  {Lee}},\ }\bibfield  {title} {\enquote {\bibinfo {title} {On the energy
  separation between the open and cyclic forms of ozone},}\ }\href
  {https://doi.org/https://doi.org/10.1016/0009-2614(90)85642-P} {\bibfield
  {journal} {\bibinfo  {journal} {Chem. Phys. Lett.}\ }\textbf {\bibinfo
  {volume} {169}},\ \bibinfo {pages} {529--533} (\bibinfo {year}
  {1990})}\BibitemShut {NoStop}%
\bibitem [{\citenamefont {Xantheas}\ \emph {et~al.}(1991)\citenamefont
  {Xantheas}, \citenamefont {Atchity}, \citenamefont {Elbert},\ and\
  \citenamefont {Ruedenberg}}]{Xantheas1991}%
  \BibitemOpen
  \bibfield  {author} {\bibinfo {author} {\bibfnamefont {S.~S.}\ \bibnamefont
  {Xantheas}}, \bibinfo {author} {\bibfnamefont {G.~J.}\ \bibnamefont
  {Atchity}}, \bibinfo {author} {\bibfnamefont {S.~T.}\ \bibnamefont
  {Elbert}},\ and\ \bibinfo {author} {\bibfnamefont {K.}~\bibnamefont
  {Ruedenberg}},\ }\bibfield  {title} {\enquote {\bibinfo {title} {Potential
  energy surfaces of ozone. i},}\ }\href {https://doi.org/10.1063/1.460140}
  {\bibfield  {journal} {\bibinfo  {journal} {J. Chem. Phys.}\ }\textbf
  {\bibinfo {volume} {94}},\ \bibinfo {pages} {8054--8069} (\bibinfo {year}
  {1991})}\BibitemShut {NoStop}%
\bibitem [{\citenamefont {Qu}, \citenamefont {Zhu},\ and\ \citenamefont
  {Schinke}(2005)}]{Qu2005}%
  \BibitemOpen
  \bibfield  {author} {\bibinfo {author} {\bibfnamefont {Z.-W.}\ \bibnamefont
  {Qu}}, \bibinfo {author} {\bibfnamefont {H.}~\bibnamefont {Zhu}},\ and\
  \bibinfo {author} {\bibfnamefont {R.}~\bibnamefont {Schinke}},\ }\bibfield
  {title} {\enquote {\bibinfo {title} {Infrared spectrum of cyclic ozone: A
  theoretical investigation},}\ }\href {https://doi.org/10.1063/1.2130709}
  {\bibfield  {journal} {\bibinfo  {journal} {J. Chem. Phys.}\ }\textbf
  {\bibinfo {volume} {123}},\ \bibinfo {pages} {204324} (\bibinfo {year}
  {2005})}\BibitemShut {NoStop}%
\bibitem [{\citenamefont {{De Vico}}\ \emph {et~al.}(2008)\citenamefont {{De
  Vico}}, \citenamefont {Pegado}, \citenamefont {Heimdal}, \citenamefont
  {S\"oderhjelm},\ and\ \citenamefont {Roos}}]{DeVico2008}%
  \BibitemOpen
  \bibfield  {author} {\bibinfo {author} {\bibfnamefont {L.}~\bibnamefont {{De
  Vico}}}, \bibinfo {author} {\bibfnamefont {L.}~\bibnamefont {Pegado}},
  \bibinfo {author} {\bibfnamefont {J.}~\bibnamefont {Heimdal}}, \bibinfo
  {author} {\bibfnamefont {P.}~\bibnamefont {S\"oderhjelm}},\ and\ \bibinfo
  {author} {\bibfnamefont {B.~O.}\ \bibnamefont {Roos}},\ }\bibfield  {title}
  {\enquote {\bibinfo {title} {The ozone ring closure as a test for multi-state
  multi-configurational second order perturbation theory (ms-caspt2)},}\ }\href
  {https://doi.org/https://doi.org/10.1016/j.cplett.2008.06.065} {\bibfield
  {journal} {\bibinfo  {journal} {Chem. Phys. Lett.}\ }\textbf {\bibinfo
  {volume} {461}},\ \bibinfo {pages} {136--141} (\bibinfo {year}
  {2008})}\BibitemShut {NoStop}%
\bibitem [{\citenamefont {Daday}\ \emph {et~al.}(2012)\citenamefont {Daday},
  \citenamefont {Smart}, \citenamefont {Booth}, \citenamefont {Alavi},\ and\
  \citenamefont {Filippi}}]{Daday2012}%
  \BibitemOpen
  \bibfield  {author} {\bibinfo {author} {\bibfnamefont {C.}~\bibnamefont
  {Daday}}, \bibinfo {author} {\bibfnamefont {S.}~\bibnamefont {Smart}},
  \bibinfo {author} {\bibfnamefont {G.~H.}\ \bibnamefont {Booth}}, \bibinfo
  {author} {\bibfnamefont {A.}~\bibnamefont {Alavi}},\ and\ \bibinfo {author}
  {\bibfnamefont {C.}~\bibnamefont {Filippi}},\ }\bibfield  {title} {\enquote
  {\bibinfo {title} {Full configuration interaction excitations of ethene and
  butadiene: Resolution of an ancient question},}\ }\href
  {https://doi.org/10.1021/ct300486d} {\bibfield  {journal} {\bibinfo
  {journal} {J. Chem. Theory Comput.}\ }\textbf {\bibinfo {volume} {8}},\
  \bibinfo {pages} {4441--4451} (\bibinfo {year} {2012})}\BibitemShut {NoStop}%
\bibitem [{\citenamefont {Ansari}\ and\ \citenamefont
  {Ali}(2018)}]{Ansari2018}%
  \BibitemOpen
  \bibfield  {author} {\bibinfo {author} {\bibfnamefont {S.~P.}\ \bibnamefont
  {Ansari}}\ and\ \bibinfo {author} {\bibfnamefont {F.}~\bibnamefont {Ali}},\
  }\enquote {\bibinfo {title} {Conjugated organic polymers for optoelectronic
  devices},}\ in\ \href {https://doi.org/10.1007/978-3-319-92067-2_21-1} {\emph
  {\bibinfo {booktitle} {Functional Polymers}}},\ \bibinfo {editor} {edited by\
  \bibinfo {editor} {\bibfnamefont {M.~A.}\ \bibnamefont {Jafar~Mazumder}},
  \bibinfo {editor} {\bibfnamefont {H.}~\bibnamefont {Sheardown}},\ and\
  \bibinfo {editor} {\bibfnamefont {A.}~\bibnamefont {Al-Ahmed}}}\ (\bibinfo
  {publisher} {Springer International Publishing},\ \bibinfo {year} {2018})\
  pp.\ \bibinfo {pages} {1--40}\BibitemShut {NoStop}%
\bibitem [{\citenamefont {Oka}, \citenamefont {Winther-Jensen},\ and\
  \citenamefont {Nishide}(2021)}]{Oka2021}%
  \BibitemOpen
  \bibfield  {author} {\bibinfo {author} {\bibfnamefont {K.}~\bibnamefont
  {Oka}}, \bibinfo {author} {\bibfnamefont {B.}~\bibnamefont
  {Winther-Jensen}},\ and\ \bibinfo {author} {\bibfnamefont {H.}~\bibnamefont
  {Nishide}},\ }\bibfield  {title} {\enquote {\bibinfo {title} {Organic
  $\pi$-conjugated polymers as photocathode materials for
  visible-light-enhanced hydrogen and hydrogen peroxide production from
  water},}\ }\href@noop {} {\bibfield  {journal} {\bibinfo  {journal} {Adv.
  Energy Mater.}\ }\textbf {\bibinfo {volume} {11}},\ \bibinfo {pages}
  {2003724} (\bibinfo {year} {2021})}\BibitemShut {NoStop}%
\bibitem [{\citenamefont {Tavan}\ and\ \citenamefont
  {Schulten}(1987)}]{Tavan1987}%
  \BibitemOpen
  \bibfield  {author} {\bibinfo {author} {\bibfnamefont {P.}~\bibnamefont
  {Tavan}}\ and\ \bibinfo {author} {\bibfnamefont {K.}~\bibnamefont
  {Schulten}},\ }\bibfield  {title} {\enquote {\bibinfo {title} {Electronic
  excitations in finite and infinite polyenes},}\ }\href
  {https://doi.org/10.1103/PhysRevB.36.4337} {\bibfield  {journal} {\bibinfo
  {journal} {Phys. Rev. B}\ }\textbf {\bibinfo {volume} {36}},\ \bibinfo
  {pages} {4337--4358} (\bibinfo {year} {1987})}\BibitemShut {NoStop}%
\bibitem [{\citenamefont {Watts}, \citenamefont {Gwaltney},\ and\ \citenamefont
  {Bartlett}(1996)}]{Watts1996}%
  \BibitemOpen
  \bibfield  {author} {\bibinfo {author} {\bibfnamefont {J.~D.}\ \bibnamefont
  {Watts}}, \bibinfo {author} {\bibfnamefont {S.~R.}\ \bibnamefont
  {Gwaltney}},\ and\ \bibinfo {author} {\bibfnamefont {R.~J.}\ \bibnamefont
  {Bartlett}},\ }\bibfield  {title} {\enquote {\bibinfo {title}
  {Coupled-cluster calculations of the excitation energies of ethylene,
  butadiene, and cyclopentadiene},}\ }\href {https://doi.org/10.1063/1.471988}
  {\bibfield  {journal} {\bibinfo  {journal} {J. Chem. Phys.}\ }\textbf
  {\bibinfo {volume} {105}},\ \bibinfo {pages} {6979--6988} (\bibinfo {year}
  {1996})}\BibitemShut {NoStop}%
\bibitem [{\citenamefont {Starcke}\ \emph {et~al.}(2006)\citenamefont
  {Starcke}, \citenamefont {Wormit}, \citenamefont {Schirmer},\ and\
  \citenamefont {Dreuw}}]{Starcke2006}%
  \BibitemOpen
  \bibfield  {author} {\bibinfo {author} {\bibfnamefont {J.~H.}\ \bibnamefont
  {Starcke}}, \bibinfo {author} {\bibfnamefont {M.}~\bibnamefont {Wormit}},
  \bibinfo {author} {\bibfnamefont {J.}~\bibnamefont {Schirmer}},\ and\
  \bibinfo {author} {\bibfnamefont {A.}~\bibnamefont {Dreuw}},\ }\bibfield
  {title} {\enquote {\bibinfo {title} {How much double excitation character do
  the lowest excited states of linear polyenes have?}}\ }\href
  {https://doi.org/https://doi.org/10.1016/j.chemphys.2006.07.020} {\bibfield
  {journal} {\bibinfo  {journal} {Chem. Phys.}\ }\textbf {\bibinfo {volume}
  {329}},\ \bibinfo {pages} {39--49} (\bibinfo {year} {2006})},\ \bibinfo
  {note} {electron Correlation and Multimode Dynamics in Molecules}\BibitemShut
  {NoStop}%
\bibitem [{\citenamefont {Watson}\ and\ \citenamefont
  {Chan}(2012)}]{Watson2012}%
  \BibitemOpen
  \bibfield  {author} {\bibinfo {author} {\bibfnamefont {M.~A.}\ \bibnamefont
  {Watson}}\ and\ \bibinfo {author} {\bibfnamefont {G.~K.-L.}\ \bibnamefont
  {Chan}},\ }\bibfield  {title} {\enquote {\bibinfo {title} {Excited states of
  butadiene to chemical accuracy: Reconciling theory and experiment},}\ }\href
  {https://doi.org/10.1021/ct300591z} {\bibfield  {journal} {\bibinfo
  {journal} {J. Chem. Theory Comput.}\ }\textbf {\bibinfo {volume} {8}},\
  \bibinfo {pages} {4013--4018} (\bibinfo {year} {2012})}\BibitemShut {NoStop}%
\bibitem [{\citenamefont {Widmark}, \citenamefont {Malmqvist},\ and\
  \citenamefont {Roos}(1990)}]{Widmark1990}%
  \BibitemOpen
  \bibfield  {author} {\bibinfo {author} {\bibfnamefont {P.-O.}\ \bibnamefont
  {Widmark}}, \bibinfo {author} {\bibfnamefont {P.-{\AA}.}\ \bibnamefont
  {Malmqvist}},\ and\ \bibinfo {author} {\bibfnamefont {B.~O.}\ \bibnamefont
  {Roos}},\ }\bibfield  {title} {\enquote {\bibinfo {title} {Density matrix
  averaged atomic natural orbital (ano) basis sets for correlated molecular
  wave functions},}\ }\href {https://doi.org/10.1007/BF01120130} {\bibfield
  {journal} {\bibinfo  {journal} {Theor. Chim. Acta}\ }\textbf {\bibinfo
  {volume} {77}},\ \bibinfo {pages} {291--306} (\bibinfo {year}
  {1990})}\BibitemShut {NoStop}%
\bibitem [{\citenamefont {Blunt}(2018)}]{Blunt2018}%
  \BibitemOpen
  \bibfield  {author} {\bibinfo {author} {\bibfnamefont {N.~S.}\ \bibnamefont
  {Blunt}},\ }\bibfield  {title} {\enquote {\bibinfo {title} {Communication: An
  efficient and accurate perturbative correction to initiator full
  configuration interaction quantum {M}onte {C}arlo},}\ }\href
  {https://doi.org/10.1063/1.5037923} {\bibfield  {journal} {\bibinfo
  {journal} {J. Chem. Phys.}\ }\textbf {\bibinfo {volume} {148}},\ \bibinfo
  {pages} {221101} (\bibinfo {year} {2018})}\BibitemShut {NoStop}%
\bibitem [{\citenamefont {Blunt}, \citenamefont {Thom},\ and\ \citenamefont
  {Scott}(2019)}]{Blunt2019}%
  \BibitemOpen
  \bibfield  {author} {\bibinfo {author} {\bibfnamefont {N.~S.}\ \bibnamefont
  {Blunt}}, \bibinfo {author} {\bibfnamefont {A.~J.}\ \bibnamefont {Thom}},\
  and\ \bibinfo {author} {\bibfnamefont {C.~J.}\ \bibnamefont {Scott}},\
  }\bibfield  {title} {\enquote {\bibinfo {title} {Preconditioning and
  perturbative estimators in full configuration interaction quantum {M}onte
  {C}arlo},}\ }\href {https://doi.org/10.1021/acs.jctc.9b00049} {\bibfield
  {journal} {\bibinfo  {journal} {J. Chem. Theory Comput.}\ }\textbf {\bibinfo
  {volume} {15}},\ \bibinfo {pages} {3537--3551} (\bibinfo {year}
  {2019})}\BibitemShut {NoStop}%
\bibitem [{\citenamefont {Spencer}\ and\ \citenamefont
  {Thom}(2016)}]{Spencer2016}%
  \BibitemOpen
  \bibfield  {author} {\bibinfo {author} {\bibfnamefont {J.~S.}\ \bibnamefont
  {Spencer}}\ and\ \bibinfo {author} {\bibfnamefont {A.~J.~W.}\ \bibnamefont
  {Thom}},\ }\bibfield  {title} {\enquote {\bibinfo {title} {Developments in
  stochastic coupled cluster theory: The initiator approximation and
  application to the uniform electron gas},}\ }\href
  {https://doi.org/10.1063/1.4942173} {\bibfield  {journal} {\bibinfo
  {journal} {J. Chem. Phys.}\ }\textbf {\bibinfo {volume} {144}},\ \bibinfo
  {pages} {084108} (\bibinfo {year} {2016})}\BibitemShut {NoStop}%
\bibitem [{\citenamefont {Casulleras}\ and\ \citenamefont
  {Boronat}(1995)}]{Casulleras1995}%
  \BibitemOpen
  \bibfield  {author} {\bibinfo {author} {\bibfnamefont {J.}~\bibnamefont
  {Casulleras}}\ and\ \bibinfo {author} {\bibfnamefont {J.}~\bibnamefont
  {Boronat}},\ }\bibfield  {title} {\enquote {\bibinfo {title} {Unbiased
  estimators in quantum monte carlo methods: Application to liquid
  $^{4}\mathrm{He}$},}\ }\href {https://doi.org/10.1103/PhysRevB.52.3654}
  {\bibfield  {journal} {\bibinfo  {journal} {Phys. Rev. B}\ }\textbf {\bibinfo
  {volume} {52}},\ \bibinfo {pages} {3654--3661} (\bibinfo {year}
  {1995})}\BibitemShut {NoStop}%
\bibitem [{\citenamefont {Motta}\ and\ \citenamefont
  {Zhang}(2017)}]{Motta2017}%
  \BibitemOpen
  \bibfield  {author} {\bibinfo {author} {\bibfnamefont {M.}~\bibnamefont
  {Motta}}\ and\ \bibinfo {author} {\bibfnamefont {S.}~\bibnamefont {Zhang}},\
  }\bibfield  {title} {\enquote {\bibinfo {title} {Computation of ground-state
  properties in molecular systems: Back-propagation with auxiliary-field
  quantum monte carlo},}\ }\href {https://doi.org/10.1021/acs.jctc.7b00730}
  {\bibfield  {journal} {\bibinfo  {journal} {J. Chem. Theory Comput.}\
  }\textbf {\bibinfo {volume} {13}},\ \bibinfo {pages} {5367--5378} (\bibinfo
  {year} {2017})}\BibitemShut {NoStop}%
\bibitem [{\citenamefont {Overy}\ \emph {et~al.}(2014)\citenamefont {Overy},
  \citenamefont {Booth}, \citenamefont {Blunt}, \citenamefont {Shepherd},
  \citenamefont {Cleland},\ and\ \citenamefont {Alavi}}]{Overy2014}%
  \BibitemOpen
  \bibfield  {author} {\bibinfo {author} {\bibfnamefont {C.}~\bibnamefont
  {Overy}}, \bibinfo {author} {\bibfnamefont {G.~H.}\ \bibnamefont {Booth}},
  \bibinfo {author} {\bibfnamefont {N.~S.}\ \bibnamefont {Blunt}}, \bibinfo
  {author} {\bibfnamefont {J.~J.}\ \bibnamefont {Shepherd}}, \bibinfo {author}
  {\bibfnamefont {D.}~\bibnamefont {Cleland}},\ and\ \bibinfo {author}
  {\bibfnamefont {A.}~\bibnamefont {Alavi}},\ }\bibfield  {title} {\enquote
  {\bibinfo {title} {Unbiased reduced density matrices and electronic
  properties from full configuration interaction quantum {M}onte {C}arlo},}\
  }\href {https://doi.org/10.1063/1.4904313} {\bibfield  {journal} {\bibinfo
  {journal} {J. Chem. Phys.}\ }\textbf {\bibinfo {volume} {141}},\ \bibinfo
  {pages} {244117} (\bibinfo {year} {2014})}\BibitemShut {NoStop}%
\bibitem [{\citenamefont {Thomas}\ \emph {et~al.}(2015)\citenamefont {Thomas},
  \citenamefont {Opalka}, \citenamefont {Overy}, \citenamefont {Knowles},
  \citenamefont {Alavi},\ and\ \citenamefont {Booth}}]{Thomas2015analytic}%
  \BibitemOpen
  \bibfield  {author} {\bibinfo {author} {\bibfnamefont {R.~E.}\ \bibnamefont
  {Thomas}}, \bibinfo {author} {\bibfnamefont {D.}~\bibnamefont {Opalka}},
  \bibinfo {author} {\bibfnamefont {C.}~\bibnamefont {Overy}}, \bibinfo
  {author} {\bibfnamefont {P.~J.}\ \bibnamefont {Knowles}}, \bibinfo {author}
  {\bibfnamefont {A.}~\bibnamefont {Alavi}},\ and\ \bibinfo {author}
  {\bibfnamefont {G.~H.}\ \bibnamefont {Booth}},\ }\bibfield  {title} {\enquote
  {\bibinfo {title} {Analytic nuclear forces and molecular properties from full
  configuration interaction quantum {M}onte {C}arlo},}\ }\href
  {https://doi.org/10.1063/1.4927594} {\bibfield  {journal} {\bibinfo
  {journal} {J. Chem. Phys.}\ }\textbf {\bibinfo {volume} {143}},\ \bibinfo
  {pages} {054108} (\bibinfo {year} {2015})}\BibitemShut {NoStop}%
\bibitem [{\citenamefont {Blunt}, \citenamefont {Booth},\ and\ \citenamefont
  {Alavi}(2017)}]{Blunt2017density}%
  \BibitemOpen
  \bibfield  {author} {\bibinfo {author} {\bibfnamefont {N.~S.}\ \bibnamefont
  {Blunt}}, \bibinfo {author} {\bibfnamefont {G.~H.}\ \bibnamefont {Booth}},\
  and\ \bibinfo {author} {\bibfnamefont {A.}~\bibnamefont {Alavi}},\ }\bibfield
   {title} {\enquote {\bibinfo {title} {Density matrices in full configuration
  interaction quantum {M}onte {C}arlo: Excited states, transition dipole
  moments, and parallel distribution},}\ }\href
  {https://doi.org/10.1063/1.4986963} {\bibfield  {journal} {\bibinfo
  {journal} {J. Chem. Phys.}\ }\textbf {\bibinfo {volume} {146}},\ \bibinfo
  {pages} {244105} (\bibinfo {year} {2017})}\BibitemShut {NoStop}%
\bibitem [{\citenamefont {Motta}\ and\ \citenamefont
  {Zhang}(2018)}]{Motta2018}%
  \BibitemOpen
  \bibfield  {author} {\bibinfo {author} {\bibfnamefont {M.}~\bibnamefont
  {Motta}}\ and\ \bibinfo {author} {\bibfnamefont {S.}~\bibnamefont {Zhang}},\
  }\bibfield  {title} {\enquote {\bibinfo {title} {Ab initio computations of
  molecular systems by the auxiliary-field quantum {M}onte {C}arlo method},}\
  }\href {https://doi.org/10.1002/wcms.1364} {\bibfield  {journal} {\bibinfo
  {journal} {Wiley Interdiscip. Rev.: Comput. Mol. Sci.}\ ,\ \bibinfo {pages}
  {1364}} (\bibinfo {year} {2018})}\BibitemShut {NoStop}%
\bibitem [{\citenamefont {Booth}, \citenamefont {Smart},\ and\ \citenamefont
  {Alavi}(2014)}]{Booth2014}%
  \BibitemOpen
  \bibfield  {author} {\bibinfo {author} {\bibfnamefont {G.~H.}\ \bibnamefont
  {Booth}}, \bibinfo {author} {\bibfnamefont {S.~D.}\ \bibnamefont {Smart}},\
  and\ \bibinfo {author} {\bibfnamefont {A.}~\bibnamefont {Alavi}},\ }\bibfield
   {title} {\enquote {\bibinfo {title} {Linear-scaling and parallelisable
  algorithms for stochastic quantum chemistry},}\ }\href
  {https://doi.org/10.1080/00268976.2013.877165} {\bibfield  {journal}
  {\bibinfo  {journal} {Mol. Phys.}\ }\textbf {\bibinfo {volume} {112}},\
  \bibinfo {pages} {1855--1869} (\bibinfo {year} {2014})}\BibitemShut {NoStop}%
\bibitem [{\citenamefont {Umrigar}, \citenamefont {Nightingale},\ and\
  \citenamefont {Runge}(1993)}]{Umrigar1993}%
  \BibitemOpen
  \bibfield  {author} {\bibinfo {author} {\bibfnamefont {C.~J.}\ \bibnamefont
  {Umrigar}}, \bibinfo {author} {\bibfnamefont {M.~P.}\ \bibnamefont
  {Nightingale}},\ and\ \bibinfo {author} {\bibfnamefont {K.~J.}\ \bibnamefont
  {Runge}},\ }\bibfield  {title} {\enquote {\bibinfo {title} {A diffusion monte
  carlo algorithm with very small time-step errors},}\ }\href
  {https://doi.org/10.1063/1.465195} {\bibfield  {journal} {\bibinfo  {journal}
  {J. Chem. Phys.}\ }\textbf {\bibinfo {volume} {99}},\ \bibinfo {pages}
  {2865--2890} (\bibinfo {year} {1993})}\BibitemShut {NoStop}%
\end{thebibliography}%

\end{document}